%% file: multipatch_1.tex
\documentclass[prd, aps, nofootinbib, preprintnumbers, showpacs, superscriptaddress, twocolumn, floatfix]{revtex4}

\usepackage{amssymb}
\usepackage[english]{babel}
\usepackage{graphicx}
\usepackage{ifpdf}
\usepackage[latin9]{inputenc}
\usepackage{mathpazo}
\usepackage{mathrsfs}
\usepackage{float}
\usepackage{psfrag}
\usepackage{color}

\ifpdf
\usepackage{hyperref}
\fi

\newcommand{\be}{\begin{equation}}
\newcommand{\ee}{\end{equation}}

\begin{document}

\title{Multi-patch methods in general relativistic astrophysics --
I. Hydrodynamical flows on fixed backgrounds}

\author{Burkhard~Zink} 
\affiliation{Center for Computation and
Technology, Louisiana State University, Baton Rouge, LA 70803, USA}
\affiliation{Department of Physics and Astronomy,
  Louisiana State University, Baton Rouge, LA 70803, USA}

\author{Erik~Schnetter}   
\affiliation{Center for Computation and
Technology, Louisiana State University, Baton Rouge, LA 70803, USA}
\affiliation{Department of Physics and Astronomy,
  Louisiana State University, Baton Rouge, LA 70803, USA}

\author{Manuel~Tiglio}   
\affiliation{Center for Computation and
Technology, Louisiana State University, Baton Rouge, LA 70803, USA}
\affiliation{Department of Physics and Astronomy,
  Louisiana State University, Baton Rouge, LA 70803, USA}
\affiliation{Department of Physics, University of Maryland, College Park MD
  20742, USA}
\affiliation{Center for Scientific Computation and Mathematical Modeling,
  University of Maryland, College Park MD 20742, USA}

\begin{abstract}

Many systems of interest in general relativistic astrophysics, including
neutron stars, accreting compact objects in X-ray binaries and active 
galactic nuclei, core collapse, and collapsars, are assumed to be approximately
spherically symmetric or axisymmetric. In Newtonian
or fixed-background relativistic approximations it is common practice to 
use spherical polar coordinates for computational grids; however, these
coordinates have singularities and are difficult to use in \emph{fully relativistic}
models. We present, in this series of papers, a numerical technique 
 which is able to 
use \emph{effectively spherical} grids by employing multiple patches. We provide detailed
instructions on how to implement such a scheme, and present a number of code tests for
the fixed background case, including an accretion torus around a black hole.

\end{abstract}


\maketitle

\input{introduction}

\input{theory}

\input{implementation}

\input{grids}

\input{results}

\input{conclusions}

%%%%%%%%%%%%%%%%%%%%%%%%%%%%%%%%%%%%%%%%%%%%%%%%%%%%%%%%%%%%%%%%%%%%
\begin{acknowledgments}
%%%%%%%%%%%%%%%%%%%%%%%%%%%%%%%%%%%%%%%%%%%%%%%%%%%%%%%%%%%%%%%%%%%%

  We greatly enjoyed the inspiring discussions we had with M. Anderson, 
  P. Diener, H. Dimmelmeier, N. Dorband, J. Frank, I. Hawke, K. Kifonidis, L. Lehner, M. Megevand, 
  E. M{\"u}ller, P. Motl, M. Obergaulinger, I. Olabarrieta, C. D. Ott, C. Palenzuela-Luque, J. Pullin, 
  E. Seidel, N. Stergioulas, and J. Tohline.
  This research was supported in part by the National Science
  Foundation through grant PHY 0505761 to Louisiana State University
  and Teragrid allocation TG-MCA02N014.
  This research also employed resources of the Center for Computation
  \& Technology at LSU, which is supported by funding from the
  Louisiana legislature's Information Technology Initiative, and of
  the Louisiana Optical Network Initiative.
  We used the computational resources of Abe and Tungsten at the NCSA,
  and the workstation machines at CCT.

\end{acknowledgments}

%%%%%%%%%%%%%%%%%%%%%%%%%%%%%%%%%%%%%%%%%%%%%%%%%%%%%%%%%%%%%%%%%%%%%%%%%%%%%%%%

% Link this file from the bibtex directory
\bibliographystyle{apsrev-nourl}

% Link this file from the bibtex directory
%\bibliography{references}

\end{document}

%% file: introduction.tex
%%%%%%%%%%%%%%%%%%%%%%%%%%%%%%%%%%%%%%%%%%%%%%%%%%%%%%%%
\section{Introduction}
\label{sec:intro}
%%%%%%%%%%%%%%%%%%%%%%%%%%%%%%%%%%%%%%%%%%%%%%%%%%%%%%%

Numerical simulations have become one of our most valuable tools in
building and refining models of compact astrophysical objects and their 
environments, which are commonly associated with high-energy events like
gamma-ray bursts, active galactic nuclei and jets, X-ray binaries and
supernovae. These rather exotic systems form one of the arguably
most exciting physical laboratories we know of, where general relativity,
nuclear physics, transport of radiation, magnetohydrodynamics
(and so on) contribute to their dynamical properties.

The rapid advance of computing performance has made it possible to
simulate increasingly sophisticated problems.
\footnote{For some recent results in general relativistic flow models
see \cite{Zink2005a, Zink2006a, Ott06b, Shibata07a, Shibata06d,
Stephens07a, duez:031101, Dimmelmeier07a, Baiotti06b, Baiotti07,
Montero07a, Anderson:2007a, Oechslin07a}, though this list is far
from being complete.}
But even with
current high-performance supercomputers building general relativistic three-dimensional models 
 poses a formidable challenge, since
the complexity of solving an hyperbolic problem of dimension $n$ scales 
with the linear spatial resolution $h$ like $O(1/h)^{n+1}$. It is
therefore imperative to investigate the application of advanced, efficient numerical
techniques in astrophysical models.

This series of papers will focus on a particular approach in which
the computational domain is understood as a manifold covered by
several distinct coordinate maps called \emph{patches}.
Above all else, this approach admits to cover spheres with smooth,
and in particular singularity-free, coordinates, and also allows
to employ many of the advantages of spherical polar grids, like
the decoupling of radial and angular resolution and intrinsically
spherical domain boundaries, without sharing their major disadvantages.

Multi-patch techniques have been used in different contexts in
general relativity (see, for example,
\cite{Thornburg87,Thornburg93,Gomez97,Gomez98a,Bonazzola-etal-1998:spectral-methods-in-gr-astrophysics,
  Kidder01a,Gourgoulhon-etal-2000:2ns-initial-data,Grandclement-etal-2000:multi-domain-spectral-method,Pfeiffer:2002wt, 
Thornburg2004:multipatch-BH-excision,Calabrese2004:boosted-bh,Scheel-etal-2006:dual-frame,Pfeiffer:2007}
and references therein). The multi-patch
infrastructure used for this paper \cite{Schnetter06a} admits both  
overlapping patches and abutting ones (also called {\em blocks}), with the
latter typically used when solving systems of equations with smooth solutions. In
those cases  the inter-block boundary
information  is transported by the \emph{simultaneous approximation technique} 
\cite{Carpenter-etal-1999:high-order-multiple-patch-FD}, combined with 
high-order finite difference operators satisfying an algebraic property called \emph{summation by parts}
and associated dissipation operators. With these techniques  
 the evolution system at the block-interfaces is decomposed into its characteristic
structure and any numerical mismatch between modes at these interfaces is
dissipated away in time and/or with increasing resolution in a numerically
stable way (see \cite{Lehner2005a,Diener05b1} for more details).  
It has been possible to perform very accurate simulations of scalar fields
on a Kerr background \cite{Dorband05a} and fully non-linear simulations of
distorted black holes \cite{Pazos2006} with this approach.

As mentioned, these multi-patch techniques are designed for systems with smooth solutions,
as is usually the case when solving the Einstein equations in vacuum. 
To model a larger class of astrophysical objects of interest one needs to deal
with the presence of matter and the development of shocks in multi-patch
simulations, which is precisely the goal of this series of papers. 

Here we treat \emph{hydrodynamical flows} in the so-called
 test-fluid approximation, i.e., on a specified spacetime geometry. While
this has interesting applications of its own, like the dynamics of 
non-self gravitating accretion disk models around black holes \cite{Hawley84,
Hawley84a, DeVilliers03, Montero07a}
and the normal mode spectrum of isolated neutron stars \cite{Font01},
we will focus here on \emph{test cases with well-defined error functions} (either 
Riemann problems or stationary solutions), to demonstrate the ability of the code
to transport shock fronts across interfaces, and to keep stationary
solutions near equilibrium. In later papers in the series we will consider
the technique for the fully coupled evolution system and magnetized flows.

This paper is organized as follows: In Section~\ref{sec:theory} we 
give an overview of the multi-patch setup, the evolution system, and the
discrete techniques we are using. In Section~\ref{sec:implementation},
we give a specific description how to implement such a scheme, which should
be helpful for practitioners who would like to adopt this approach.
In Section~\ref{sec:grids} we discuss
a number of multi-patch systems useful for test and production simulations. In
Section~\ref{sec:results} we present the results from a number of code
tests, including shock tubes, rotating stars and an accretion torus around
a black hole. Finally, in Section~\ref{sec:conclusions}, we summarize our 
results.

%% file: theory.tex
\section{Theory and discrete techniques}
\label{sec:theory}

%%%%%%%%%%%%%%%%%%%%%%%%%%%%%%%%%%%%%%%%%%%%%%%%%%%%%%%%
\subsection{The multi-patch setup in general relativity}
\label{sec:multipatch-setu}
%%%%%%%%%%%%%%%%%%%%%%%%%%%%%%%%%%%%%%%%%%%%%%%%%%%%%%%
In general relativity (GR) one starts with a spacetime manifold $(M, g)$, where
$M$ denotes the set of events and $g$ the metric tensor field. Since the manifold
is endowed with a differentiable structure $D$, an obvious choice for a continuum 
multi-patch model is, given an atlas $[A] \in D$, a subset $A^0 \subseteq A$ 
of charts $\phi: M \rightarrow \mathbb{R}^4$ which cover the domain of
interest (see,
e.g., \cite{Wald84}). 

In our computational setup we will have, for practical reasons, an additional
element. We will assume that we are interested in solving our system of equations in a 
spacetime region $M_0 \subseteq M$ which can be covered by a \emph{single}
chart $\phi_G$. That is, we will assume that there is a  \emph{global coordinate system}. This requirement is
not fundamental to using multi-patch techniques, but makes it easier to
set up initial data, visualize results,
and transport information between local patches, as
discussed below.  If needed, the assumption of
such global system can be eliminated without any of the techniques of this
paper changing. On the other hand, whenever we assume the existence of such
global system, for definiteness we will typically choose it to be of
``Cartesian'' type. 

\begin{figure}
\psfrag{M}{$(M,D,g)$}
\psfrag{M0}{$M_0$}
\psfrag{PhiG}{$\phi_G$}
\psfrag{R4DL}{$(\mathbb{R}^4,D_L)$}
\psfrag{R4}{$\mathbb{R}^4$}
\psfrag{PhiGL1}{$\phi^{(1)}_{GL}$}
\psfrag{PhiGL2}{$\phi^{(2)}_{GL}$}
\psfrag{PhiL1}{$\phi^{(1)}_{GL} \circ \phi_G \equiv \phi^{(1)}_L$}
\psfrag{PhiL2}{$\phi^{(2)}_{GL} \circ \phi_G \equiv \phi^{(2)}_L$}
\includegraphics[width=\columnwidth]{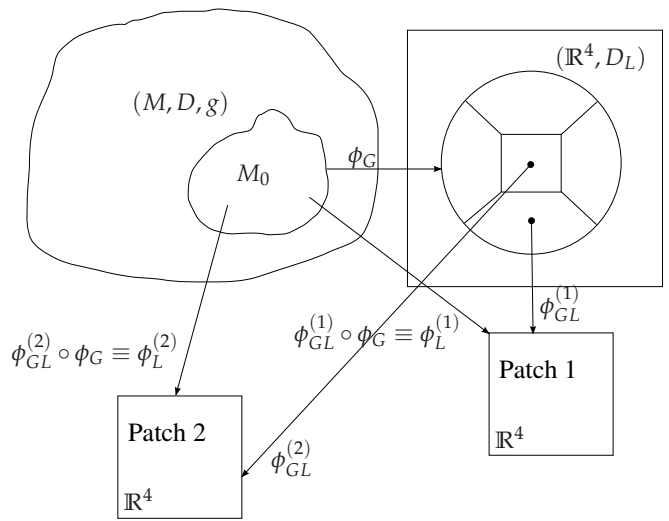}
\caption{The general-relativistic multi-patch model used in this paper. 
The space-time sub-domain $M_0 \subset M$ of interest is assumed to
be covered by a single \emph{global coordinate system} $\phi_G$.
The target domain $\phi_G(M_0) \subset \mathbb{R}^4$ is then further
covered by several patches associated with the charts $\phi^n_{GL}$,
which give rise to the \emph{local coordinate systems} $\phi^n_L$.}
\label{fig:multipatch}
\end{figure}

In differential geometry language, a differentiable
structure $D_L$ can be associated with the global region of interest
$\phi_G(M_0)$. A multi-patch setup, given an atlas $[A_L] \in D_L$,
is then a subset of $A_L$ covering $\phi_G(M_0)$. For any $\phi_{GL} \in A_L$, 
the product chart $\phi_L \equiv \phi_{GL} \circ \phi_G$ defines a \emph{local coordinate system}
on that local patch. Fig.~\ref{fig:multipatch} illustrates this relationship.

A point of some technical relevance is the use of local or global tensor bases.
At each point in $M_0$, tensorial quantities can be written in terms of global
or local coordinates (in differential geometry language, the tangent space
bases associated with $\phi_G$ or $\phi_L$, respectively, can be used). The latter
appears to be the more natural choice and will be employed to represent the
hydrodynamical variables in this paper to enforce conservation when this is
available. A global system is still useful (though not strictly necessary) to
serve as a ``reference'' frame. On the other hand, there is no obvious conservation law for the
metric sector, and therefore there is no advantage in using local
coordinates. In fact, it is easier to use global ones, and therefore when we
solve for the metric we do so for its global components (see, e.g.,
\cite{Schnetter06a, Pazos2006}) . In such case, a standard finite-difference discretization delivers
partial derivatives consistent with $\phi_L$ which are then transformed by the
Jacobians associated with $\phi_{GL}$.

\begin{figure}
\includegraphics[width=\columnwidth]{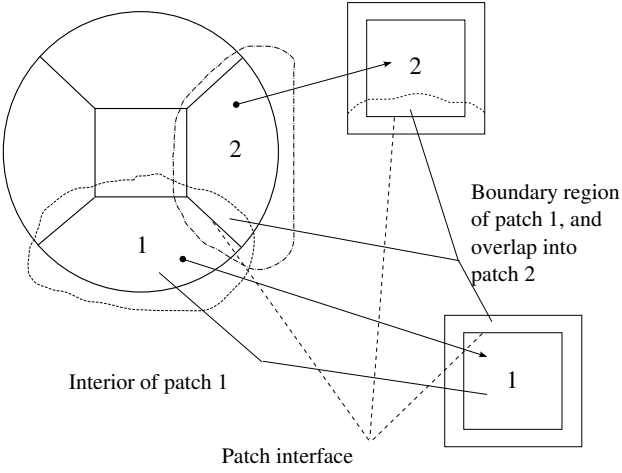}
\caption{The notions of interior, interface and boundary region associated with
a patch. The global coordinate domain $\phi_G(M_0)$ is decomposed into the 
patch interiors as indicated in the figure, with interfaces adjoining two
interiors. The charts $\phi^n_{GL}$ therefore define an interior and a boundary
region in $\phi^n_L (M_0)$. }
\label{fig:multipatch_int_bnd}
\end{figure}

Once the multi-patch system has ben set up, the system of equations to be
 solved can be discretized on each patch and
 two types of boundary conditions appear: those associated with the outer
boundaries of the whole domain $M_0$, and those connecting neighboring
 patches. In our applications each point in $M_0$ will in general be associated with the
\emph{interior} of exactly one patch, and with the boundary region of
 several (two or more) patches (see
 Fig.~\ref{fig:multipatch_int_bnd}). 
Multi-patch techniques can generally make use of the overlap
region to define \emph{ghost zones}, or they use the interfaces to establish
a common surface for the communication of variables. 

%%%%%%%%%%%%%%%%%%%%%%%%%%%%%%%%%%%%%%%%%%%%%%%%%%%%%%%%
\subsection{General relativistic hydrodynamics}
\label{sec:grhd}
%%%%%%%%%%%%%%%%%%%%%%%%%%%%%%%%%%%%%%%%%%%%%%%%%%%%%%%

\subsubsection{Evolution system}
We assume an energy-momentum tensor of 
the usual form \cite{Misner73}

\begin{equation}
T^{ab} = (\rho + u + P) u^a u^b + P g^{ab}
\label{eqn:fluid_tab}
\end{equation}
where $\rho$ is the fluid's rest-frame mass density, $u$ is the internal energy density,
$u^a$ the four-velocity, $P$ the isotropic pressure and $g^{ab}$ the contravariant components of the
spacetime metric. Given a patch with coordinate system $\phi_L$ the conservation laws of mass and energy-momentum

\begin{eqnarray}
\nabla_a (\rho u^a) & = & 0 \\
\nabla_a T^{ab} & = & 0
\end{eqnarray}
are transformed into an evolution system by a 3+1 split (see, e.g., \cite{Gammie03})

\begin{eqnarray}
\partial_t (\sqrt{-g} \rho u^t) + \partial_i (\sqrt{-g} \rho u^i) & = & 0 \\
\partial_t (\sqrt{-g} {T^t}_a) + \partial_i (\sqrt{-g} {T^i}_a) & = & \sqrt{-g} {T^c}_d {\Gamma^d}_{ac}
\label{eqn:grhd}
\end{eqnarray}
where $t$ is the time coordinate, $i$ is a {\em local} space coordinate, $g \equiv det(g_{ab})$
is the determinant of the spacetime metric, and ${\Gamma^d}_{ac}$ are the Christoffel symbols 
associated with it. We therefore promote the expressions $D \equiv \sqrt{-g} \rho u^t$ and
$Q_a \equiv \sqrt{-g} {T^t}_a$ to evolution variables, and collect them into the tuple 
$w = (D,Q_a)$. The equations can then be expressed
in the flux conservative form

\begin{equation}
\partial_t w + \partial_i F^i(v(w)) = s(v(w)).
\label{eqn:conservation_law}
\end{equation}
Here, the tupel $v = (\rho, u, u^x, u^y, u^z, P)$ denotes a choice of 
\emph{primitive variables}, which, in general, are an implicit function of the 
\emph{conserved variables} $w$. The \emph{fluxes} $F^i(v)$ and \emph{sources} $s(v)$ are 
expressed in terms of the primitive variables, which  makes it necessary to
obtain the primitive from the conserved variables at each evolution step.

For calculating the primitive variables, we make use of the ``2D scheme'' from \cite{Noble2006}, since it 
easily lends itself to the inclusion of magnetic fields. The conversion is a non-linear 
root-finding problem, which is being solved using a Newton-Raphson scheme, and
as such is one of the most delicate
parts of the implementation. The Newton-Raphson
scheme only converges given a sufficiently close initial guess: We use the
value of the primitive variables at the last sub-step for this purpose. Also, in 
some cases the conserved variables could potentially obtain values which are not
compatible with any set of physical primitive variables and, finally, the round-off
errors may lead to subtle problems, e.g.\ a Lorentz factor of $\approx 1 - 10^{-16}$. 
Most of this is discussed in \cite{Noble2006} and in the accompanying source code 
examples.

The pressure function $P = P(\rho,u)$ is chosen according to the physical properties of
the fluid, which is in general determined by the equation of state. For purposes 
of this publication, we will assume the pressure to be 
obtained from the \emph{gamma law} $P = (\Gamma - 1) u$.

\subsubsection{Discretization}
\label{sec:discretization}

On any patch, we use a standard finite-volume scheme \cite{Font00} to update
the evolution variables. The computational domain is broken into blocks,
and each grid point is assumed to coincide with the center of a finite cell. The primitive
variables associated with a grid point are naturally interpreted in terms of volume
averages over the cell. The evolution system eqn.~\ref{eqn:conservation_law} can then be integrated
over the cell to yield the weak form of the equations, which represents the update of
the volume averages in terms of fluxes across the cell interfaces and source terms:

\begin{eqnarray}
 & & \partial_t \int_{(a_1,b_1,c_1)}^{(a_2,b_2,c_2)} w \, da \, db \, dc \\ 
  & = & -\int_{(b_1,c_1)}^{(b_2,c_2)} \left( F^1(v)|_{a_2} - F^1(v)|_{a_1} \right) \, db \, dc \nonumber \\
  & & -\int_{(a_1,c_1)}^{(a_2,c_2)} \left( F^2(v)|_{b_2} - F^2(v)|_{b_1} \right) \, da \, dc \nonumber \\
  & & -\int_{(a_1,b_1)}^{(a_2,b_2)} \left( F^3(v)|_{c_2} - F^3(v)|_{c_1} \right) \, da \, db \nonumber \\
  & & + \int_{(a_1,b_1,c_1)}^{(a_2,b_2,c_2)} s(v) \, da \, db \, dc \nonumber
\end{eqnarray}

The fluxes across the cell interfaces are obtained in two steps: First, the primitive
variables are extrapolated to their (left and right) interface values using a
\emph{reconstruction} algorithm, and then the thus-defined local Riemann problem
is solved with an approximate method. For purposes of reconstruction, we use the
MC (monotonized central) algorithm. If $u_i$ is the value of the primitive 
variables at zone $i$, we define the reconstructed quantities $v_L$ and $v_R$
at the interface location $i+\frac{1}{2}$ as \cite{vanLeer77}

\begin{eqnarray}
v_L & \equiv & v_i + \bar{\Delta}_i / 2 \\
v_R & \equiv & v_{i+1} - \bar{\Delta}_{i+1} / 2 \nonumber \\
\bar{\Delta}_i & \equiv & \left\{
  \begin{array}{ll}
    \mathrm{sgn}(\Delta_{i+1}) \min(2|\Delta_i|, 2|\Delta_{i+1}|, & \\ 
      \; |\Delta_i + \Delta_{i+1}|/2) 
      & \Delta_i \Delta_{i+1} \geq 0 \\
    0 & \textrm{otherwise}
  \end{array} \right. \nonumber \\
\Delta_i & \equiv & v_i - v_{i-1} \nonumber.
\label{eqn:reconstruction_mc}
\end{eqnarray}

The resulting local Riemann problem defined by the states $v_L$ and $v_R$ is approximated
using the HLL (Harten, Lax, van Leer) flux formula \cite{Harten83}

\begin{eqnarray}
F^i & \equiv & (c_{min} F^i(v_R) + c_{max} F^i(v_L) \\
  & & - c_{min} c_{max} (w_R - w_L)) (c_{min} c_{max})^{-1} \nonumber \\
c_{min} & \equiv & -\min(0, c^-_R, c^-_L) \nonumber \\
c_{max} & \equiv & \max(0, c^+_R, c^+_L) \nonumber
\end{eqnarray}.

Here $c^-$ and $c^+$ are the minimal and maximal characteristic speeds associated with
the variables $v_{L,R}$, and $w_{L,R}$ are the conserved variables obtained from 
$v_{L,R}$. This flux formula has the advantage that it does not require the full
characteristic decomposition of eqn. \ref{eqn:grhd}, and is thus well-suited to
be extended to the more complicated case involving magnetic fields \cite{Anton05}.
The characteristic speeds are obtained from the rest-frame sound speed of the fluid
by a Lorentz transformation to the local coordinate frame \cite{Duez05MHD0}.

A particular difficulty in solving general relativistic hydrodynamics problems
are regions of very low density \cite{Baiotti03a, Duez:2002bn}, e.g.\ those outside
a neutron star. In those regions, we use an artificial atmosphere of low density
to make the problem tractable. This atmosphere effectively acts as a boundary
condition on the stellar material.

\subsubsection{Boundary treatment}
\label{sec:boundaries}

\begin{figure}
\includegraphics[width=\columnwidth]{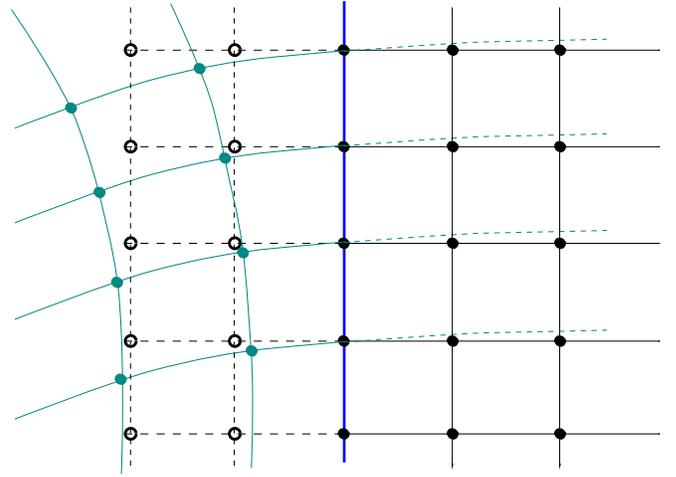}
\caption{Illustration of the patch interface boundary treatment. The interface, here
represented by the vertical thick blue line in the center, separates patch L (\emph{left})
and patch R (\emph{right}). The coordinate lines of both patches are represented in the local
coordinates of patch R, which makes the lines of L appear curved in general. To evolve
the system on patch R, \emph{boundary ghost zones}, here represented by non-filled circles,
are introduced as a linear extrapolation
of patch R's coordinate lines, and, before the time update is performed, receive data
from a data interpolation operation on patch L. Afterwards, the data is transformed
to the tensor basis defined by patch R's coordinate system via the transformation map.}
\label{fig:interface}
\end{figure}

To use an unmodified finite-volume scheme also at the boundaries of the interior
patch domain, we follow the same approach typically used in adaptive
mesh refinement implementations: we introduce \emph{ghost zones} in the boundary region which are updated
using an interpolation scheme. The results in this paper have been obtained with a first-order 
operator, which does not introduce new extrema and is compatible to the (at most) second-order 
accurate MC reconstruction technique. The patch interface boundary treatment is illustrated 
in Fig.~\ref{fig:interface}.

Since the patches have different local coordinate systems, we need to use the transformation maps
between the patches to obtain the local representation of interpolated quantities. Internally,
first a transformation to the global coordinate system is performed, followed by a separate
transformation to the appropriate local system.  For the case of general relativistic hydrodynamics 
on specified backgrounds, we interpolate the
set of primitive variables for purposes of interface reconstruction. These quantities are components
of tensors and therefore subject to trivial transformation laws \cite{Wald84}.

The decomposition of each patch into sub-domains, for purposes of implementing a distributed
computing model, introduces additional boundaries which are treated by introducing ghost zones.
No interpolation or coordinate transformation is needed in this case, and the synchronization
operation copies boundary data between these sub-domains.

The outer boundaries of the computational domain are also handled with ghost zones. For purposes
of imposing an outflow boundary condition, we copy data from the first cell inside the
computational domain to the ghost zones \cite{Hawke04}, though we will typically prefer
to select a grid setup where material ejected from it is entirely contained on the grid during 
the course of the evolution. In this case, the artificial atmosphere is the effective boundary
for the fluid. If material should leave the computational domain, the exact form of the boundary
condition is important only in those cases where there is a significant back-reaction of the
material on the central object, or other regions of interest.

\subsubsection{Setup of initial data}

For convenience, initial data is always expressed first in terms of the 
\emph{global} coordinate system, and mapped to the patch locations in this basis.
Afterwards, the coordinate transformation to the appropriate local basis is performed
if the quantity is not a scalar field. All initial data specifications in this paper
will therefore also be written in an appropriate global system.

\subsection{Computational framework}

The code has been implemented as a module into the {\tt Cactus} computational framework 
\cite{Goodale02a, url:cactus}, and using the {\tt Carpet} 
driver \cite{Schnetter-etal-03b, url:carpetcode} to administrate the multi-patch infrastructure.
In our implementation (as stated above), the {\tt Carpet} driver reserves a number of 
\emph{boundary ghost zones} beyond each patch boundary, which are then filled using interpolation
and tensor transformation. For a first order interpolation scheme, each boundary ghost zone
is located in exactly one cell in the interior of another patch. However, if the scheme
was to be extended to higher order interpolation operators, a non-central stencil
would need to be applied.

\subsection{Limitations of our approach}

In this section, we collect the most important known limitations of the
approach followed in this paper.

The most obvious limitations are clearly those related to the physical model:
the assumption of a fixed spacetime neglects motions of the gravitational 
field and therefore suppresses all gravitational instabilities, and is only
strictly justified in code tests involving equilibrium systems. Our fluid model,
besides making the assumptions common for all hydrodynamical approximations,
models \emph{one} fluid component without tangential stresses, and our
choice of the pressure relation neglects microphysical properties. This paper
does not include magnetic fields either, although we will turn to full GRMHD
in a subsequent publication.

On the numerical side, we have a number of restrictions.  While many of 
these are common to many approaches in the field,  we find it useful
to give a detailed list here:
\begin{itemize}
\item The scheme that we use is at best second order accurate. 
In particular, it
drops to local first order near maxima and stellar surfaces. The
global convergence order might be less than two in practical applications. This
approach is very robust, but may be too inaccurate for very long term
simulations (hundreds of dynamical timescales), or turbulent phenomena
like the MRI \cite{Balbus91a}. 
There are higher order schemes available \cite{Tchekhovskoy07},
and we are evaluating using such techniques.\footnote{Higher order reconstruction
operators like PPM \cite{Woodward84} do not increase the overall accuracy
but may decrease local error levels. We do have a simplified implementation of
PPM in our code (without flattening), but we did not find it superior in the particular test
cases discussed here.}
\item The inter-patch boundary treatment does not trivially lend itself
to higher order implementations, since we need a non-oscillatory higher
order interpolation operator to fill the boundary ghost zones. A compromise
would be to use Lagrangian operators, but drop to trilinear interpolation
if unphysical variables are produced, or use an ENO scheme.
\item The outer boundary conditions are only approximate, and may lead to 
unphysical variables (see \cite{Gammie03} for an improved scheme with 
densitized variables). This may have practical and fundamental consequences.
On a practical level, undesirable artifacts may appear at the boundaries, 
though in stationary systems, those are only related to artefactual flows
leaving the compact support of the equilibrium object. However, if the
system does show physical outflows across the boundary, those will be
modeled only approximately. For supersonic or even causally disconnected 
flows, however, this restriction may not be dynamically relevant
\item The artificial atmosphere is used to employ a unified computational scheme.
However, to produce an exact discrete representation of an equilibrium star
or disk, 
the discrete scheme would need to be modified
near the surface. The addition of atmospheric fluid usually
introduces error levels which are significantly smaller than errors from
the internal dynamics of the star in real applications, but for equilibrium
problems, which may be dominated by the surface errors, the code may not
show convergence to the correct solution even when reducing the atmospheric
density. 
\item The transformation from conserved to primitive variables works well
in most cases, but can be a cause of considerable difficulty in some situations.
This problem may be more severe when using a tabulated equation of state and
coupled magnetic fields; in
fact, the technique we currently use may need a considerable extension to
also operate in these cases \cite{Mignone07a}.
\item The multi-patch setup, for all its advantages mentioned above, 
requires a somewhat sophisticated software infrastructure, and a
certain reduction in computational efficiency is
introduced by the boundary communication.
However, a multi-patch setup scales more favourably when compared to
Cartesian mesh refinement when one considers increases in radial
resolution or an increase of the computational domain.  The
computational cost of a multi-patch simulation scales as $O(N^2)$ and
$O(N)$, respectively, in these two cases, while a mesh refinement
simulation scales as $O(N^4)$ and $O(N^3)$, respectively.  The
difference comes from the fact that the radial resolution remains
unchanged in these cases in a multi-patch setup, which is not possible
in a Cartesian mesh-refinement simulation.
\item The patch interface grid points from both sides must match, i.e.,
the grid topology introduces constraints on the possible choices of 
patch cell numbers. While this could be relaxed for the hydrodynamical
scheme alone, the SBP/penalty techniques used in the generalized harmonic code 
for solving Einstein's equations \cite{Schnetter06a} require these
constraints, and we are ultimately interested in the full evolution
system. 
\end{itemize}

%% file: implementation.tex
%%%%%%%%%%%%%%%%%%%%%%%%%%%%%%%%%%%%%%%%%%%%%%%%%%%%%%%%
\section{Implementing a multi-patch scheme}
\label{sec:implementation}
%%%%%%%%%%%%%%%%%%%%%%%%%%%%%%%%%%%%%%%%%%%%%%%%%%%%%%%

In this section we describe how to implement a multi-patch 
scheme, either from scratch or by extending an existing unigrid code. 
Multi-patch techniques 
can be useful for any
numerical code which evolves systems with a certain symmetry (for example,  single stars
or stars with accretion disks), mostly because the grid is adapted to the problem,
and the angular and radial resolutions are decoupled. Because of these and
other advantages,   
this section should be of interest to practitioners in the field of computational
astrophysics, whether Newtonian or relativistic.

What amount of work can one expect, and what benefits can be gained in practical 
terms? We will answer these questions in comparison to the two main alternatives to multi-patch grids:
\begin{itemize}

\item In comparison to \emph{spherical polar grids}, the multi-patch technique
avoids the need to exclude the axis of symmetry and impose artificial boundary
conditions there. This is important, in particular, for systems where outflows
are generated and collimated on or near the axis. Also, the typical
finite-difference or finite-volume Courant limitations to the
time step near the poles of each surface $r = const$ are avoided. Three capabilities
need to be added: geometric terms which are not hard-coded into the equations,
several grids instead of one, and a linear interpolation and transformation at the boundaries
(see below). The finite volume scheme, Riemann solver, and local physics are unaffected.

\item In comparison to \emph{mesh refinement}, the multi-patch technique offers decoupled
radial and angular resolution, which is of particular use far away from the central object
(e.g.\ to extract gravitational radiation at large distances). However, mesh refinement is superior for processes
which do not have explicit approximate symmetries, e.g.\ violent instabilities or binary mergers.
A mesh refinement code can already handle several grids and boundary interpolation, so 
the only capabilities needed in addition are the coordinate transformations at the boundary
and the geometric terms for the local coordinates. The finite volume scheme, Riemann solver, 
and local physics are again unaffected.
\end{itemize}

Given a certain finite volume/finite difference code, these steps need to be 
performed:
\begin{itemize}

\item The code needs to be able to handle several separate grids. These grids are all logically
Cartesian and independent, so in many cases this might just be an additional loop statement. 
The local evolution scheme (e.g.\ flux calculation, finite differences, update) is logically 
unaffected.

\item Each grid cell needs to store (or calculate during run-time) its location in global
coordinates, typically Cartesian $x^i = (x,y,z)$, and in local coordinates, called
$a^i = (a,b,c)$ below. In addition, the Jacobian ${J^i}_j = \partial x^i/\partial a^j$ and its
inverse are needed; these can be obtained from the equations in Section~\ref{sec:grids}.

\item The patches need to use local coordinates in a certain range; say $[-1,+1]$ for 
the interior, with $a^i=\pm 1$ on the boundary, and a number of grid points extruding beyond
the boundary for setting interpolation data. This depends on the stencil; for the monotonized
central scheme used here, we use two ghost cells.

\item Before each time update, and after the conversion from conserved to primitive variables,
the boundary ghost cells need to be set by a (for example, trilinear) interpolation operation. For the patch
systems presented in Section~\ref{sec:grids}, it is known, for each patch, which other patch
needs to be interpolated on,\footnote{In general, this is not the case. Given the location
of the ghost point in global coordinates, a generic approach is then to transform the point
to \emph{all} local coordinate systems, and check whether it is in the appropriate interior
range, say $[-1,+1]^3$.  We perform this step once during initialization.} so a simple list of six entries for each patch is enough to store
the information. Given the target patch, we transform the location of the ghost point into
\emph{local coordinates on the target patch} using the equations in Section~\ref{sec:grids},
and use traditional trilinear interpolation to get the data. 

\item Although boundary data has now been set, it is still expressed in the coordinate system
of the target patch. Say, for relativistic hydrodynamics, we are interpolating boundary data 
for patch 4, and the data is coming from patch 5. After the interpolation, patch 4 has one side
of its boundary stencil (the side adjoining to patch 5) filled with interpolated values for
$(\rho,u,u_{(5)}^a,u_{(5)}^b,u_{(5)}^c)$. As indicated, the 3-velocity components are still
expressed in the local coordinate system of patch 5. Therefore, a simple transition map 
$u_{(4)}^i = (\partial a_{(4)}^i / \partial x^j) (\partial x^j / \partial a_{(5)}^k)  u_{(5)}^k$ 
is applied at each ghost cell, where $(\partial x^j / \partial a_{(5)}^k)$ is the Jacobian
local $\rightarrow$ global to transform $u_{(5)}^k$ to global coordinates, and
$(\partial a_{(4)}^i / \partial x^j)$ is the Jacobian global $\rightarrow$ local on patch 4.

\item The numerical scheme needs to be able to work on a general background metric.
For relativistic codes that is already contained in the covariant form, but a Newtonian
code requires an addition of the geometrical terms.

\item It is convenient to describe the initial data in terms of the global Cartesian
coordinate system. That is, at each grid point all tensorial data can be represented in global
coordinates $(x,y,z)$ and  afterwards transformed to the patch-local coordinate systems
using the standard tensor transformation laws. In general relativistic hydrodynamics on fixed backgrounds, 
we need to transform the 3-velocity $u_{(G)}^i$ and the 4-metric $g^{(G)}_{\mu\nu}$ from global into
local coordinates on patch $m$: $u_{(m)}^i = (\partial a_{(m)}^i / \partial x^j) u_{(G)}^j$,
and $g^{(m)}_{0i} = (\partial x^j / \partial a_{(m)}^i) g^{(G)}_{0j}$, 
$g^{(m)}_{ij} = (\partial x^k / \partial a_{(m)}^i) (\partial x^l / \partial a_{(m)}^j) 
g^{(G)}_{kl}$.

\item To parallelize the code, a simple domain decomposition technique can be used for each
patch. In addition, it may be useful to have each patch distributed to as few processes
as possible. For example, for a setup with $N$ patches 
one might want to use $N\times M$ processes and assign each patch onto
$M$ processes, in order to reduce communication overhead.\footnote{Our
load distribution scheme is in fact more complex and can also
efficiently handle patches of different sizes.}
\end{itemize}

%% file: grids.tex
%%%%%%%%%%%%%%%%%%%%%%%%%%%%%%%%%%%%%%%%%%%%%%%%%%%%%%%%
\section{Multi-patch systems}
\label{sec:grids}
%%%%%%%%%%%%%%%%%%%%%%%%%%%%%%%%%%%%%%%%%%%%%%%%%%%%%%%

In this section we specify the particular set of multi-patch setups used
in this paper. A general introduction to the kind of systems useful for
single stars or  black holes can also be found in \cite{Lehner2005a, Diener05b1}.

For convenience and clarity, we will denote those 3-coordinates associated 
with the global coordinate system as $x^i, i = 1 \ldots 3$ or $(x,y,z)$, and those 
associated with a particular local coordinate system as $a^i, i = 1 \ldots 3$.
If necessary, we indicate the particular patch $p$ with a subscript, as in
$a_p^i$.

\subsection{The uni-patch system}
\label{sec:unipatch}

We refer to a setup with one patch and a coordinate transformation with
a constant diagonal Jacobian of the form 

\be
{J^i}_j \equiv \frac{\partial x^i}{\partial a^j} = c \, {\delta^i}_j
\label{eqn:unipatch_jacobian}
\ee
as a \emph{uni-patch} system, where ${\delta^i}_j$ is the Kronecker symbol and $c \in \mathbb{R}$ a 
number. The associated transformation between local and global coordinates 
is the affine map 

\be
x^i = c \, a^i + b^i, i = 1 \ldots 3
\ee
and therefore we only need to specify the scale $c$ and the 3-vector location of
the origin $b^i$.

This patch system most closely matches the unigrid setups used in some
three-dimensional simulations, and since it contains no patch interfaces it is 
used to test the performance of the code in the patch interior.\footnote{It is also
easily possible to make this system toroidal by performing a topological 
identification between some boundaries, though we will not make use of this
option in the examples below.}

\subsection{The distorted two patches system}
\label{sec:two_patches_distorted}

The \emph{distorted two patches} system consists of two patches with coordinates

\begin{eqnarray}
\textrm{Patch 0:} & & x^i = a_1^i, \; i = 1 \ldots 3 \\
\textrm{Patch 1:} & & x^j = -\frac{1}{2} (a_2^j + 1)^2 + 3, \; j \in \mathbb{R} \label{eq:generalized_map} \\
& & x^i = a_2^i, \; i = 1 \ldots 3, \, i \neq j \nonumber.
\end{eqnarray}

\begin{figure}
\includegraphics[width=\columnwidth]{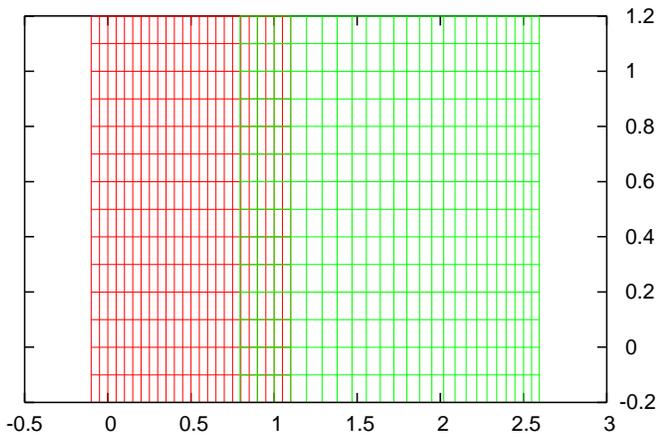}
\caption{An illustration of the distorted two patches system. The left patch is
constructed using an identity map between local and global coordinates, whereas 
the right patch follows from eqn.~\ref{eq:generalized_map}. The common interface
is located at $x=1$, which also coincides with the local coordinates 
$a_1^1 = 1$ and $a_2^1 = 1$. Note the ghost zones extending beyond the interface
(cf. Fig.~\ref{fig:interface}).}
\label{fig:two_patches_distorted}
\end{figure}

The internal coordinate range of each patch is chosen to be $(0,1)^3$, and
the common interface is then located at $x^j = a_1^j = a_2^j = 1$. The second
patch is quadratically distorted and matches the first one at the \emph{upper}
boundary of coordinate $j$. An example with $j=1$ is illustrated in 
Fig.~\ref{fig:two_patches_distorted}.

\subsection{The cubed-sphere six patches system}
\label{sec:six_patches}

Whereas the two multi-patch systems discussed above are useful mostly for purposes
of code testing, the \emph{cubed-sphere six patches} system \cite{Lehner2005a} is a setup 
which can be
directly used to efficiently model black holes and accretion disks around black
holes. The system covers a spherical region with a central sphere cut out (\emph{excised}).
The central excised sphere can be used to exclude part
of the interior of a black hole, since it is causally disconnected from the exterior
region. In a fully relativistic setting, a symmetric hyperbolic formulation of Einstein's
equations with only physical speeds of propagation admits to use the interior spherical
surface as an outflow boundary, since the \emph{continuum} property of causal disconnection
is then also represented on the \emph{discrete} level. We will make use of this fact in 
the coupled evolutions in a later paper in this series.

The domain is covered by a family of spheres with different ``radial'' coordinates
$r = (\sum_i (x^i)^2)^{1/2}$. A simple approach
would be to cover almost the entire sphere by spherical polar coordinates, which, unfortunately, also
introduces coordinate singularities. Therefore, either one opts to cover the sphere using 
at least to patches with properly rotated spherical polar coordinate systems (\emph{yin-yang grid}
\cite{kageyama2004}), or one uses a patch system which does not derive from spherical polar
coordinates. We will use the latter approach here.

The \emph{cubed sphere} patch system on a sphere is conceptually constructed by imagining
a cube con-central with the sphere, introducing the canonical Cartesian restrictions on
each surface of the cube, and projecting these coordinate lines onto the sphere. When mapping
the radial coordinate via the identity, this
produces six patches with coordinate transformations \cite{Lehner2005a}

\begin{eqnarray}
\textrm{Patch 0:} & & x = \frac{G_0}{E_0}, \; y = \frac{G_0 a_0^2}{E_0}, \; z = \frac{G_0 a_0^1}{E_0} \\
\textrm{Patch 1:} & & x = \frac{-G_1 a_1^2}{E_1}, \; y = \frac{G_1}{E_1}, \; z = \frac{G_1 a_1^1}{E_1} \nonumber \\
\textrm{Patch 2:} & & x = \frac{-G_2}{E_2}, \; y = \frac{-G_2 a_2^2}{E_2}, \; z = \frac{G_2 a_2^1}{E_2} \nonumber \\
\textrm{Patch 3:} & & x = \frac{G_3 a_3^2}{E_3}, \; y = \frac{-G_3}{E_3}, \; z = \frac{G_3 a_3^1}{E_3} \nonumber \\
\textrm{Patch 4:} & & x = \frac{-G_4 a_4^1}{E_4}, \; y = \frac{G_4 a_4^2}{E_4}, \; z = \frac{G_4}{E_4} \nonumber \\
\textrm{Patch 5:} & & x = \frac{G_5 a_5^1}{E_5}, \; y = \frac{G_5 a_5^2}{E_5}, \; z = \frac{-G_5}{E_5} \nonumber \\
& & E_i \equiv 1 + (a_i^1)^2 + (a_i^2)^2 \nonumber \\
& & G_i \equiv \frac{1}{2} [ r_0 (1 - a_i^3) + r_1 (1 + a_i^3) ] \nonumber \\
& & a_i^j \in [-1,1] \nonumber
\label{eqn:six_patches}
\end{eqnarray}

\begin{figure}
\includegraphics[width=\columnwidth]{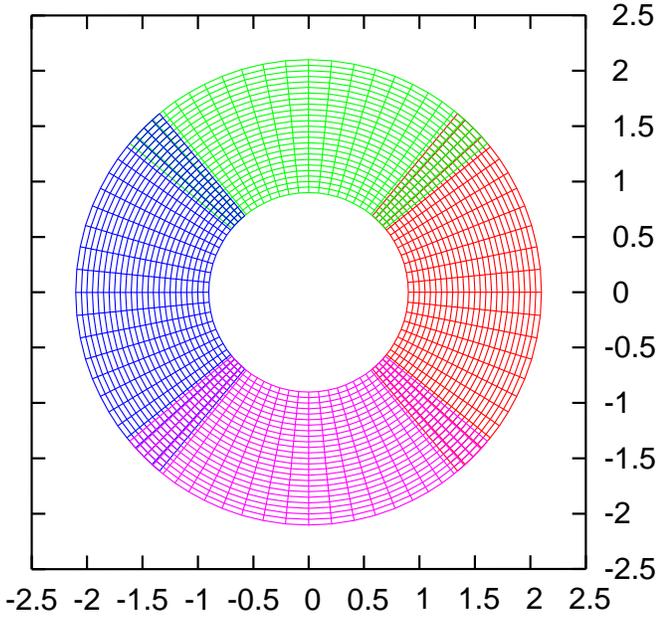}
\caption{An illustration of the cubed-sphere six patches system. The diagram shows
a cut of grid lines with the equatorial plane $z = 0$. In the notation of eqn.~\ref{eqn:six_patches},
patch 0 is to the right (in red), counterclockwise continuing with patch 1 (in green),
patch 2 (in blue) and patch 3 (in violet). The inner spherical boundary can be used
as an \emph{excision} surface for modeling black holes.}
\label{fig:six_patches}
\end{figure}

Here, $r_0$ and $r_1$ are free parameters which determine the location of the two outer boundary
spheres in terms of global coordinates. Patches 0 to 3 are constructed in the equatorial plane counterclockwise
starting from the positive x axis, and patches 4 and 5 intersect with the z axis. An illustration
of this patch system in the equatorial plane $z = 0$ is given in Fig.~\ref{fig:six_patches}.

\subsection{The cubed-sphere seven patches system}
\label{sec:seven_patches}

While the six patches system is well-suited to excise the interior of black holes from the
computational domain, there are also situations (for example, stellar
simulations) where we would like to cover the
entire interior of a sphere with regular coordinates. One approach to achieve this is
to cover the central part of the computational domain with Cartesian coordinates (i.e.,
a uni-patch as described above), describe the outer boundary as a sphere, and decompose
the region between the surface of the Cartesian patch and the outer boundary into six
additional patches \cite{Diener05b1}. Then, on the outer boundary, spherical
coordinates are fixed as in the six patches system, whereas the intermediate surfaces
of constant coordinate $a^3$ (which are spheres in the six patches system) are deformed
in a way to interpolate between a spherical section and a flat face of the central cube.

In detail, the transformations between local and global coordinates which specify
the cubed-sphere seven patches system are as follows (please note that we introduce
the central cube as patch 6 to retain the counting convention used in the six patches
system):

\begin{eqnarray}
\textrm{Patch 0:} & & x = \frac{G_0}{F_0}, \; y = \frac{G_0 a_0^2}{F_0}, \; z = \frac{G_0 a_0^1}{F_0} \\
\textrm{Patch 1:} & & x = \frac{-G_1 a_1^2}{F_1}, \; y = \frac{G_1}{F_1}, \; z = \frac{G_1 a_1^1}{F_1} \nonumber \\
\textrm{Patch 2:} & & x = \frac{-G_2}{F_2}, \; y = \frac{-G_2 a_2^2}{F_2}, \; z = \frac{G_2 a_2^1}{F_2} \nonumber \\
\textrm{Patch 3:} & & x = \frac{G_3 a_3^2}{F_3}, \; y = \frac{-G_3}{F_3}, \; z = \frac{G_3 a_3^1}{F_3} \nonumber \\
\textrm{Patch 4:} & & x = \frac{-G_4 a_4^1}{F_4}, \; y = \frac{G_4 a_4^2}{F_4}, \; z = \frac{G_4}{F_4} \nonumber \\
\textrm{Patch 5:} & & x = \frac{G_5 a_5^1}{F_5}, \; y = \frac{G_5 a_5^2}{F_5}, \; z = \frac{-G_5}{F_5} \nonumber \\
\textrm{Patch 6:} & & x = r_0 a_6^1, \; y = r_0 a_6^2, \; z = r_0 a_6^3 \nonumber \\
& & F_i \equiv \left( \frac{(r_1 - G_i) + (G_i - r_0) E_i}{r_1 - r_0} \right)^{1/2} \nonumber \\
& & E_i \equiv 1 + (a_i^1)^2 + (a_i^2)^2 \nonumber \\
& & G_i \equiv \frac{1}{2} [ r_0 (1 - a_i^3) + r_1 (1 + a_i^3) ] \nonumber \\
& & a_i^j \in [-1,1] \nonumber
\label{eqn:seven_patches}
\end{eqnarray}

\begin{figure}
\includegraphics[width=\columnwidth]{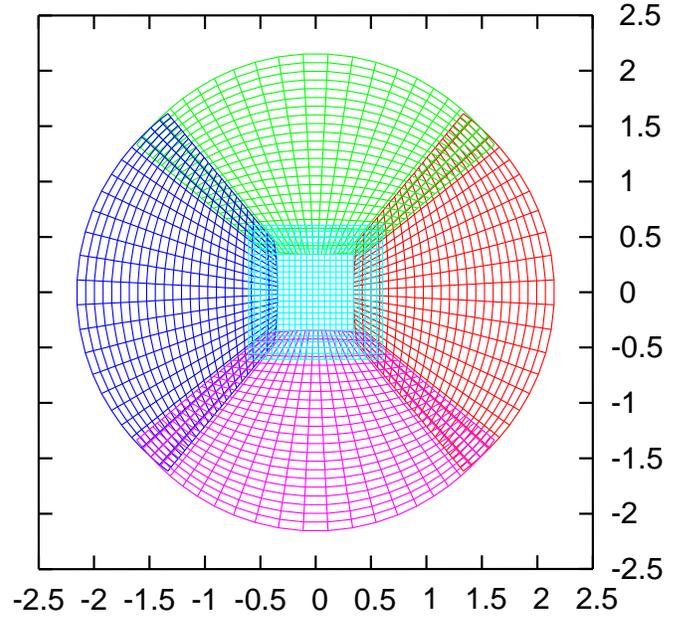}
\caption{An illustration of the cubed-sphere seven patches system. The diagram shows
a cut of grid lines with the equatorial plane $z = 0$. In the notation of eqn.~\ref{eqn:seven_patches},
patch 0 is to the right (in red), counterclockwise continuing with patch 1 (in green),
patch 2 (in blue) and patch 3 (in violet). The central, Cartesian patch is patch 6 (light blue).
The parameters for the patch system were chosen to be $r_0 = 0.5$ and $r_1 = 2$, so the outer
boundary is a sphere of radius 2 in this diagram. The outermost two grid lines, however, are 
ghost zones required for the MC reconstruction algorithm (see Section~\ref{sec:boundaries}).}
\label{fig:seven_patches}
\end{figure}

Here, $r_0$ and $r_1$ are free parameters which determine, in terms of global coordinates,
the location of the outer boundary ($r_1$) and the extent of the central cube ($r_0$). 
An illustration of the cubed-sphere seven patch system in the equatorial plane $z = 0$ 
is given in Fig.~\ref{fig:seven_patches}.

\subsection{The cubed-sphere thirteen patches system}
\label{sec:thirteen_patches}

\begin{figure}
\includegraphics[width=\columnwidth]{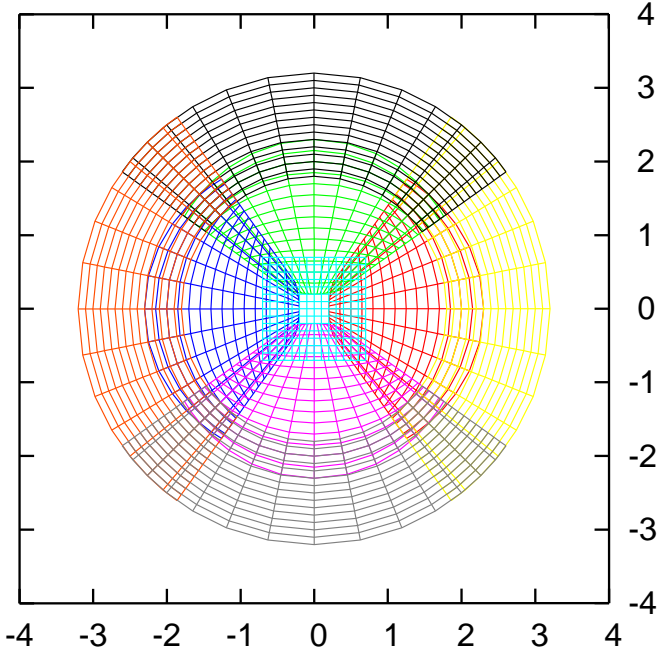}
\caption{An illustration of the cubed-sphere thirteen patches system. This system
consist of an outer six patches system, where all surfaces of constant coordinate
$a^3$ are spheres, and an inner seven patches system matched to the inner boundary
of the outer system to cover the origin in a regular manner.}
\label{fig:thirteen_patches}
\end{figure}

A (minor) disadvantage of the cubed-sphere seven patches system is that, beyond the Cartesian
patch around the origin in global coordinates, all surfaces of constant $a_i^3$
for $i = 0 \ldots 5$ up to the outer boundary at $a_i^3 = 1$ are not spherical, but rather
``intermediate'' between spheres and cube surfaces. To cover a star with spherical surfaces,
a different approach is to combine the advantages of the six and seven patches systems by
covering a large region of the star with a six patches system, and making the origin
regular by introducing a seven patches system in a way that the parameters $r_1$ of 
the seven patches system and $r_0$ of the six patches system match. This construction
can also be used, for example, to  extract gravitational waves without needing
to interpolate to spheres in order to decompose the solution into its different
multipole components. A cut of the coordinate lines with
the equatorial plane $z = 0$ is illustrated in Fig.~\ref{fig:thirteen_patches}

%% file: results.tex
%%%%%%%%%%%%%%%%%%%%%%%%%%%%%%%%%%%%%%%%%%%%%%%%%%%%%%%%
\section{Results}
\label{sec:results}
%%%%%%%%%%%%%%%%%%%%%%%%%%%%%%%%%%%%%%%%%%%%%%%%%%%%%%%

\subsection{Shock tubes on the uni-patch system}

To investigate the code's ability to evolve special relativistic Riemann problems 
without the additional complications of non-trivial Jacobians and inter-patch
interfaces, we perform three kinds of standard tests specified by the fluid
states

\begin{eqnarray}
\textrm{Sod test:} & & \rho_L = 1, \; u_L = 1.5, \; u_L^i = 0 \label{eqn:sod} \\
 & & \rho_R = 0.125, \; u_R = 0.15, \; u_R^i = 0 \nonumber \\
\textrm{``Simple'' test:} & & \rho_L = 10, \; u_L = 20, \; u_L^i = 0 \label{eqn:simple} \\
 & & \rho_R = 1, \; u_R = 10^{-6}, \; u_R^i = 0 \nonumber \\
\textrm{Blast wave test:} & & \rho_L = 1, \; u_L = 1.5 \cdot 10^3, \; u_L^i = 0 \label{eqn:blast} \\
 & & \rho_R = 1, \; u_R = 1.5 \cdot 10^{-2}, \; u_R^i = 0 \nonumber 
\end{eqnarray}

In all cases we assume an equation of state of the form $P = (\Gamma - 1) u$ with
$\Gamma = 4/3$, and the metric has the Cartesian Minkowski form 
$g_{\mu\nu} = \eta_{\mu\nu}$. The Sod test, in particular, will also be used for 
shock tube experiments on several patches below. 

For the uni-patch system, we choose 
the free parameter $c$ in eqn.~\ref{eqn:unipatch_jacobian} to be unity, and 
we restrict attention to a one-dimensional problem where the initial contact
surface is located at $x = 0.5$. The grids, therefore, are only specified by
the number of cells in the x direction, where we choose values from 50 to 800.

To generate a measure of error, we calculate the exact solution
to the the special relativistic problems using the {\tt riemann} code 
\cite{Marti99}. The error function for each primitive variable is simply taken
to be a norm over the difference function between the exact solution and 
the discrete result. The boundaries are set to the exact solution produced 
with {\tt riemann}.

\begin{figure*}
\psfrag{x}{$x$}
\psfrag{rho}{$\rho$}
\psfrag{u}{$u$}
\psfrag{ua}{$u^1$}
\psfrag{t}{$t$}
\psfrag{rhonorm}{\hspace{-0.5cm}${||\rho - \rho_{exact}||}_1$}
\psfrag{unorm}{\hspace{-0.5cm}${||u - u_{exact}||}_1$}
\psfrag{u1norm}{\hspace{-0.5cm}${||u^1 - u^1_{exact}||}_1$}
\begin{tabular}{cc}
\includegraphics[width=\columnwidth]{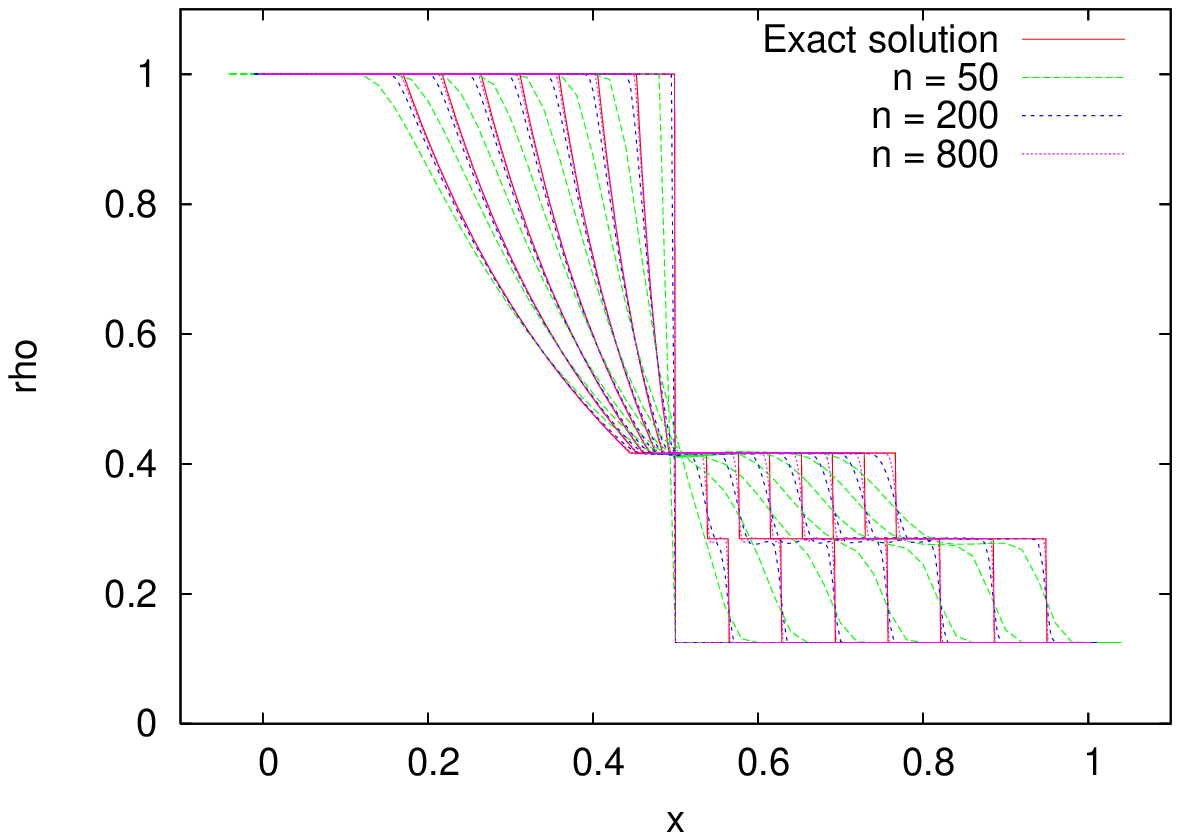} & 
\includegraphics[width=\columnwidth]{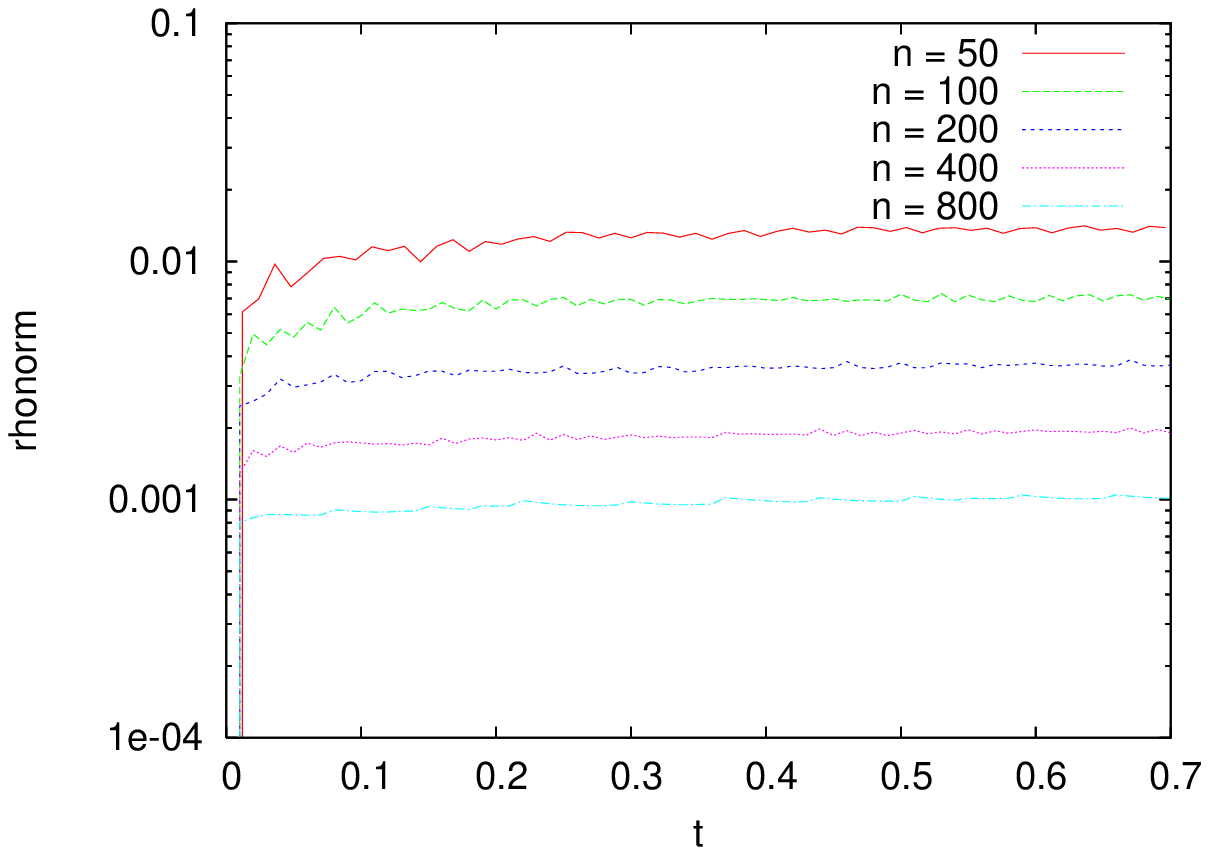} \\ 
\includegraphics[width=\columnwidth]{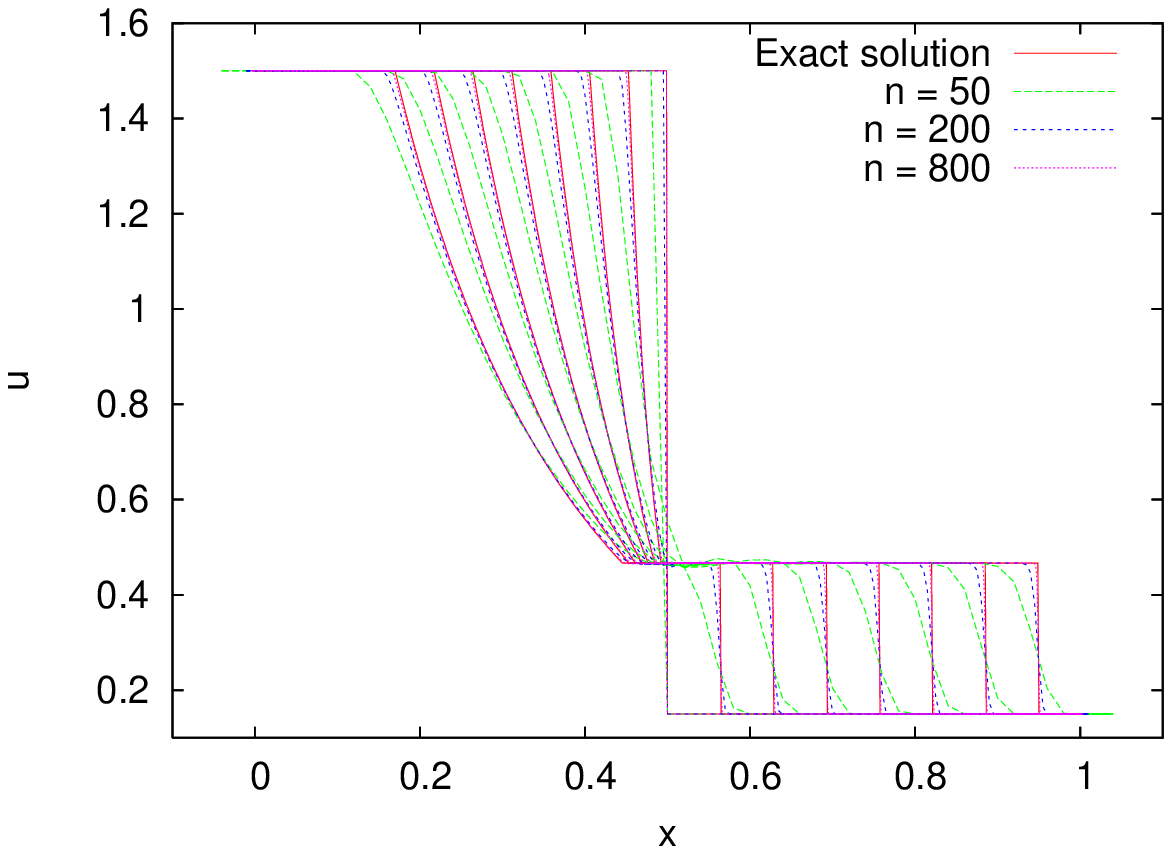} &
\includegraphics[width=\columnwidth]{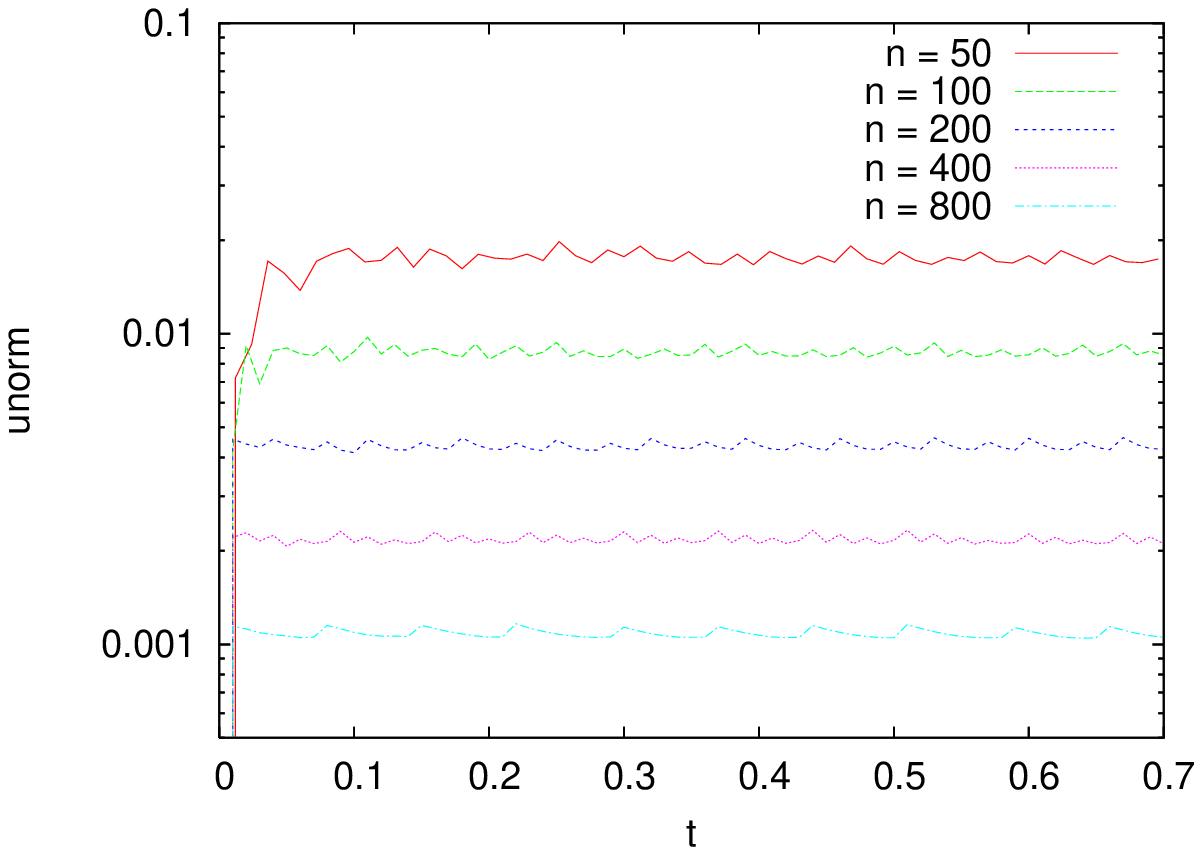} \\
\includegraphics[width=\columnwidth]{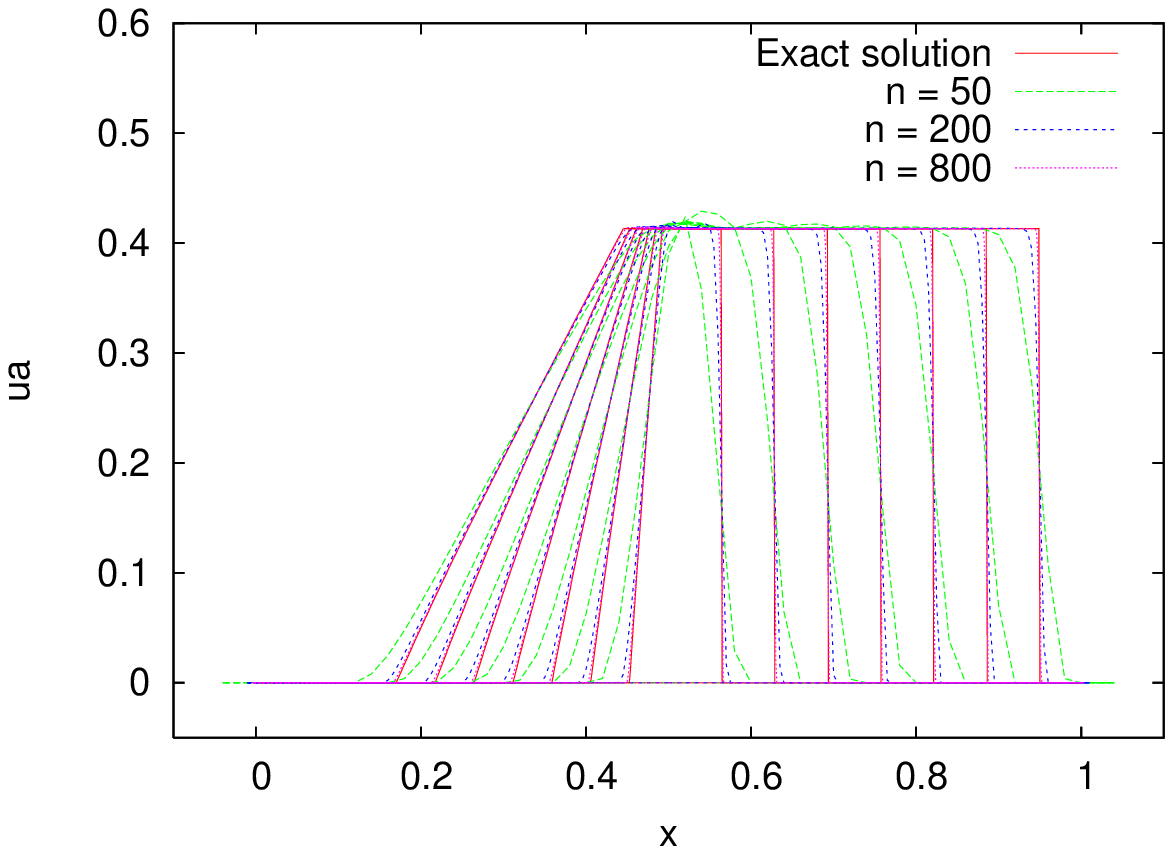} &
\includegraphics[width=\columnwidth]{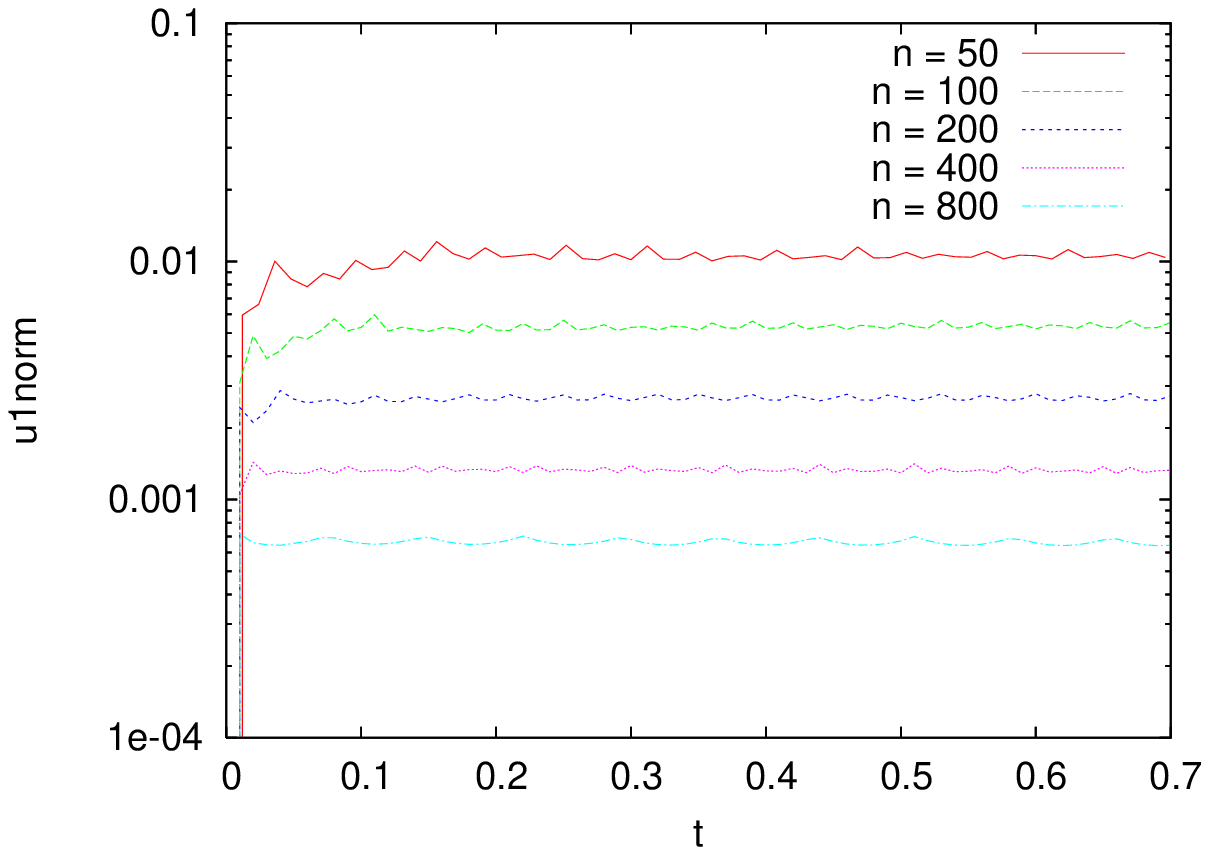}
\end{tabular}
\caption{Sod test on the uni-patch system. For each of the primitive
variables $\rho$, $u$, and $u^1$, the left plots show the comparison
of the numerical result with the exact solution for different times
(spaced in units of $\Delta t = 0.1$), and right plots show the 
error function ($l_1$ norm of the difference between the
exact solution and the discrete result) over time.}
\label{fig:sod_cart_x}
\end{figure*}

\begin{figure*}
\psfrag{x}{$x$}
\psfrag{rho}{$\rho$}
\psfrag{u}{$u$}
\psfrag{u1}{$u^1$}
\psfrag{t}{$t$}
\psfrag{rhonorm}{\hspace{-0.5cm}${||\rho - \rho_{exact}||}_1$}
\psfrag{unorm}{\hspace{-0.5cm}${||u - u_{exact}||}_1$}
\psfrag{u1norm}{\hspace{-0.5cm}${||u^1 - u^1_{exact}||}_1$}
\begin{tabular}{cc}
\includegraphics[width=\columnwidth]{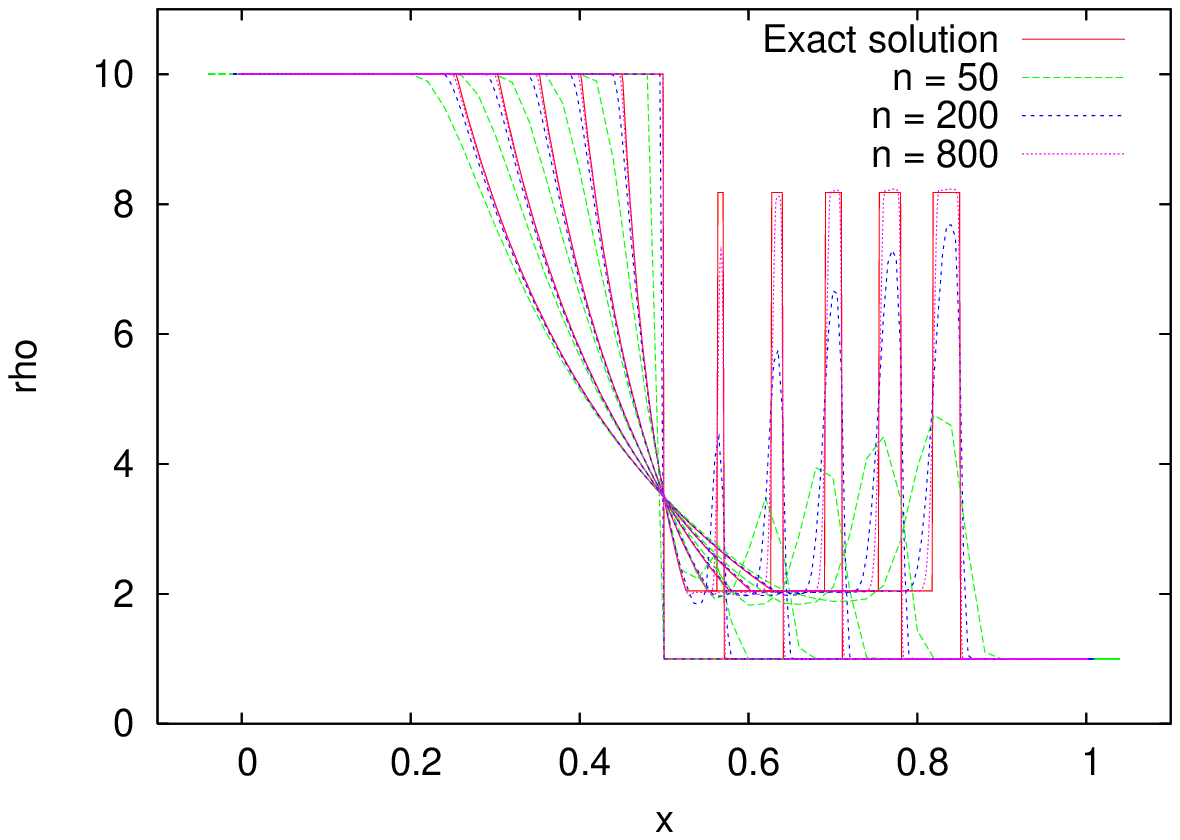} &
\includegraphics[width=\columnwidth]{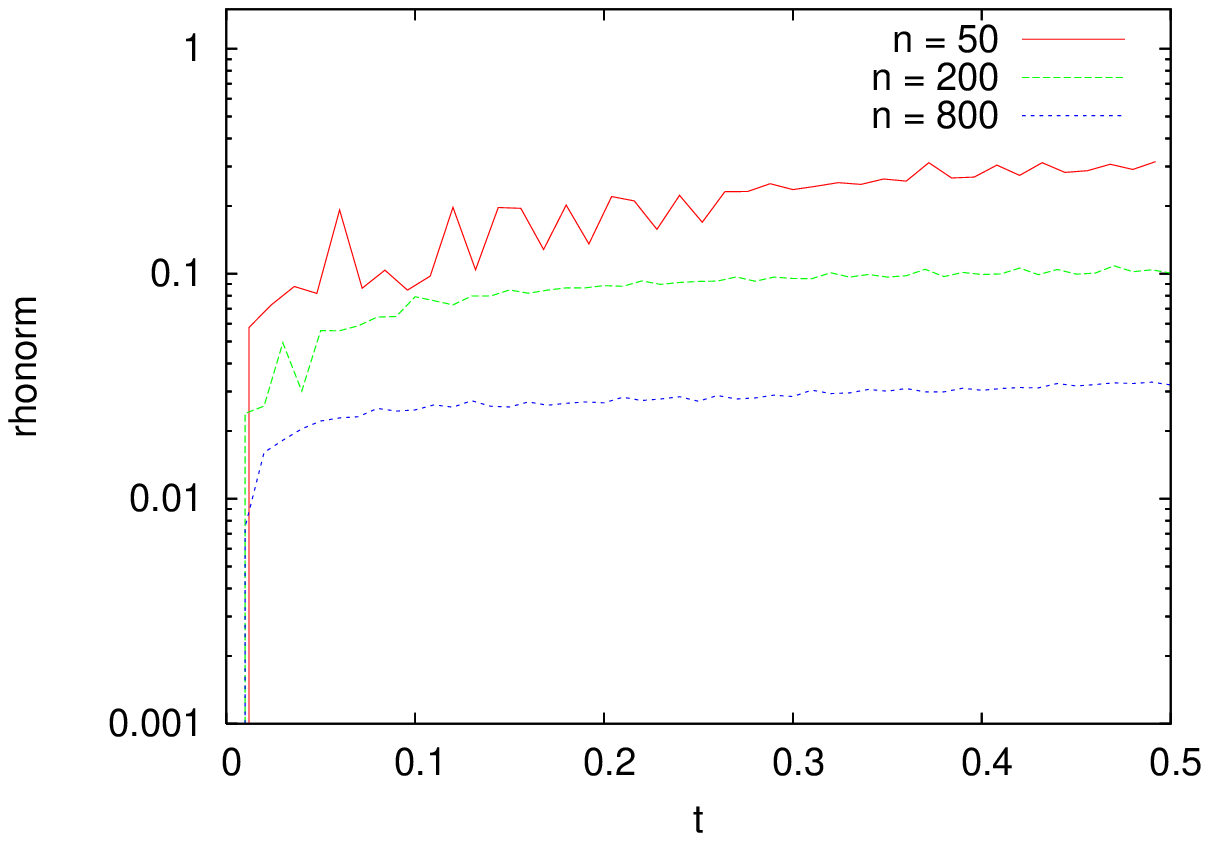} \\
\includegraphics[width=\columnwidth]{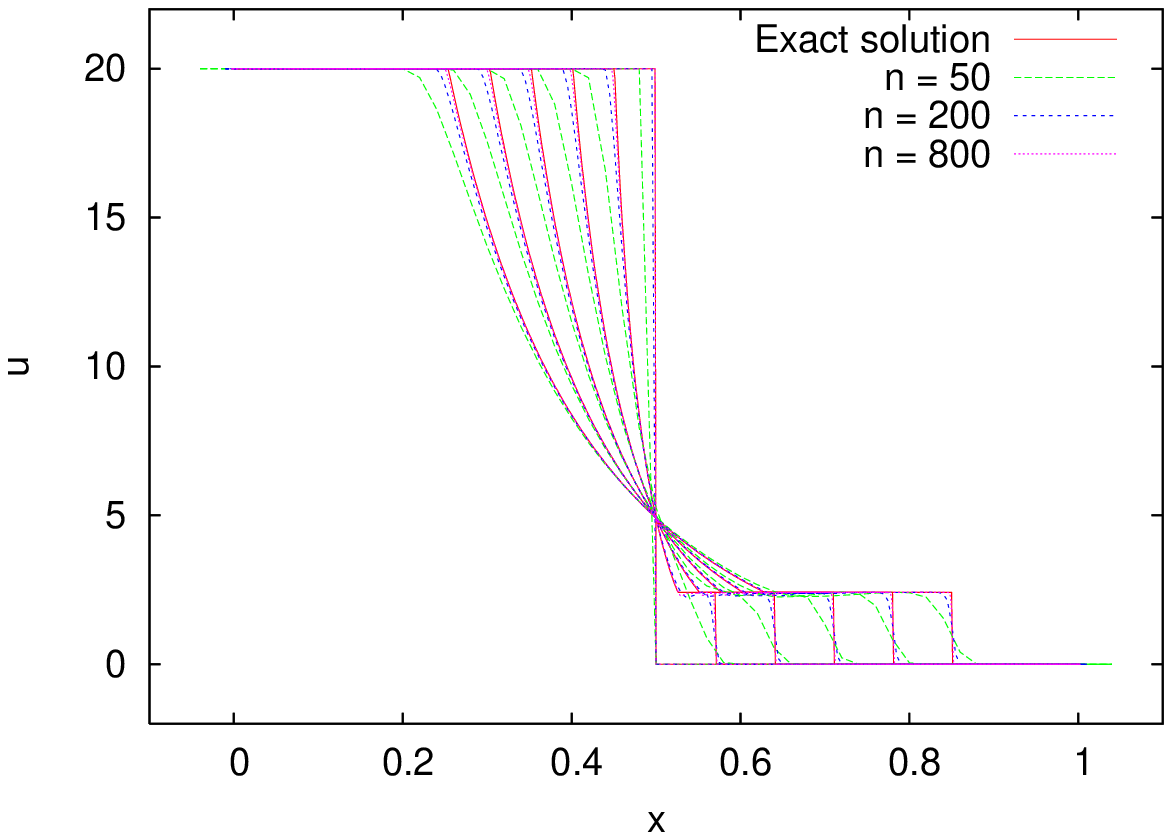} &
\includegraphics[width=\columnwidth]{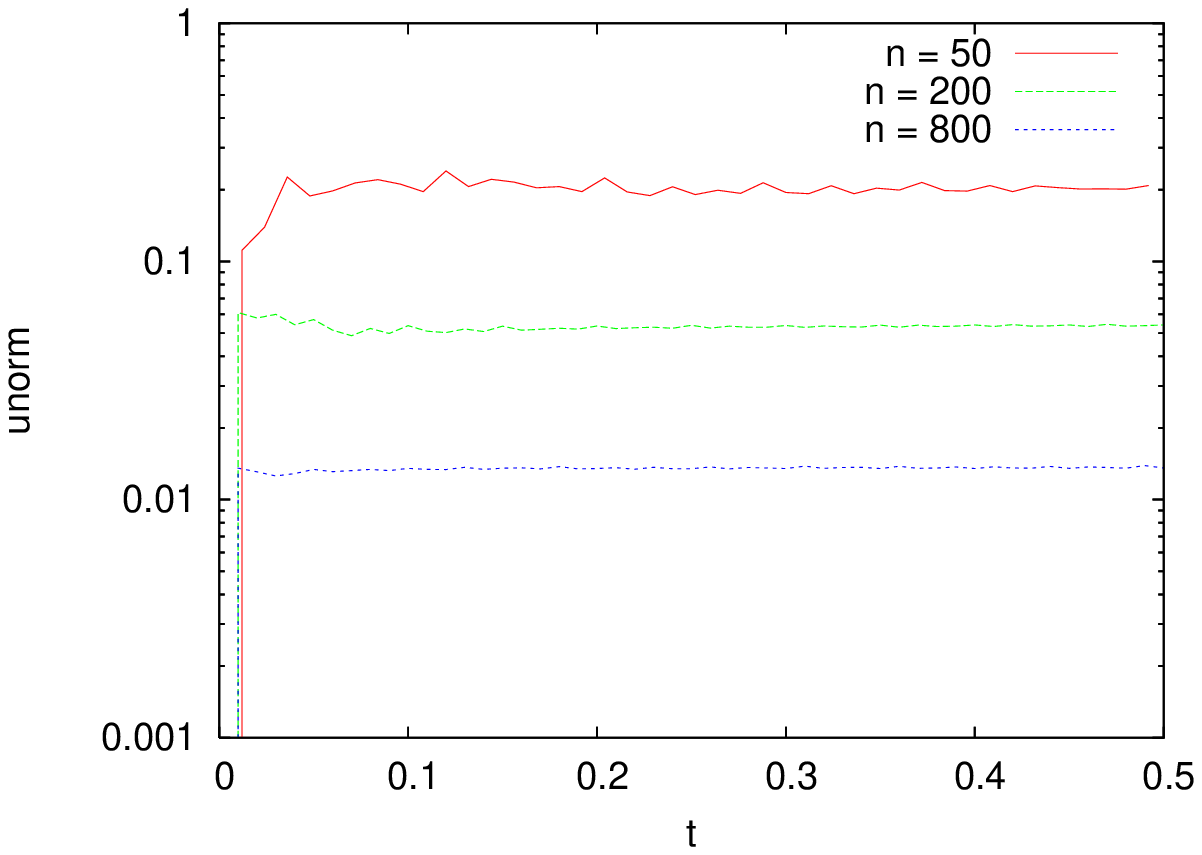} \\ 
\includegraphics[width=\columnwidth]{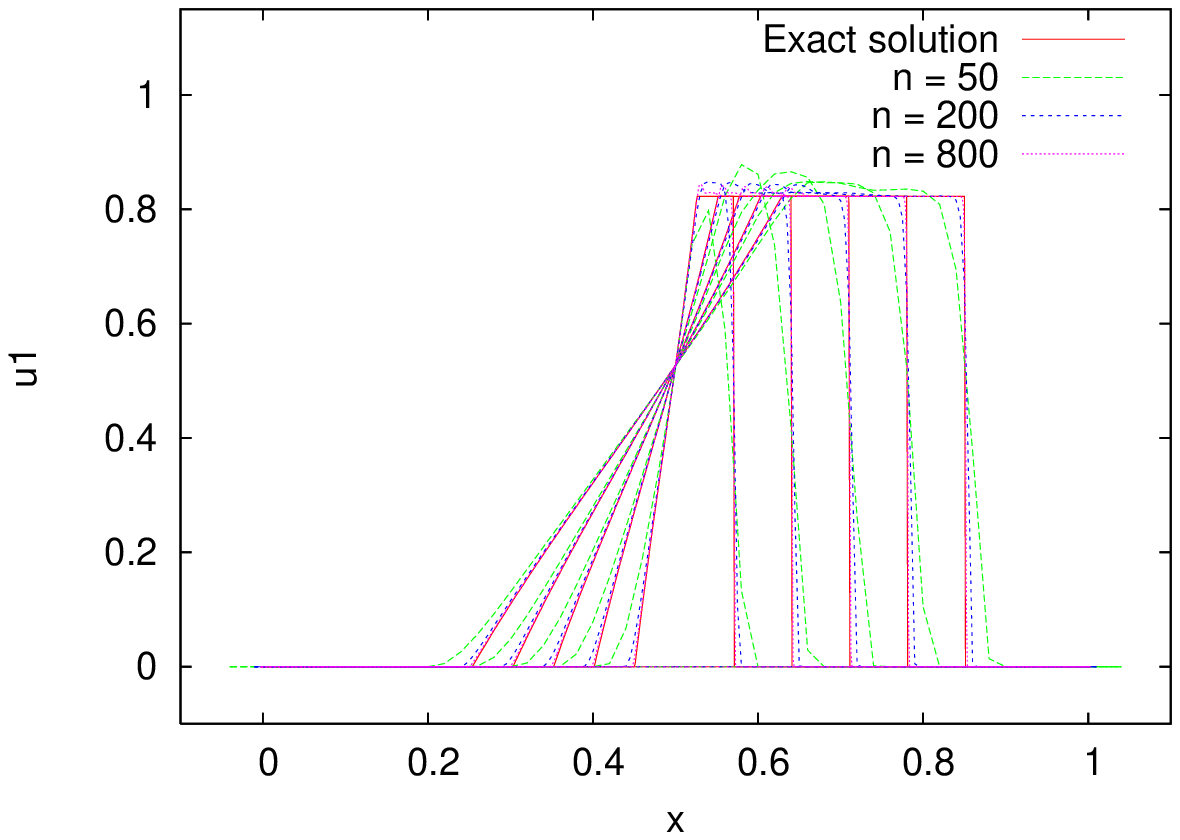} &
\includegraphics[width=\columnwidth]{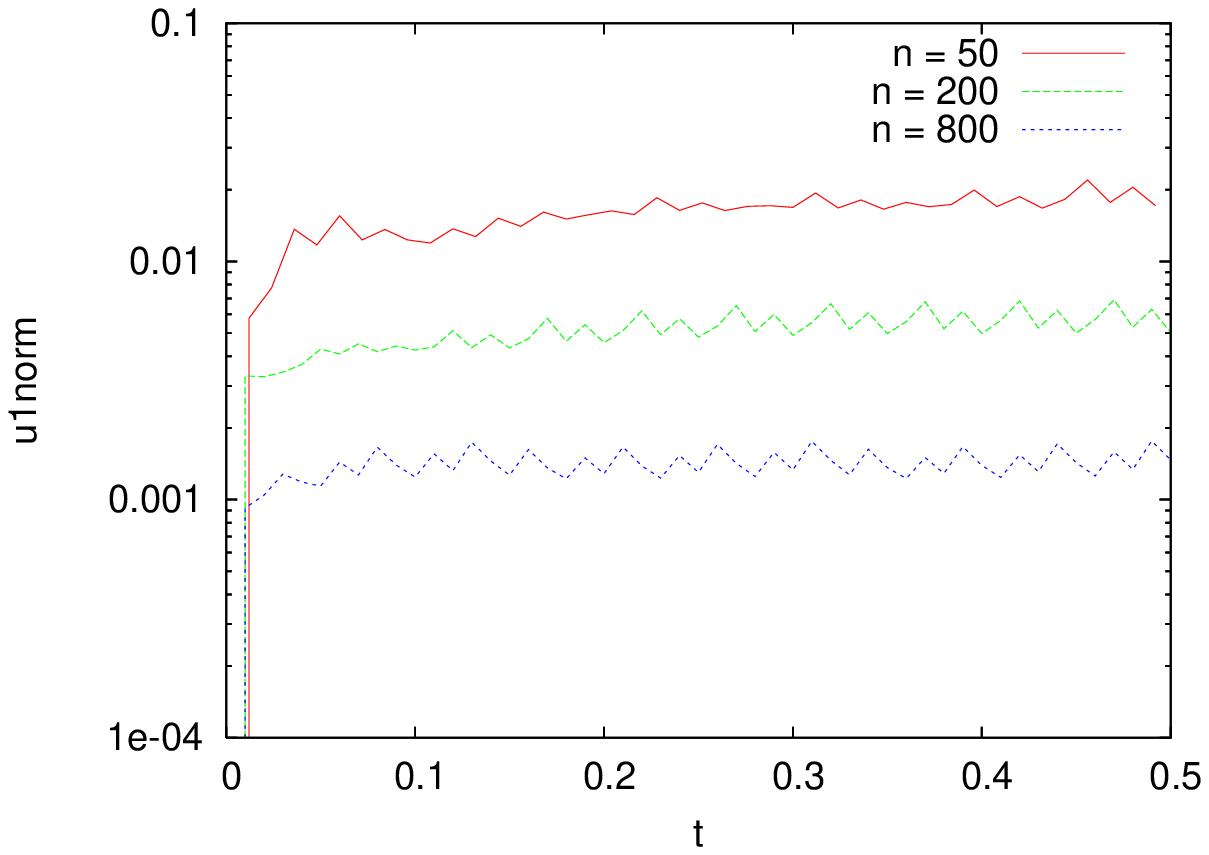}
\end{tabular}
\caption{``Simple'' shock test on the uni-patch system. Same as in
Fig.~\ref{fig:sod_cart_x}.}
\label{fig:simple_cart_x}
\end{figure*}

\begin{figure*}
\psfrag{x}{$x$}
\psfrag{rho}{$\rho$}
\psfrag{u}{$u$}
\psfrag{u1}{$u^1$}
\psfrag{t}{$t$}
\psfrag{rhonorm}{\hspace{-0.5cm}${||\rho - \rho_{exact}||}_1$}
\psfrag{unorm}{\hspace{-0.5cm}${||u - u_{exact}||}_1$}
\psfrag{u1norm}{\hspace{-0.5cm}${||u^1 - u^1_{exact}||}_1$}
\begin{tabular}{cc}
\includegraphics[width=\columnwidth]{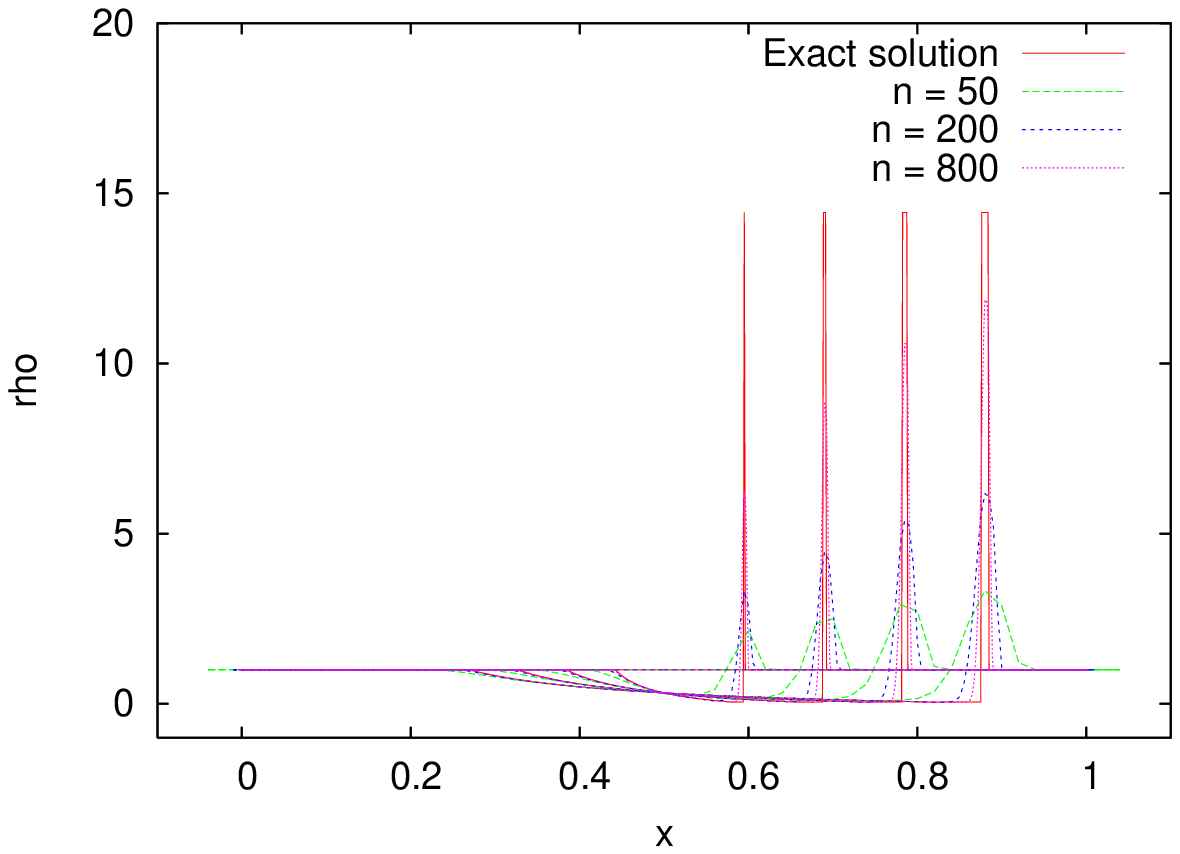} &
\includegraphics[width=\columnwidth]{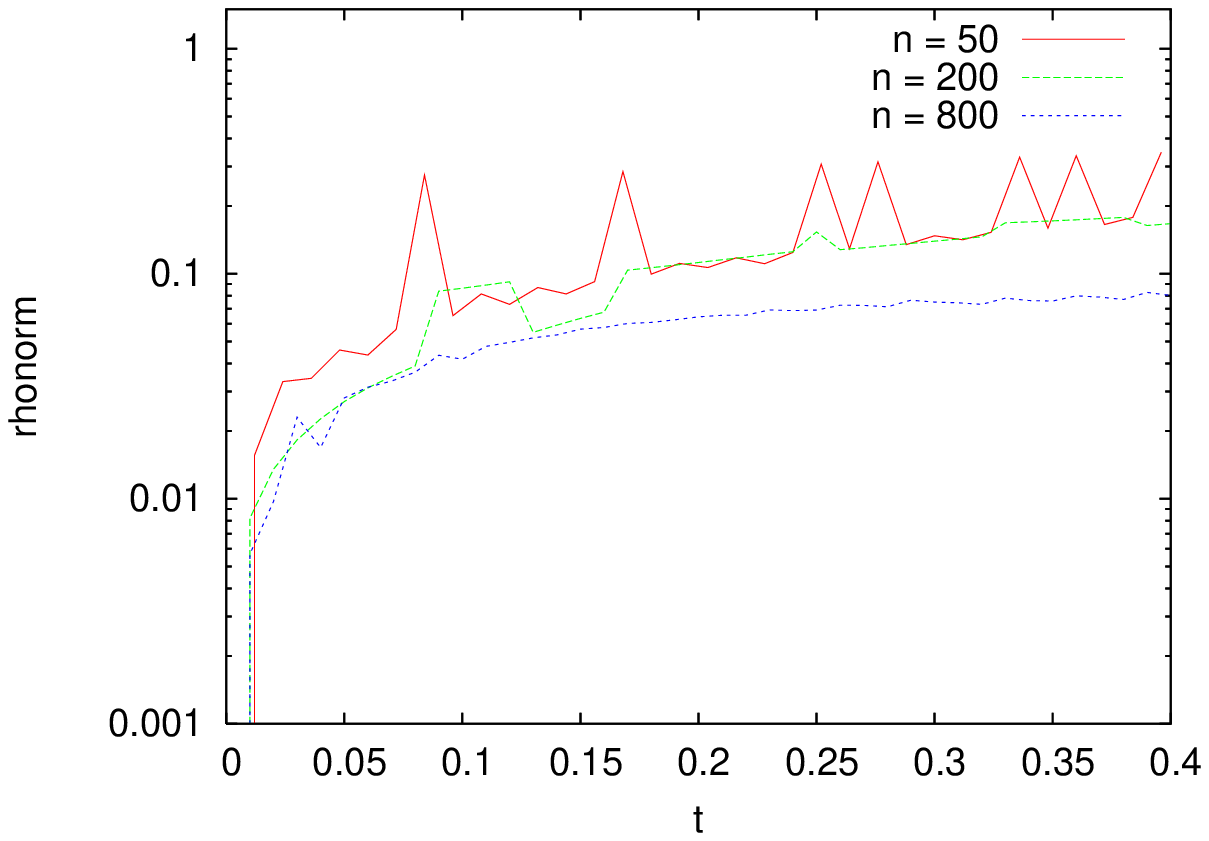} \\
\includegraphics[width=\columnwidth]{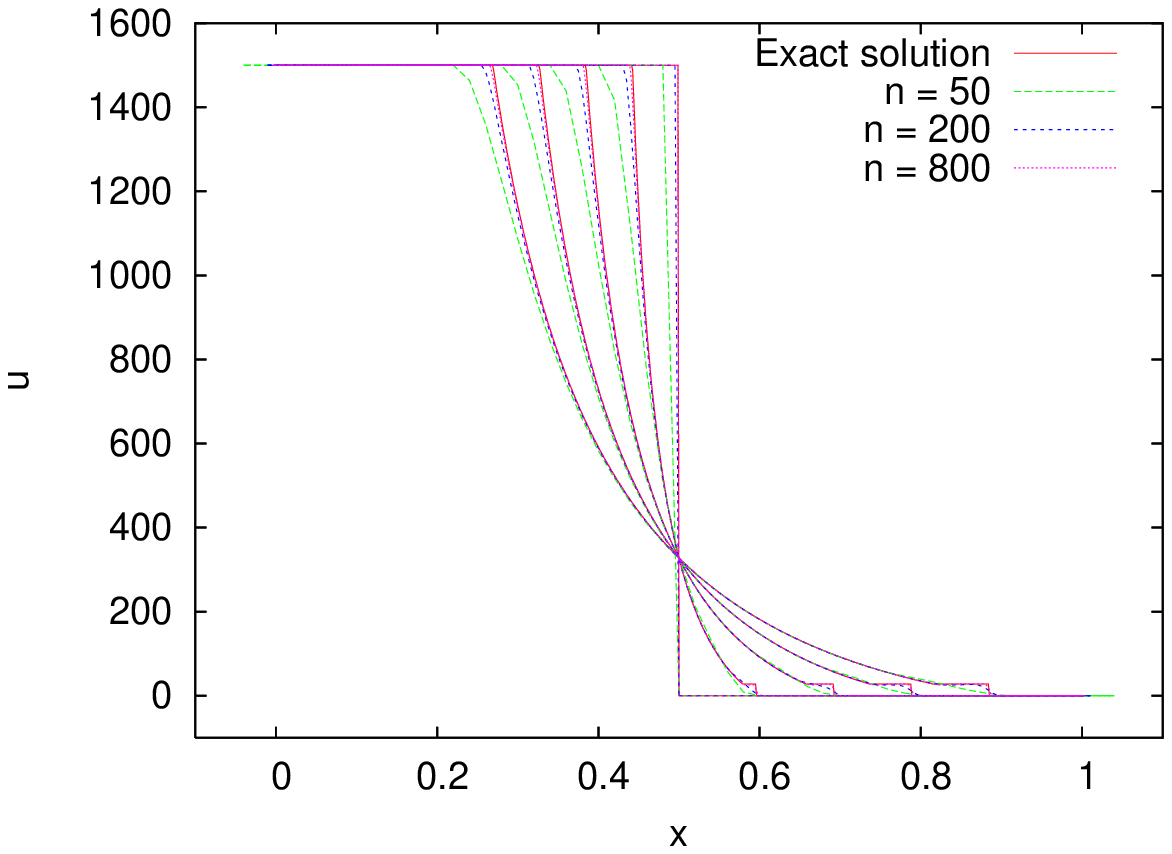} &
\includegraphics[width=\columnwidth]{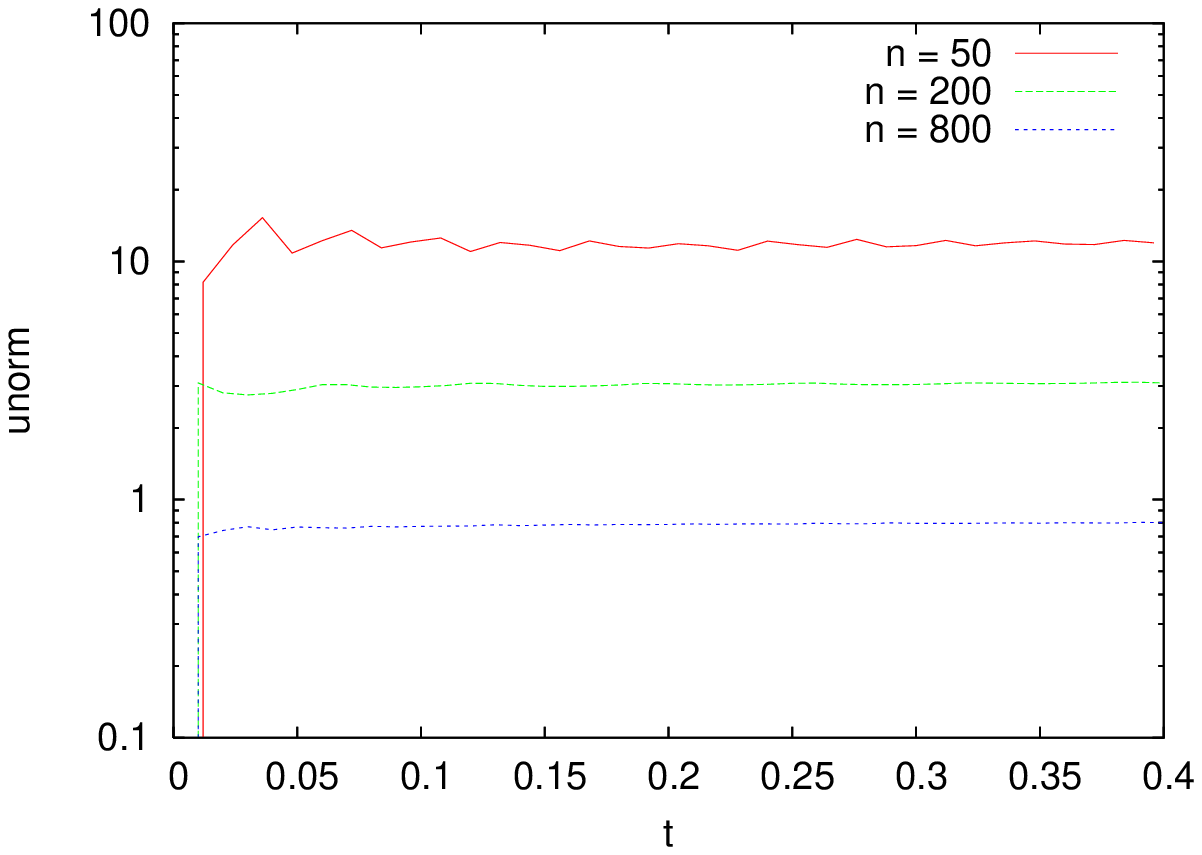} \\ 
\includegraphics[width=\columnwidth]{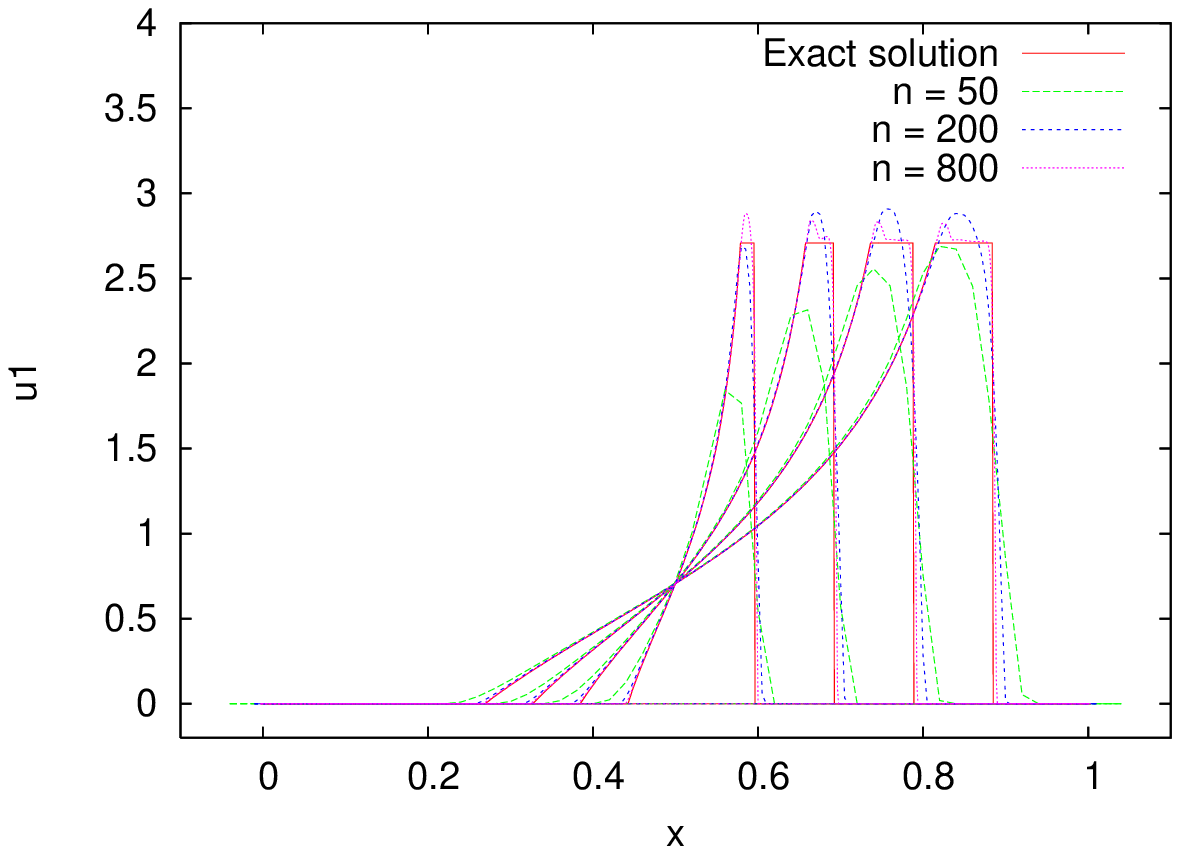} &
\includegraphics[width=\columnwidth]{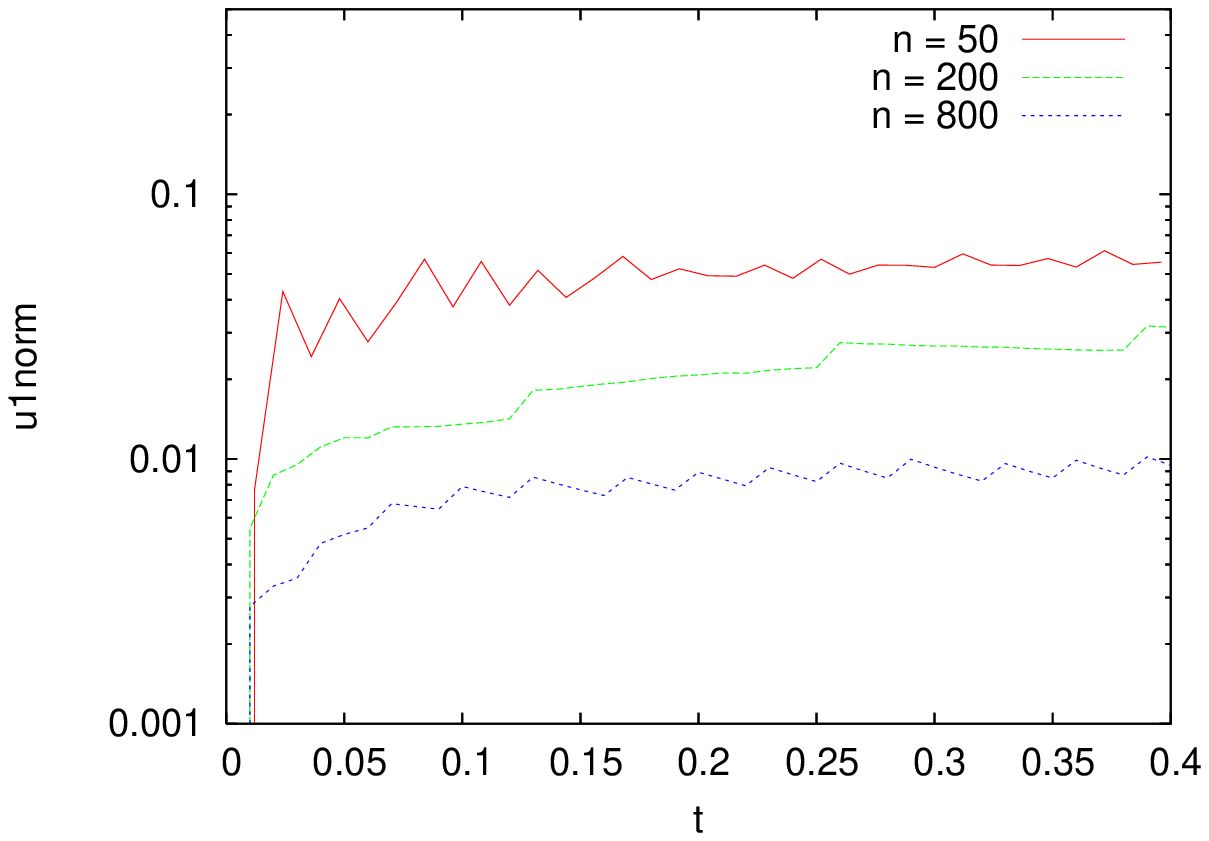}
\end{tabular}
\caption{Blast wave test on the uni-patch system. Same as in
Fig.~\ref{fig:sod_cart_x}.}
\label{fig:blast_cart_x}
\end{figure*}

The results of evolution of the Sod, ``Simple'' and blast wave problem
are illustrated in Figs.~\ref{fig:sod_cart_x}, \ref{fig:simple_cart_x}
and \ref{fig:blast_cart_x}. The code has some inherent dissipation due
to the use of the MC limiter, but it converges to the exact solution.

\subsection{Sod test on the distorted two patches system}

We have performed the Sod test (eqn.~\ref{eqn:sod}) on the two
distorted two patches system described in 
Section~\ref{sec:two_patches_distorted}. The first and second 
patch both have $n \times 1 \times 1$ cells, where $n = 50 \ldots 800$,
and the patch interface is located at $x = 1$ in terms of global coordinates.
As before, the boundary ghost cells are set to the exact solution.

\begin{figure*}
\psfrag{x}{$x$}
\psfrag{rho}{$\rho$}
\psfrag{u}{$u$}
\psfrag{u1}{$u^1$}
\psfrag{t}{$t$}
\psfrag{rhonorm}{\hspace{-0.5cm}${||\rho - \rho_{exact}||}_1$}
\psfrag{unorm}{\hspace{-0.5cm}${||u - u_{exact}||}_1$}
\psfrag{u1norm}{\hspace{-0.5cm}${||u^1 - u^1_{exact}||}_1$}
\begin{tabular}{cc}
\includegraphics[width=\columnwidth]{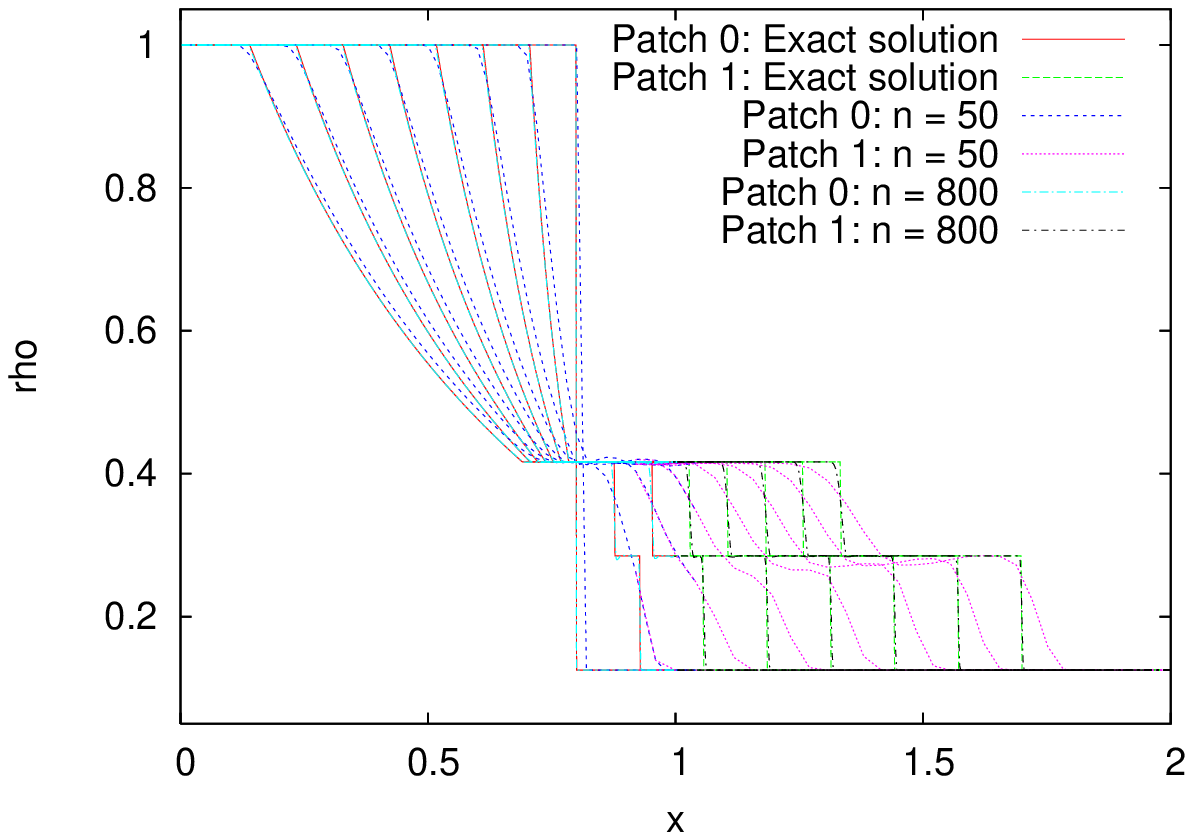} &
\includegraphics[width=\columnwidth]{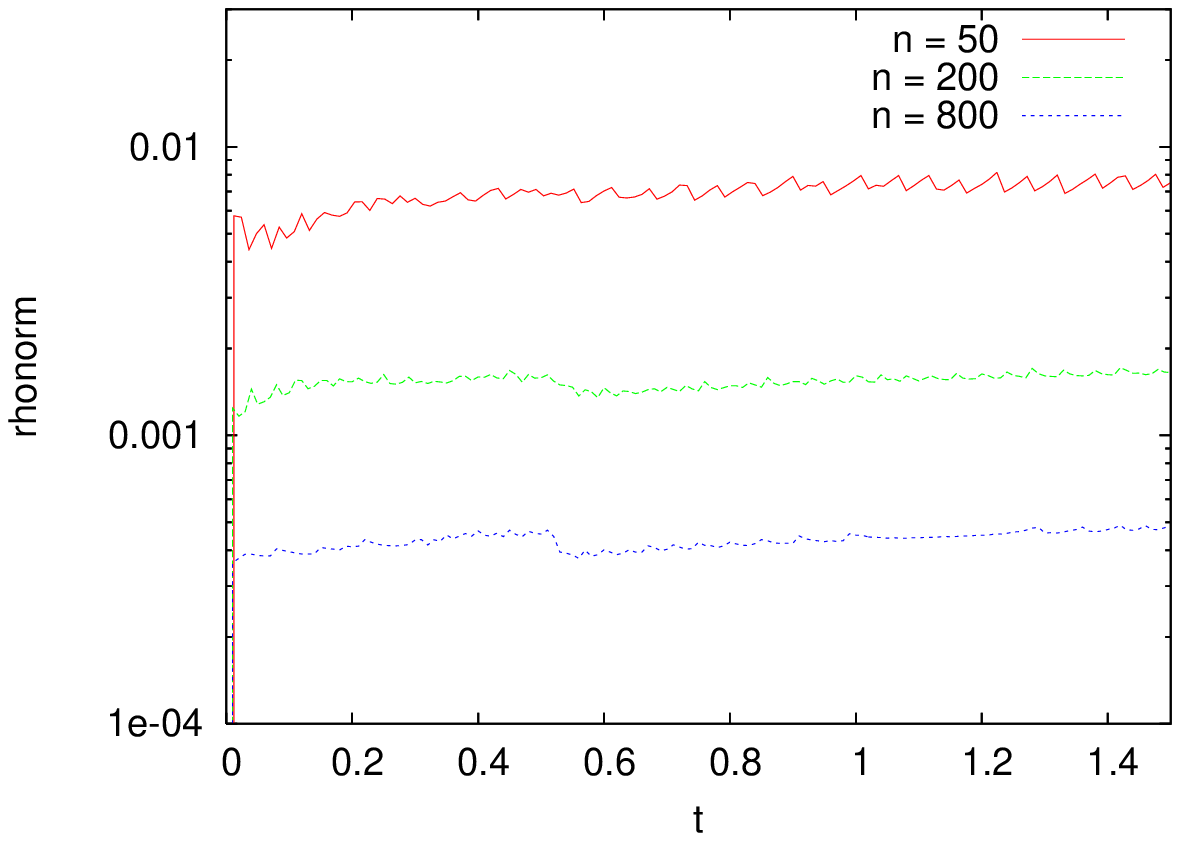} \\
\includegraphics[width=\columnwidth]{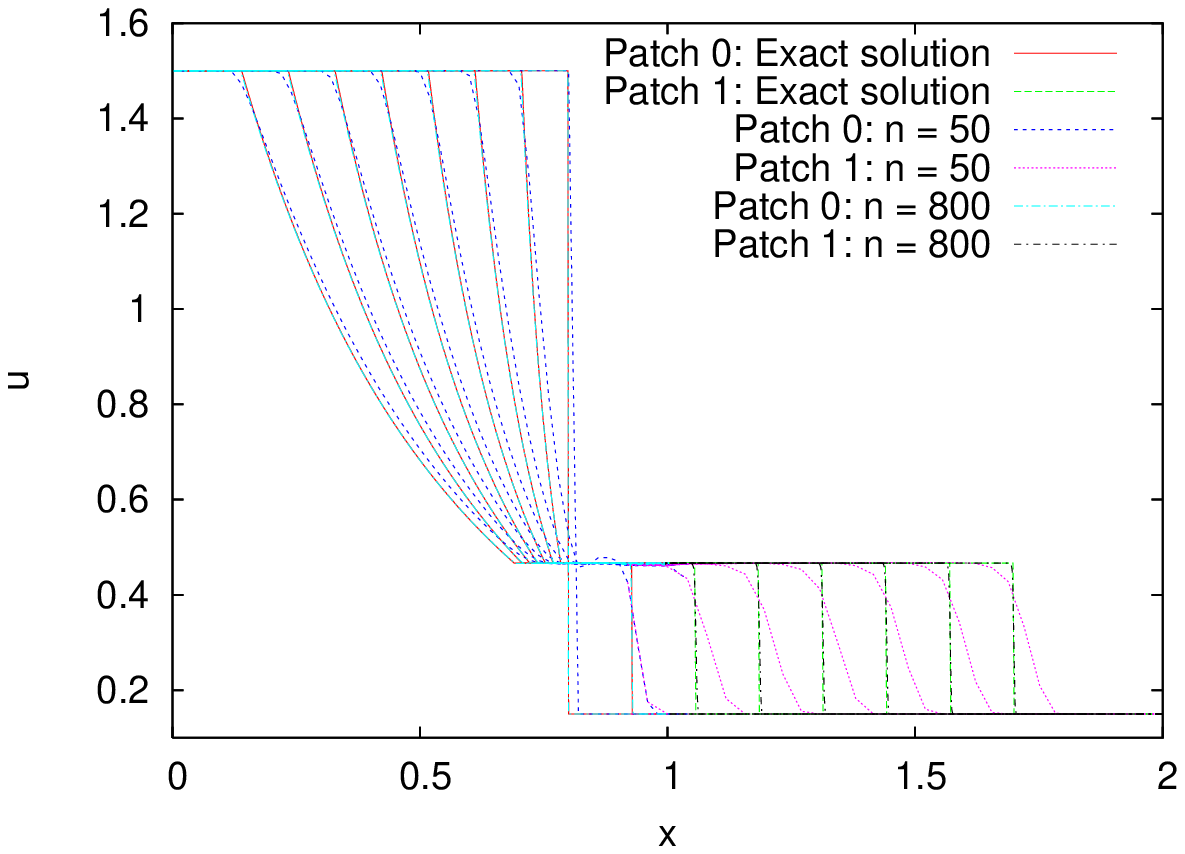} &
\includegraphics[width=\columnwidth]{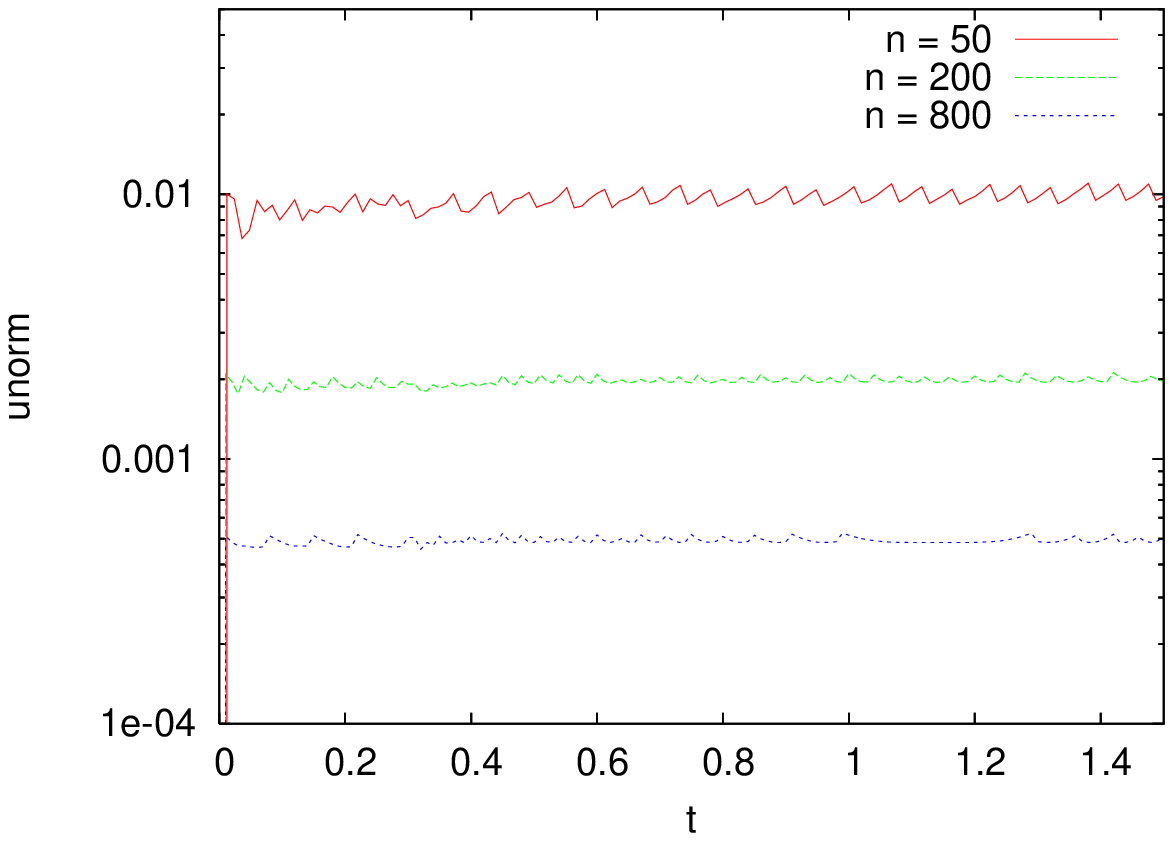} \\
\includegraphics[width=\columnwidth]{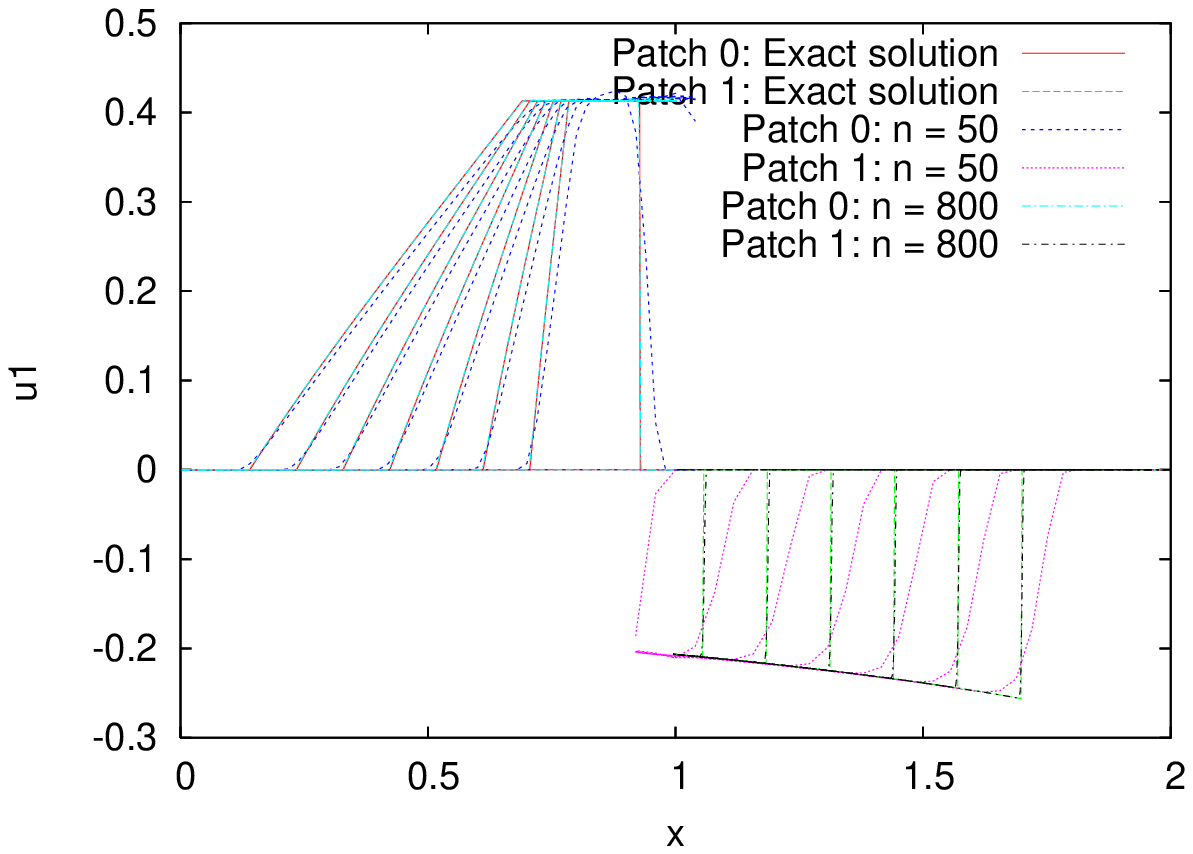} &
\includegraphics[width=\columnwidth]{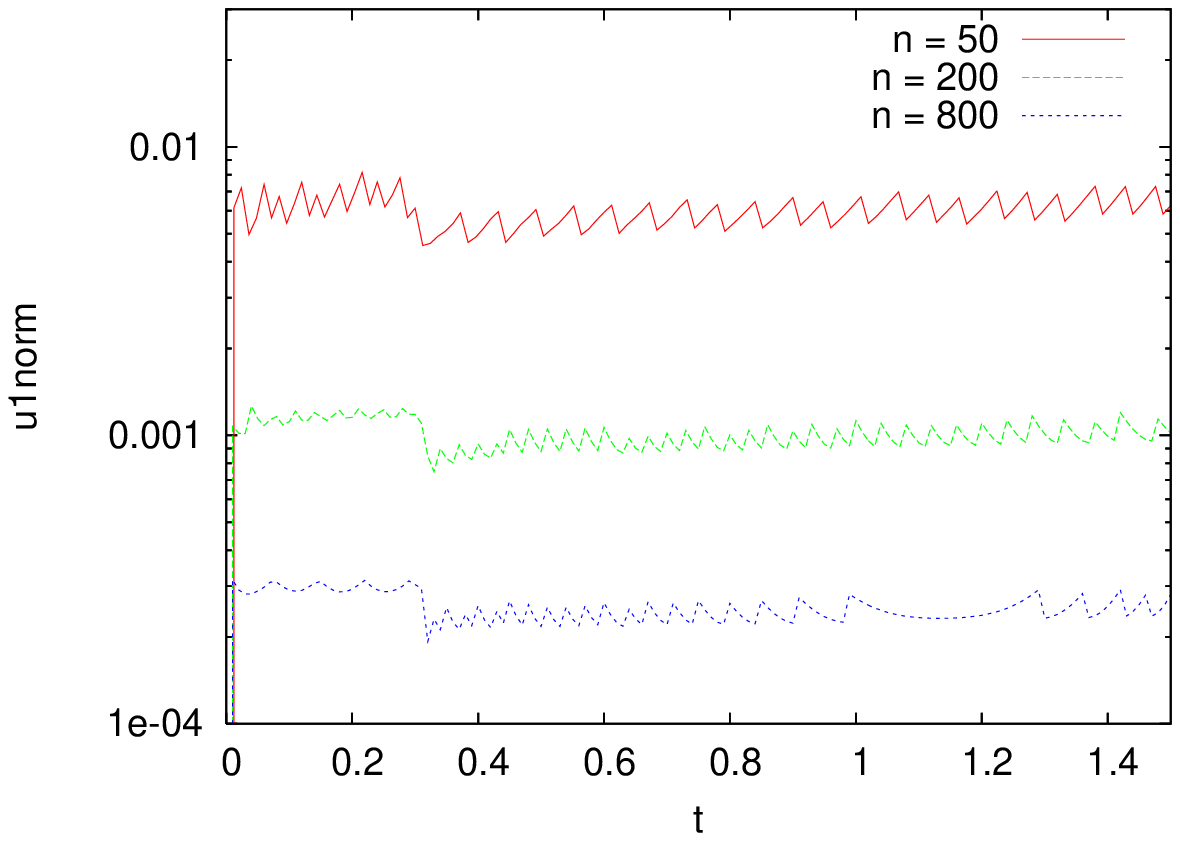} 
\end{tabular}
\caption{Sod test on the distorted two patches system. This system, defined
in Section~\ref{sec:two_patches_distorted}, consists of a left patch (\emph{patch 0})
which is undistorted, and a right patch (\emph{patch 1}) which has a non-trivial
Jacobian associated with the transformation from local to global coordinates. The left
panel shows the primitive variables $\rho$, $u$ and $u^1$ in dependence on the global
coordinate $x$, for different coordinate times ($\Delta t = 0.2$). The interface between
the patches is located at $x = 1$. In the lower left diagram, note how the velocity
component $u^1$ appears discontinuous across the interface since it is represented in
different tensor bases on each patch. The right panel shows global convergence with 
respect to the $l_1$ norm. (The left panel does not contain the graphs for $n = 200$ to
enhance the visual clarity of the plots.)}
\label{fig:sod_two_distorted}
\end{figure*}

Fig.~\ref{fig:sod_two_distorted} demonstrates that the multi-patch approximation
leads to a convergent transmission of the shock across the interface, although
both patches use different local tensor bases, and a non-trivial Jacobian is 
associated with the second patch. 

\subsection{Sod test on the cubed sphere six patches system}
\label{sec:sod_six_patches}

To perform the Sod test (eqn.~\ref{eqn:sod}) on the cubed sphere six patches system
described in Section~\ref{sec:six_patches}, we prepared a setup with $n \times n
\times n$ cells per patch, where $n = 20, 40, 80$, and chose the free parameters
$r_0$ and $r_1$, which specify the inner and outer boundary radius in terms of
the global coordinates, to be given by $r_0 = 1$ and $r_1 = 2$. The outer boundaries
of the domain are set to the exact solution produced by {\tt riemann}. 

\begin{figure*}
\begin{tabular}{cc}
\includegraphics[width=\columnwidth]{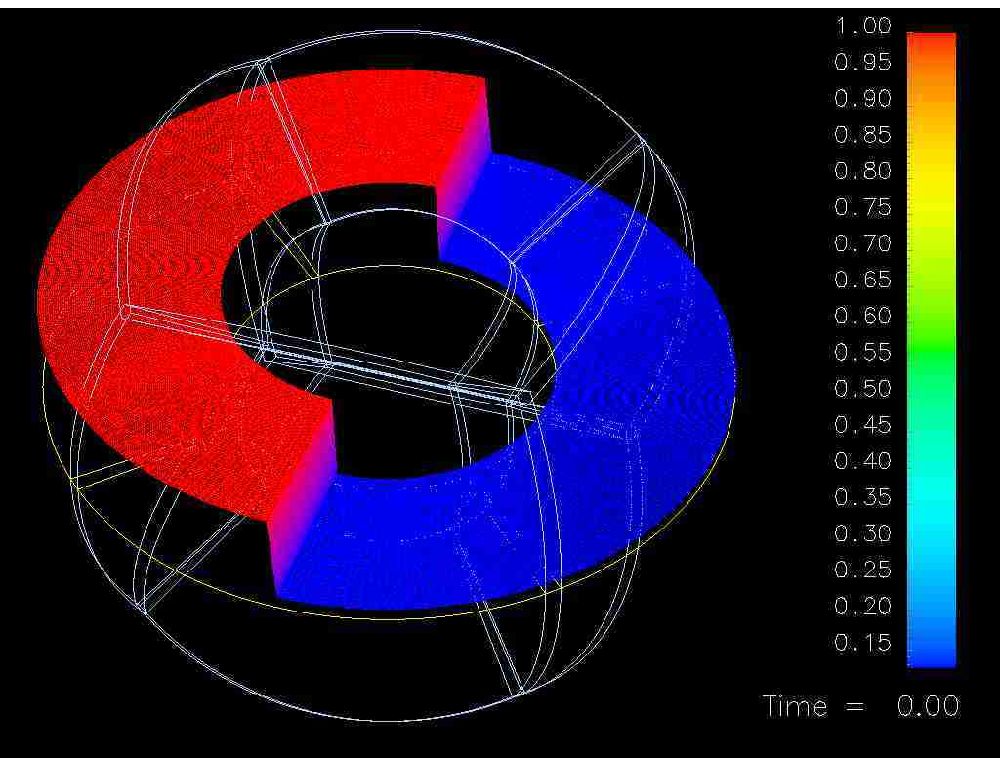} &
\includegraphics[width=\columnwidth]{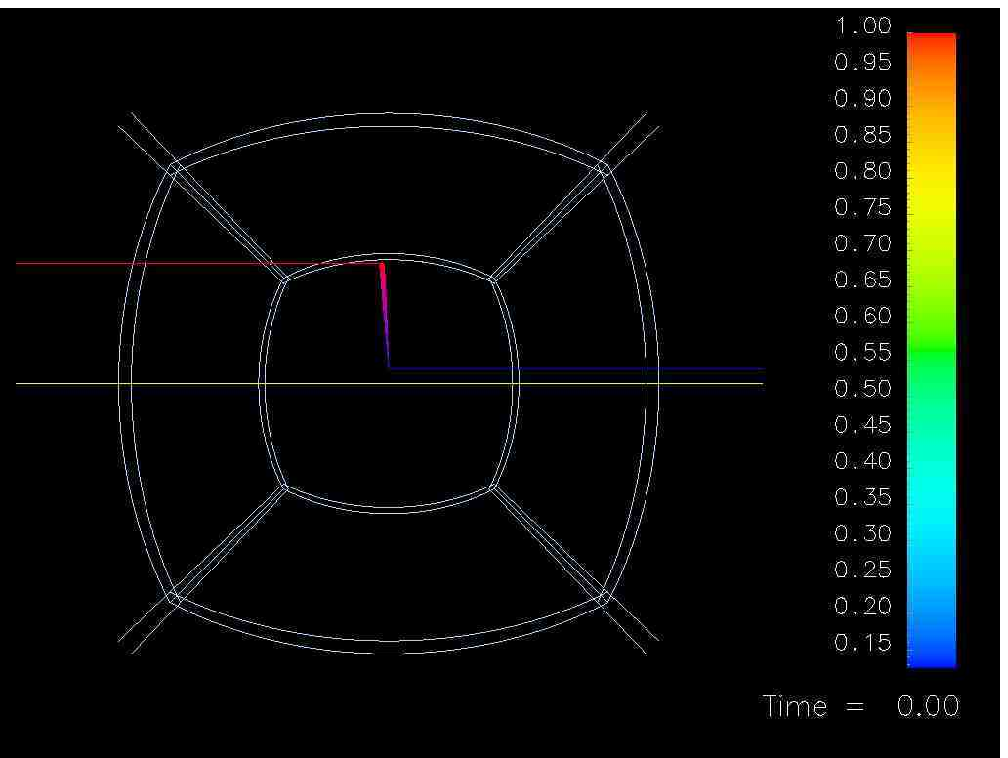} \\
\includegraphics[width=\columnwidth]{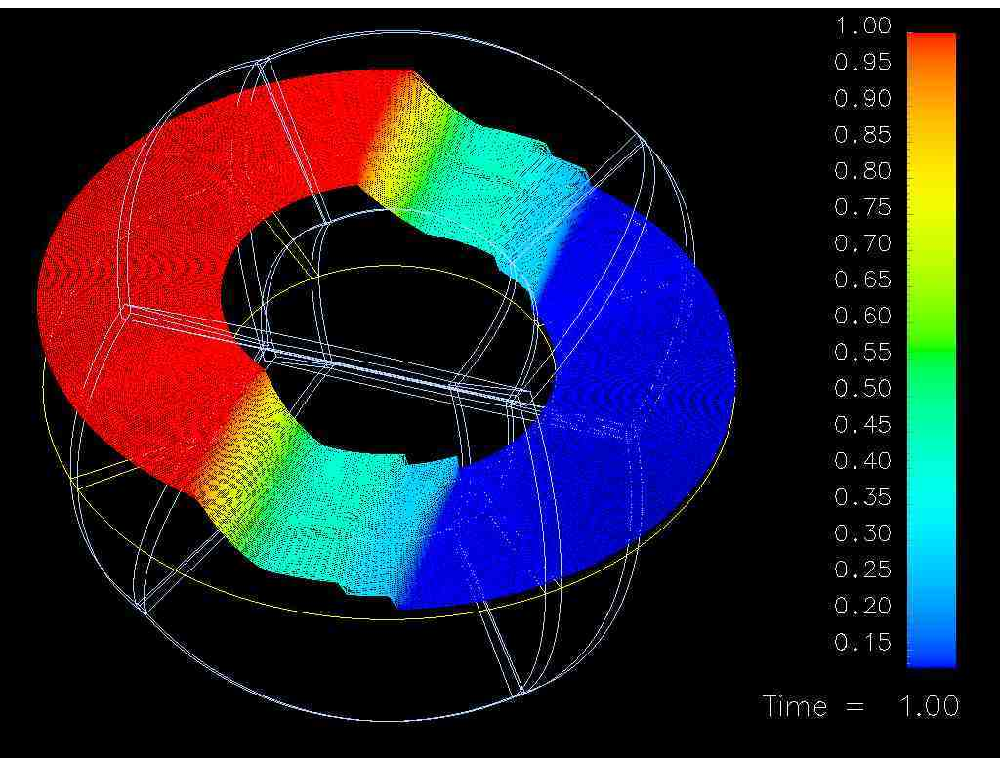} &
\includegraphics[width=\columnwidth]{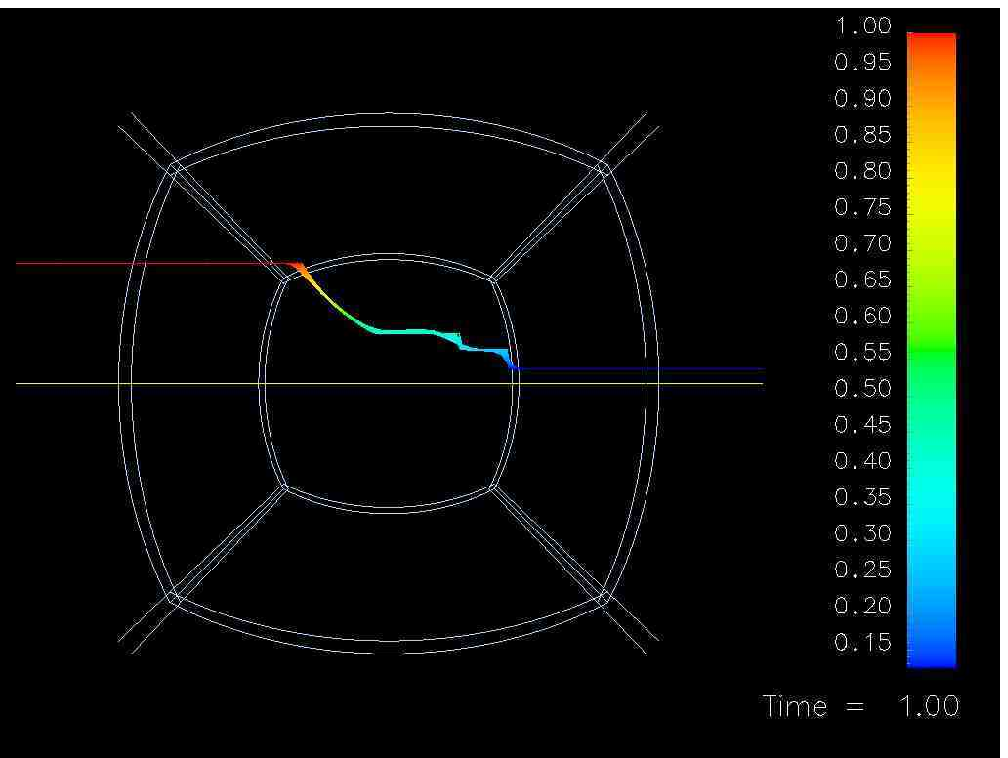} \\
\includegraphics[width=\columnwidth]{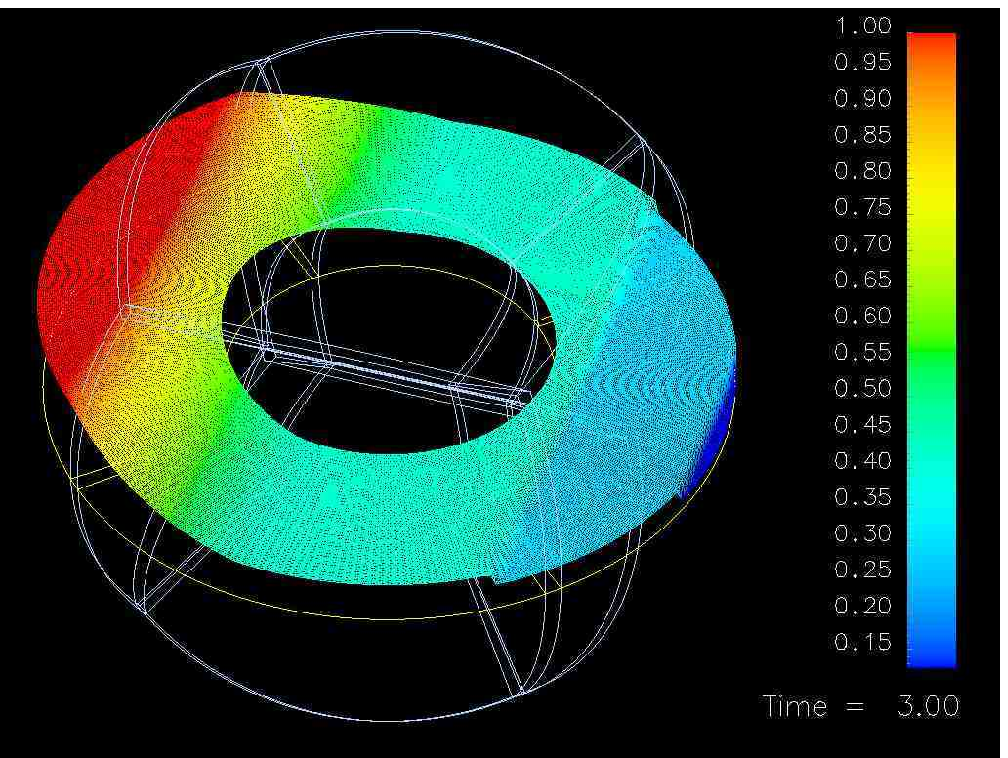} &
\includegraphics[width=\columnwidth]{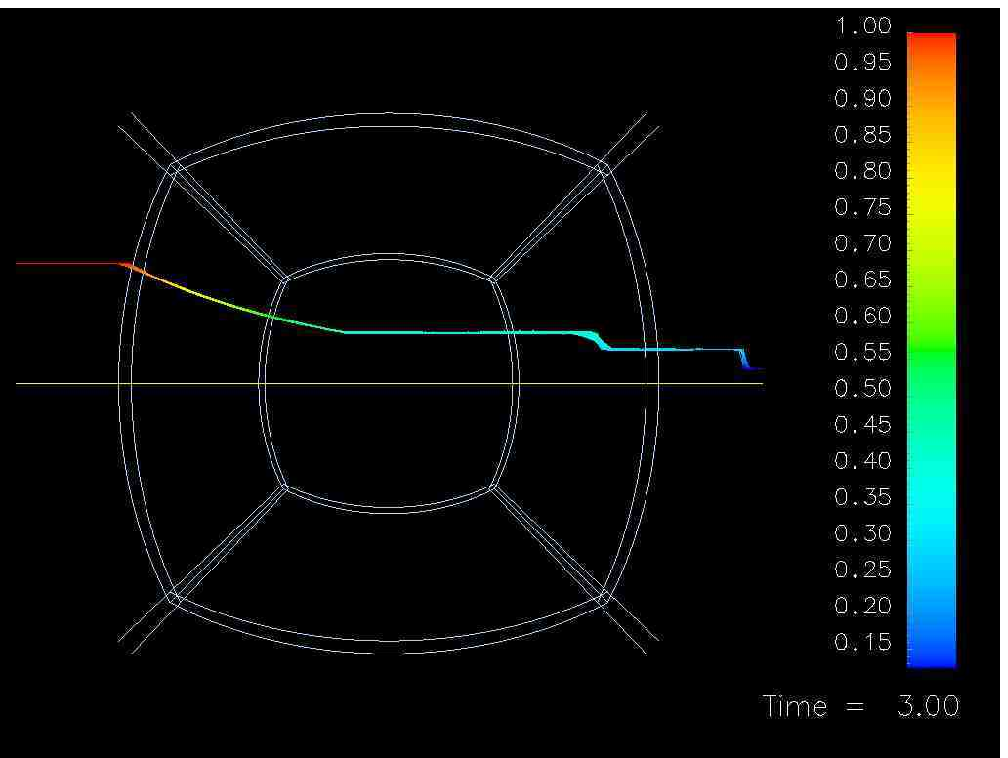} 
\end{tabular}
\caption{Sod test on the six patches system: evolution of the density for the case
where each of the patches has resolution $80 \times 80 \times 80$. The
white lines indicate the boundaries of the six patches which are used to
cover the computational domain. The yellow surface is the cut of the 
plane $z = 0$ with the grid boundaries, and the density function is shown
at $z = 0$. The left panel shows a perspective view of the evolution for 
coordinates times $0$, $1$ and $3$, whereas the right panel shows an orthogonal
projection from the negative y axis at the same times.}
\label{fig:six_patches_sod_rho}
\end{figure*}

\begin{figure}
\psfrag{x}{$x$}
\psfrag{rho}{$\rho$}
\psfrag{u}{$u$}
\psfrag{ua}{$u^1$}
\psfrag{t}{$t$}
\psfrag{rhonorm}{\hspace{-0.5cm}${||\rho - \rho_{exact}||}_1$}
\psfrag{unorm}{\hspace{-0.5cm}${||u - u_{exact}||}_1$}
\psfrag{uinorm}{\hspace{-0.5cm}${||u^i - u^i_{exact}||}_1$}
\includegraphics[width=\columnwidth]{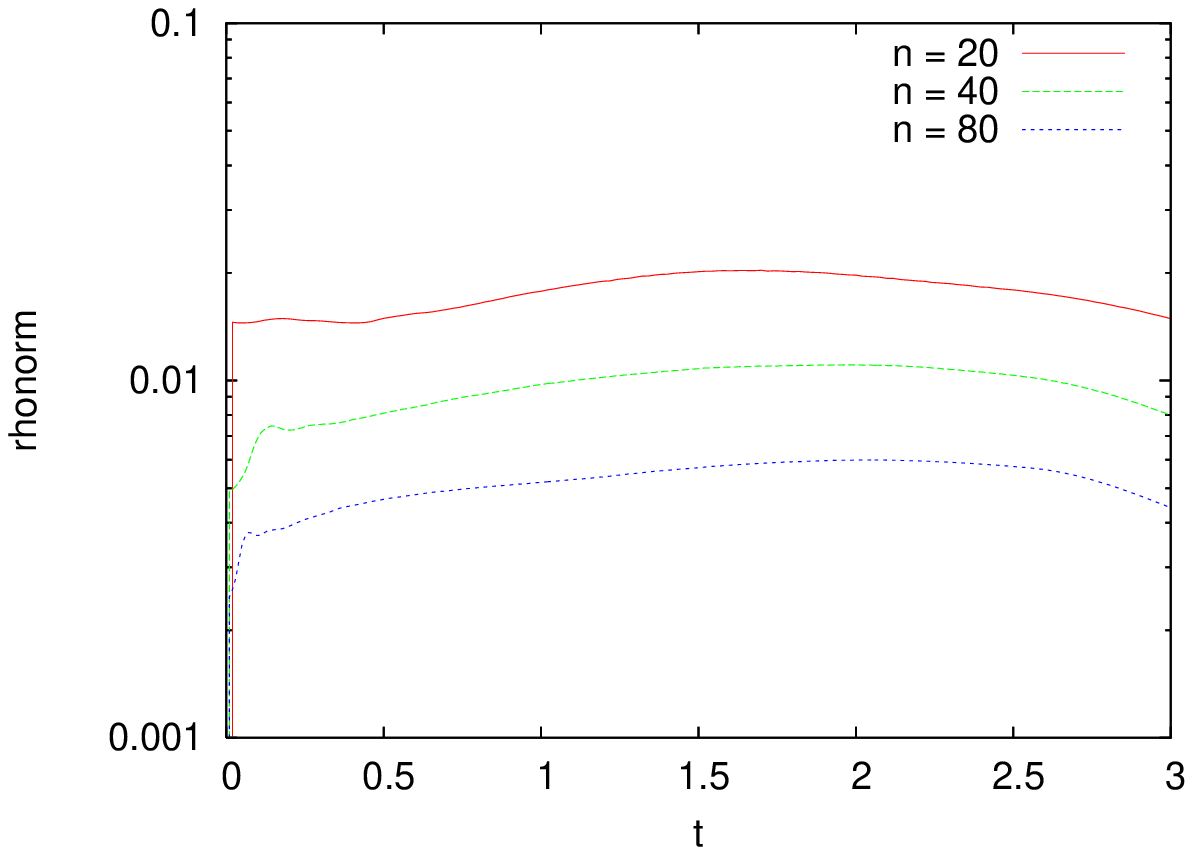} 
\includegraphics[width=\columnwidth]{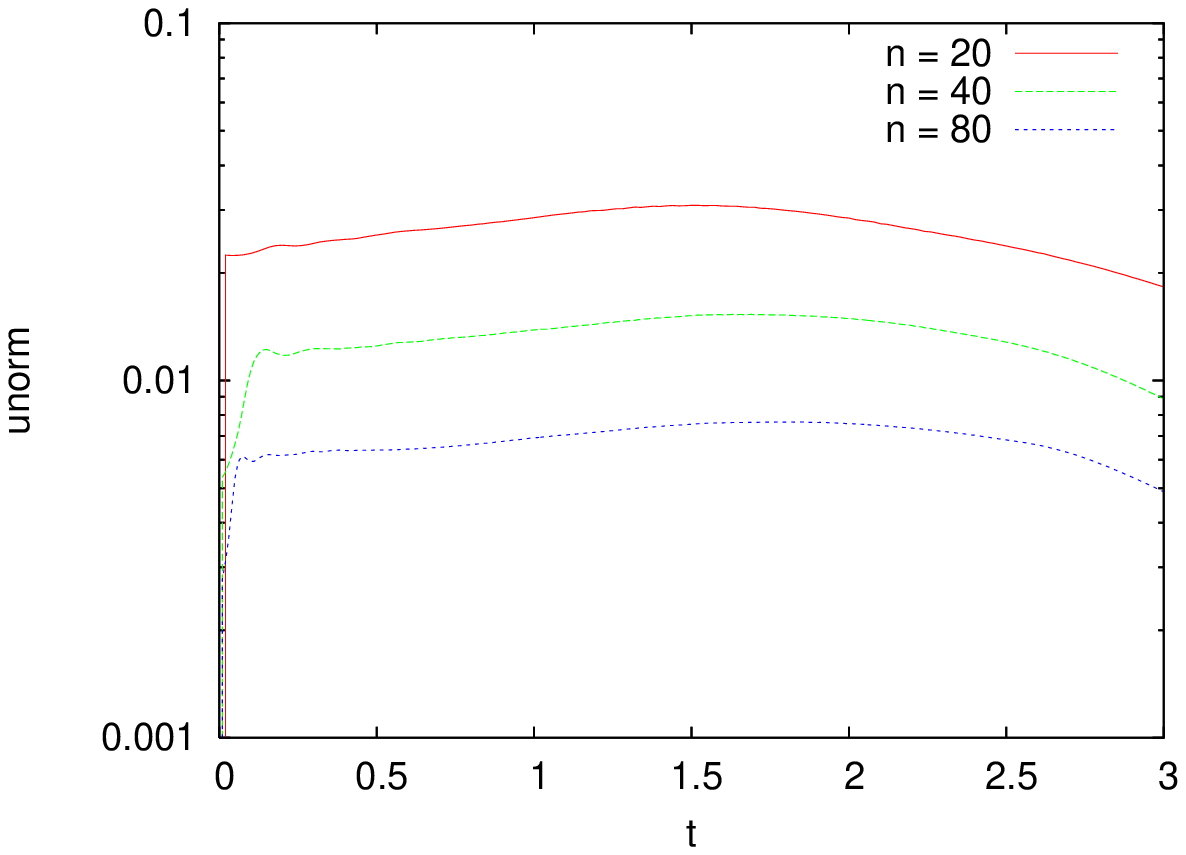}
\includegraphics[width=\columnwidth]{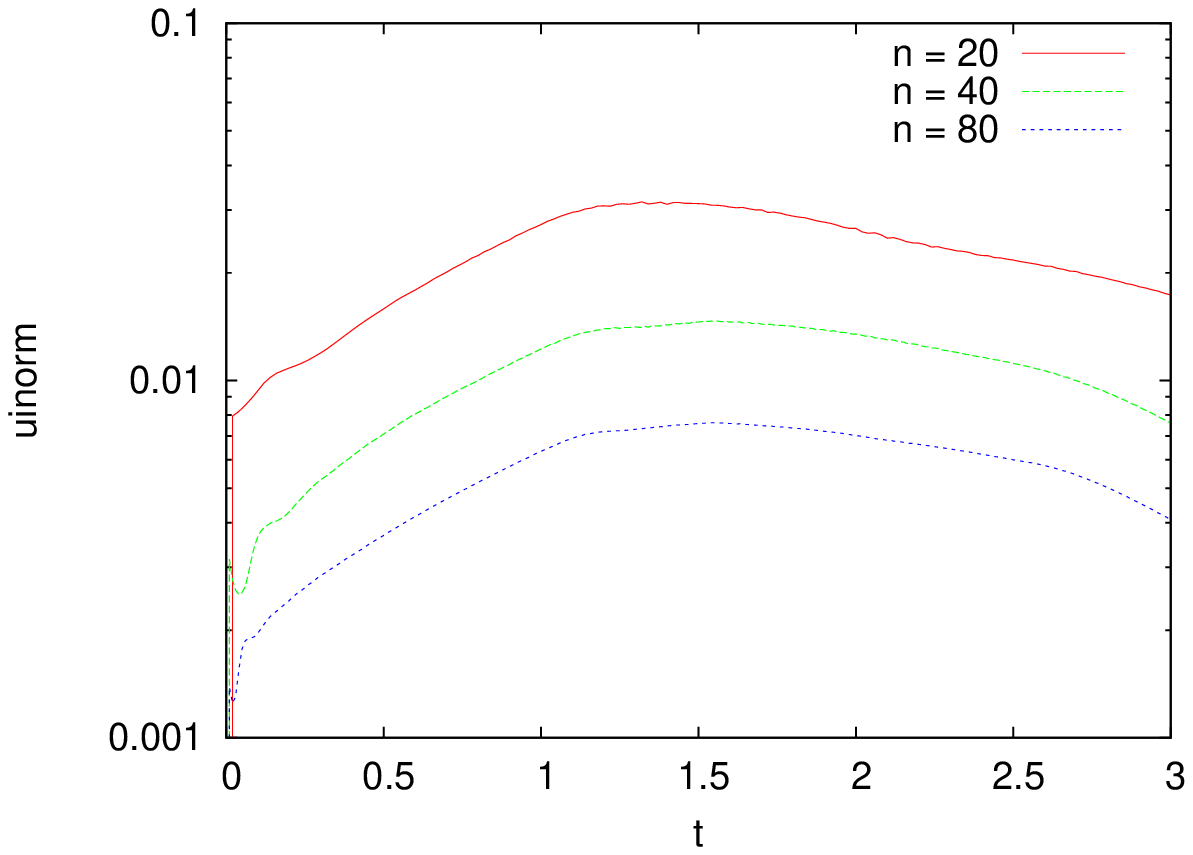} 
\caption{Sod test on the six patches system. These series of plots
demonstrate the convergence of the primitive variables to the solution 
produced by the {\tt riemann} code.
Each of the six patches has the same resolution defined by the parameter
$n$, which determines the number of cells by $n \times n \times n$.}
\label{fig:six_patches_sod_conv}
\end{figure}

The evolution of the density for the case $n = 80$ is shown in 
Fig.~\ref{fig:six_patches_sod_rho}. This plot
shows the density function in the equatorial plane, and the motion of the
waves resulting from the Riemann problem. Global convergence to the exact solution,
which is again constructed using {\tt riemann} and then transformed to the
local coordinate systems as required, is demonstrated in 
Fig.~\ref{fig:six_patches_sod_conv}.

\begin{figure}
\includegraphics[width=\columnwidth]{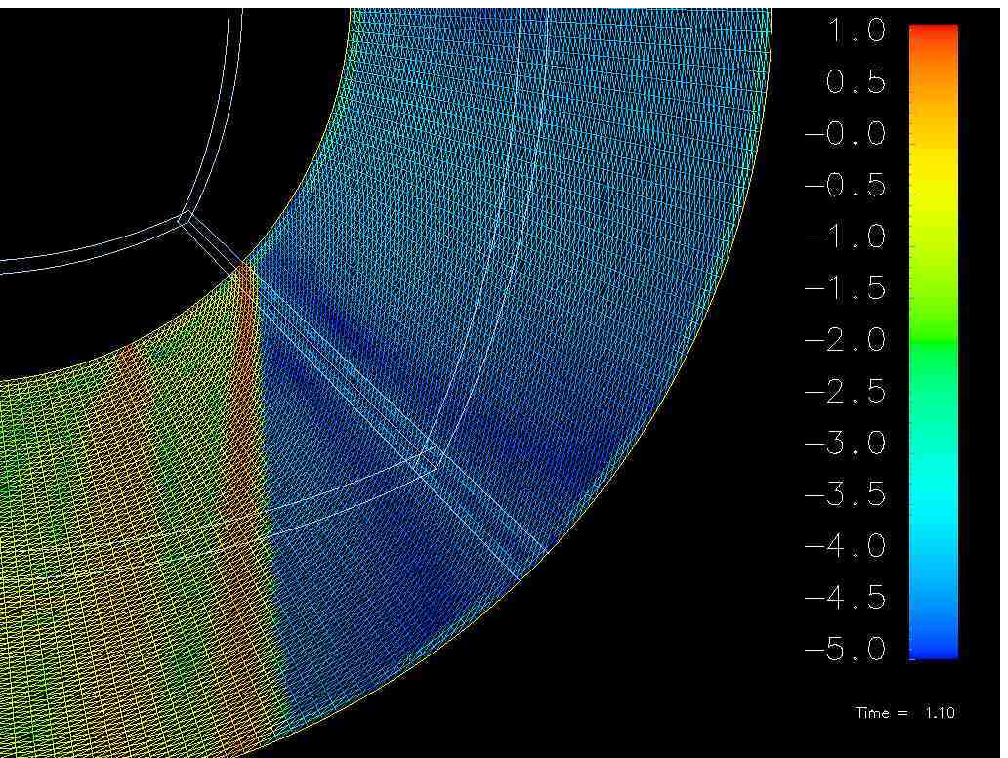}
\includegraphics[width=\columnwidth]{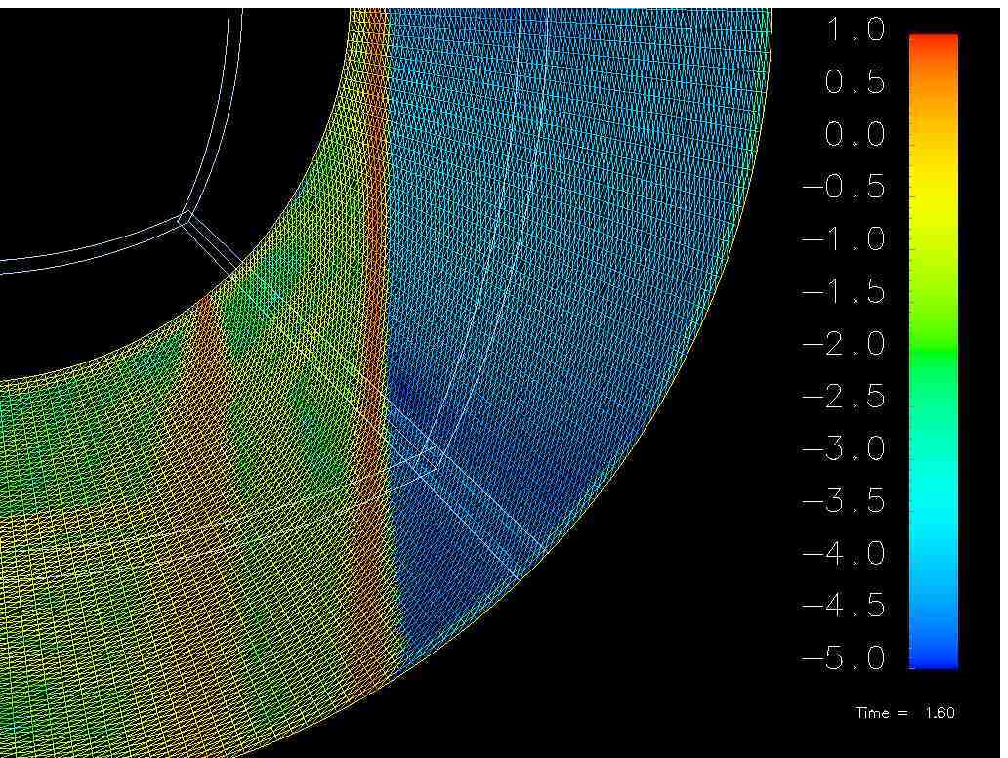}
\includegraphics[width=\columnwidth]{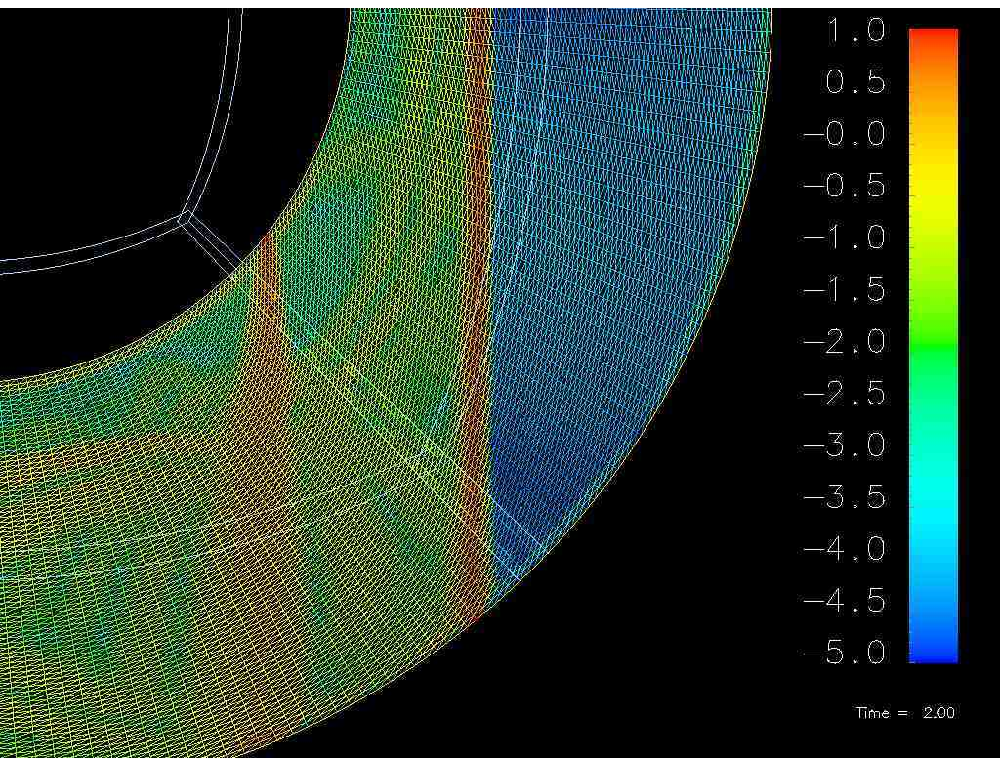}
\caption{Sod test on the six patches system. The plots show a close-up
view of the transmission of the shock front across one of the interfaces,
using a \emph{pseudo-Schlieren} plot: The function displayed is the
norm of the density gradient, in logarithmic scale. The interface is between 
the patch boundaries indicated by white lines.}
\label{fig:six_patches_sod_int}
\end{figure}

As explained in Section~\ref{sec:boundaries}, the boundaries are treated
by setting boundary ghost zones using an interpolation operation
before each time update, followed by a suitable coordinate transformation. 
Fig.~\ref{fig:six_patches_sod_int} shows a pseudo-Schlieren plot of the
transmission of the shock front across an interface. The code produces
a number of minor reflections as expected, but the overall shape of the
shock front remains intact.

\subsection{Sod test on the cubed sphere seven patches system}
\label{sec:sod_seven_patches}

The setup of this test is very similar to Section~\ref{sec:sod_six_patches},
but uses the cubed sphere seven patches system described in 
Section~\ref{sec:seven_patches}. The free parameters $r_0$ and $r_1$,
which specify the extent of the central cube and the location of the
outer spherical boundary, are set to $r_0 = 0.5$ and $r_1 = 2$. Again,
we use $n \times n \times n$ cells, with $n = 20, 40, 80$, on
each patch, and set the outer boundary ghost zones to the exact solution.

\begin{figure*}
\psfrag{x}{$x$}
\psfrag{rho}{$\rho$}
\psfrag{u}{$u$}
\psfrag{ua}{$u^1$}
\psfrag{t}{$t$}
\psfrag{rhonorm}{\hspace{-0.5cm}${||\rho - \rho_{exact}||}_1$}
\psfrag{unorm}{\hspace{-0.5cm}${||u - u_{exact}||}_1$}
\psfrag{uinorm}{\hspace{-0.5cm}${||u^i - u^i_{exact}||}_1$}
\begin{tabular}{cc}
\includegraphics[width=\columnwidth]{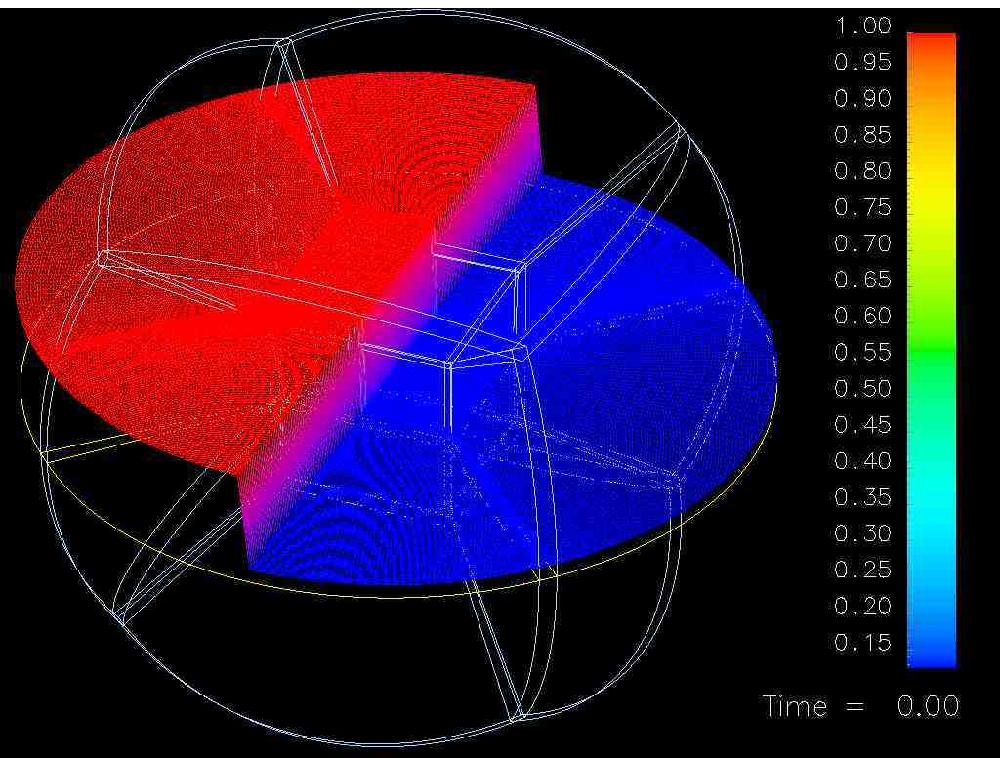} &
\includegraphics[width=\columnwidth]{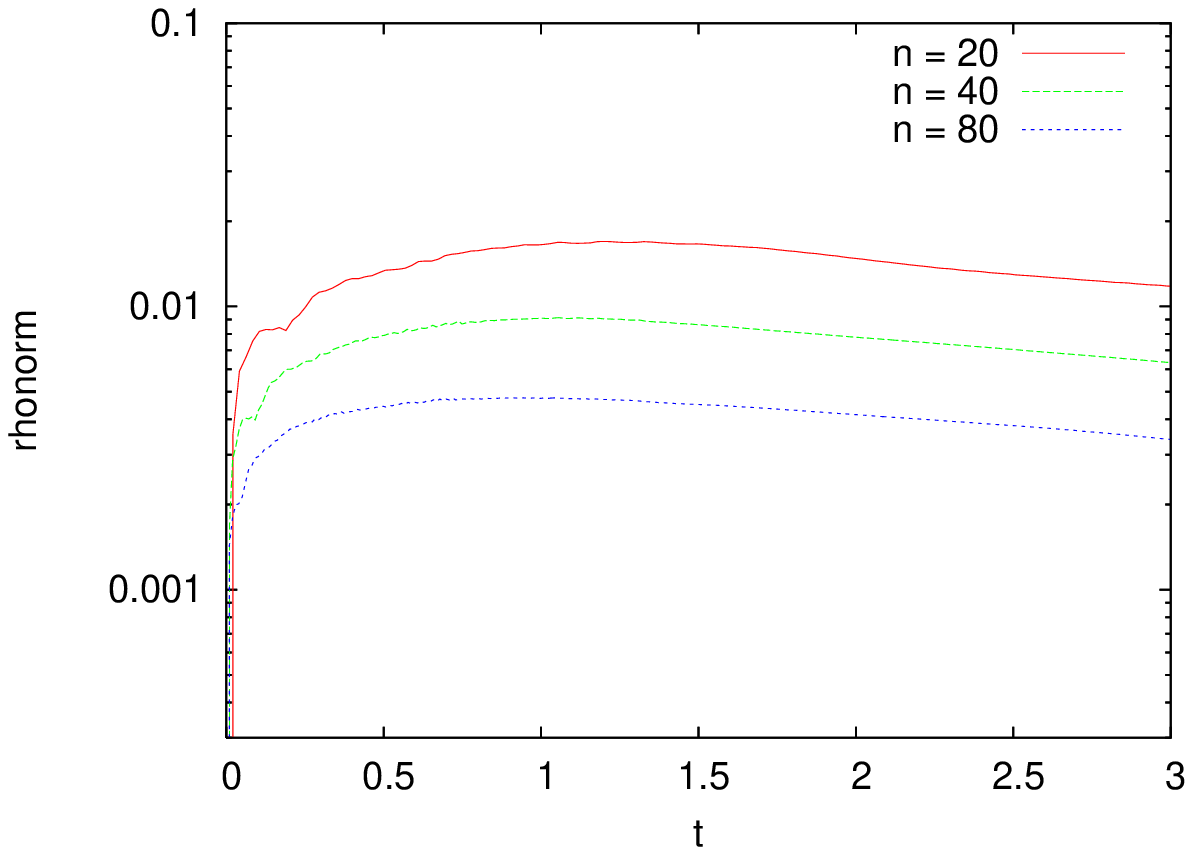} \\
\includegraphics[width=\columnwidth]{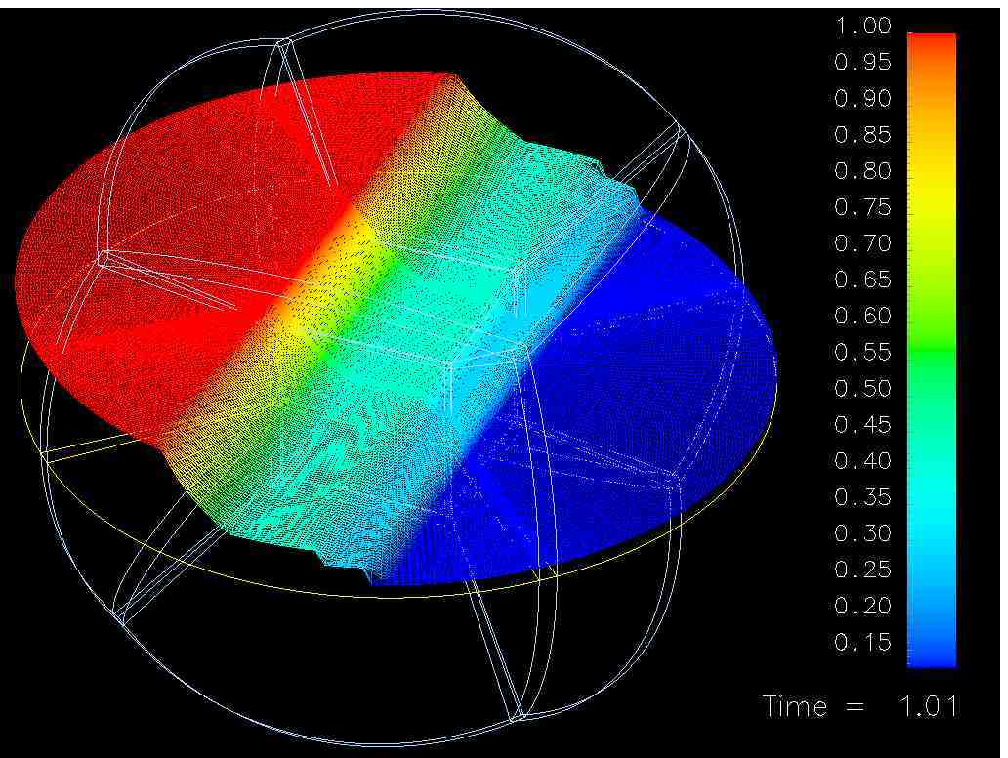} &
\includegraphics[width=\columnwidth]{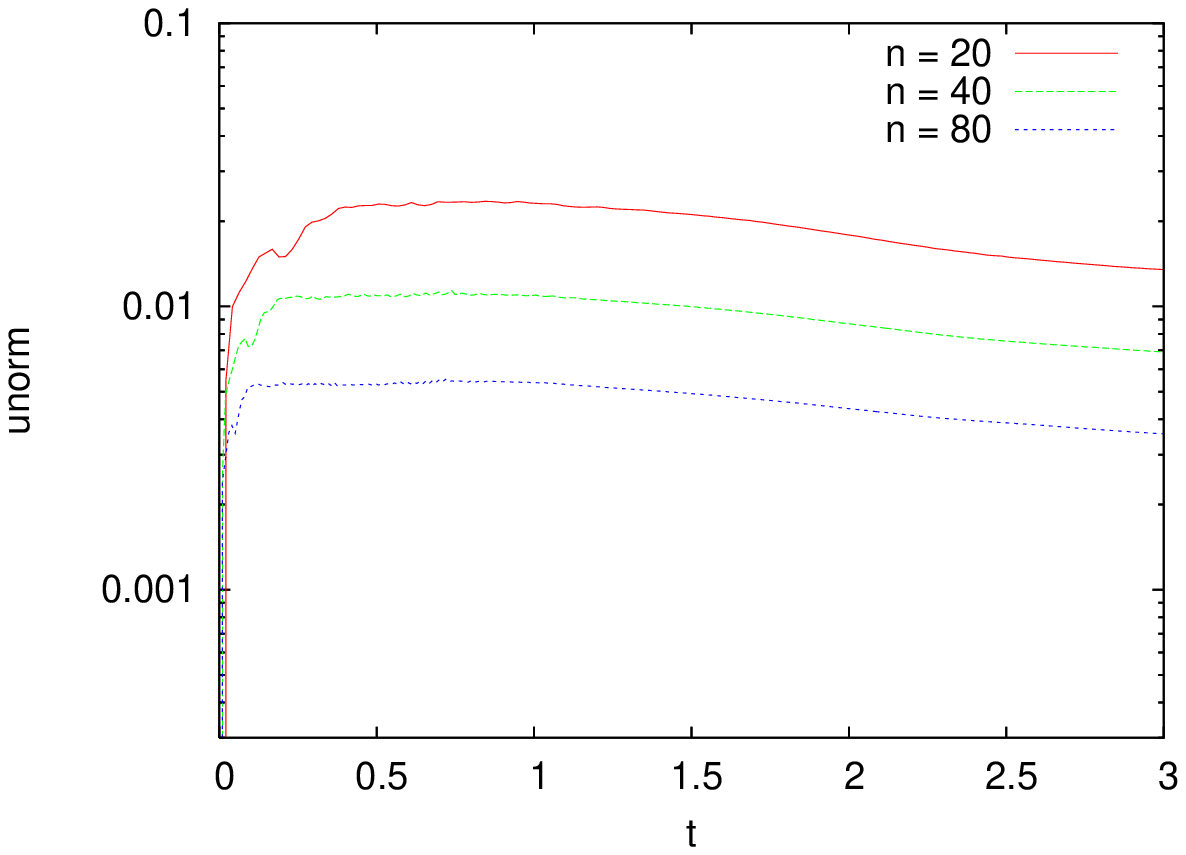} \\
\includegraphics[width=\columnwidth]{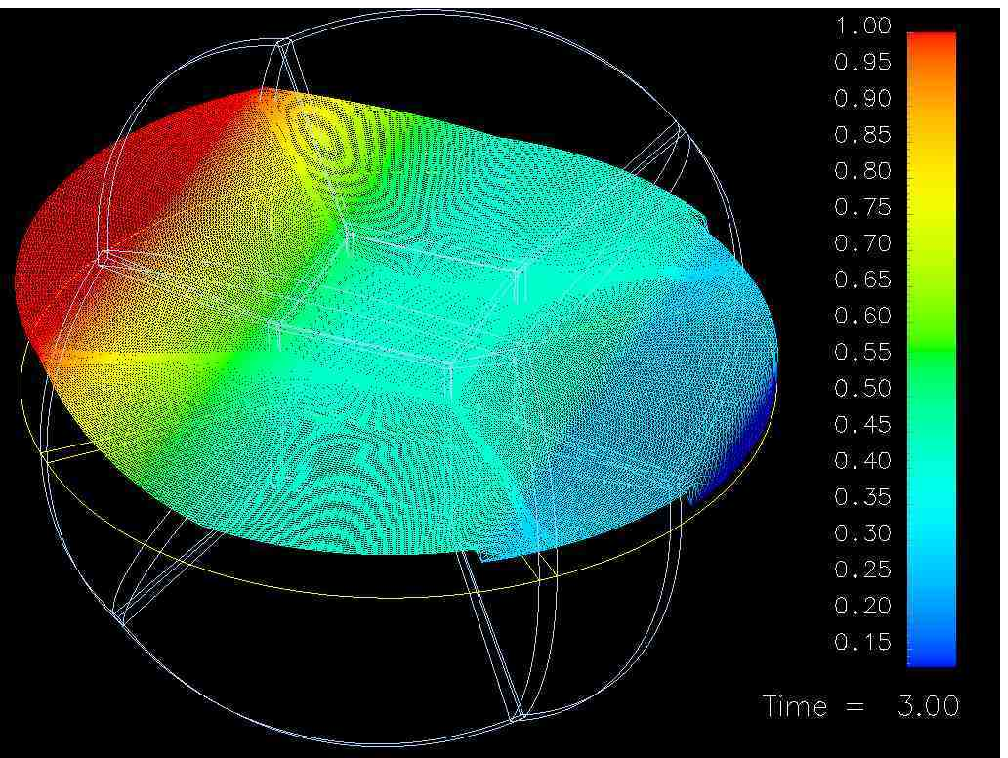} &
\includegraphics[width=\columnwidth]{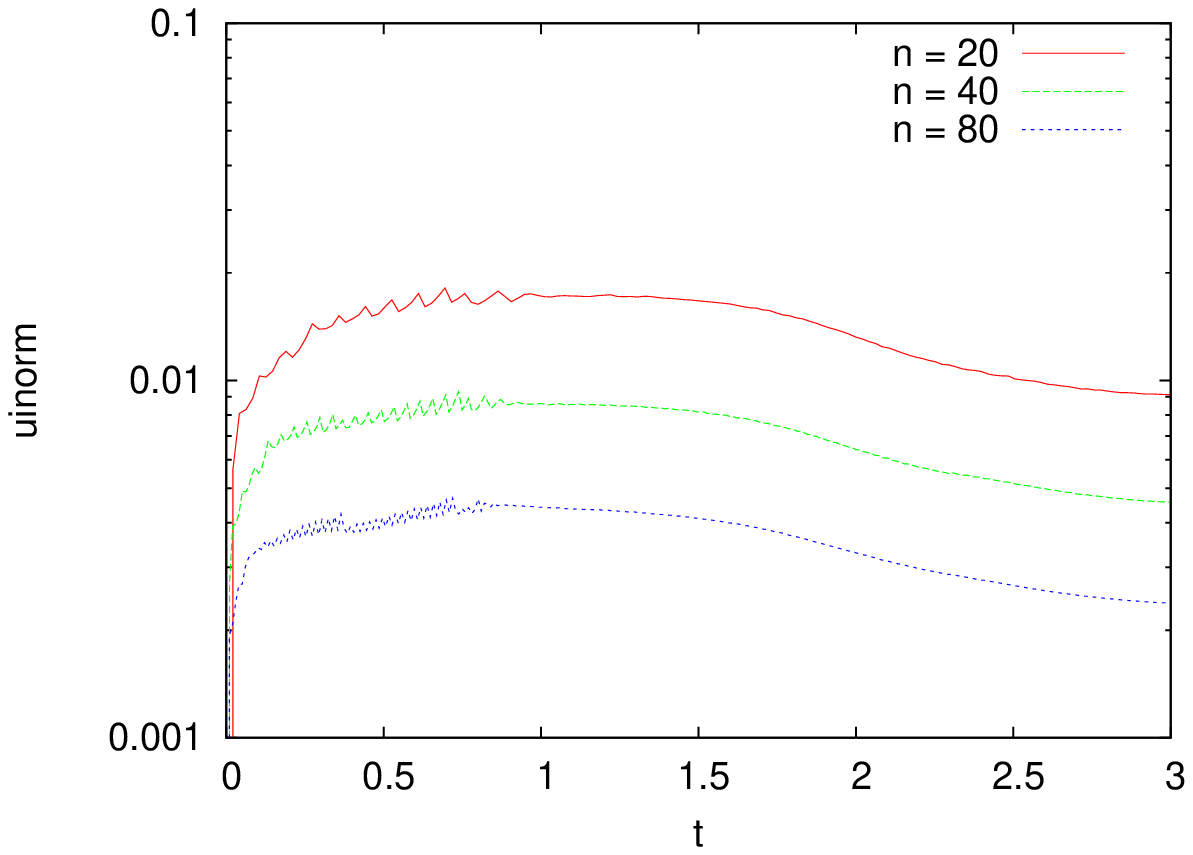} 
\end{tabular}
\caption{Sod test on the seven patches system. The left panel shows the
evolution of the density for the case
where each of the patches has resolution $80 \times 80 \times 80$. The
white lines indicate the boundaries of the seven patches which are used to
cover the computational domain. The yellow surface is the cut of the 
plane $z = 0$ with the grid boundaries, and the density function is shown
at $z = 0$. The right panel shows convergence in the primitive
variables.}
\label{fig:seven_patches_sod}
\end{figure*}

As Fig.~\ref{fig:seven_patches_sod} shows, the evolution proceeds very similar
to the six patches system and is convergent to the exact solution. 

\subsection{Sod test on the cubed sphere thirteen patches system}
\label{sec:sod_thirteen_patches}

\begin{figure*}
\psfrag{x}{$x$}
\psfrag{rho}{$\rho$}
\psfrag{u}{$u$}
\psfrag{ua}{$u^1$}
\psfrag{t}{$t$}
\psfrag{rhonorm}{\hspace{-0.5cm}${||\rho - \rho_{exact}||}_1$}
\psfrag{unorm}{\hspace{-0.5cm}${||u - u_{exact}||}_1$}
\psfrag{uinorm}{\hspace{-0.5cm}${||u^i - u^i_{exact}||}_1$}
\begin{tabular}{cc}
\includegraphics[width=\columnwidth]{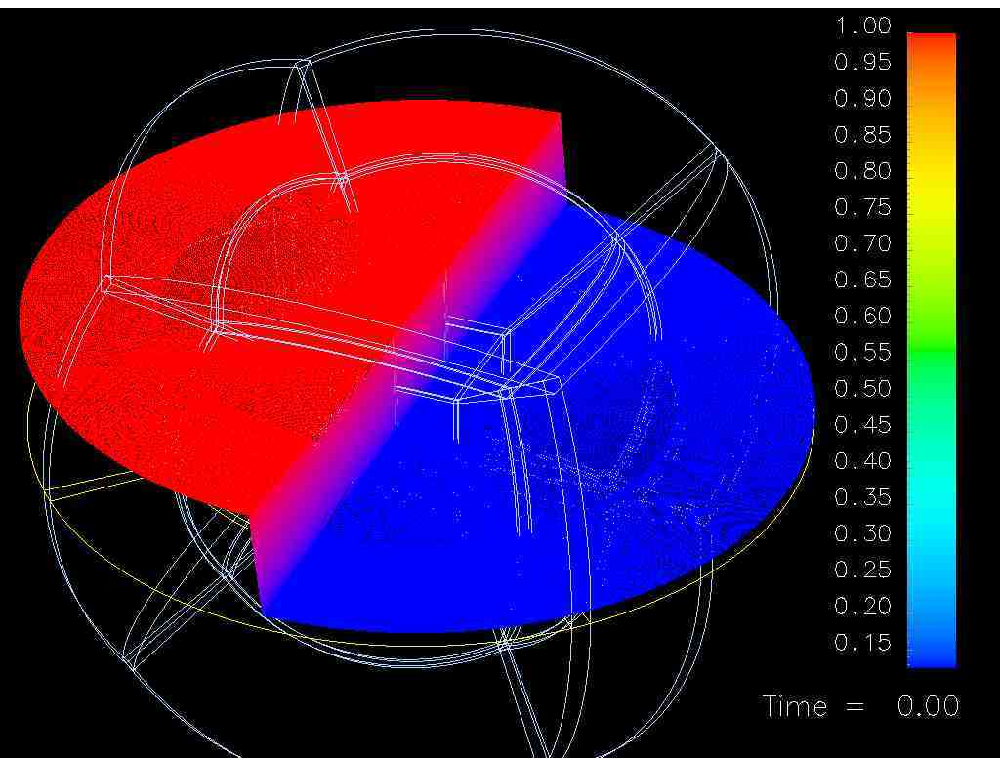} &
\includegraphics[width=\columnwidth]{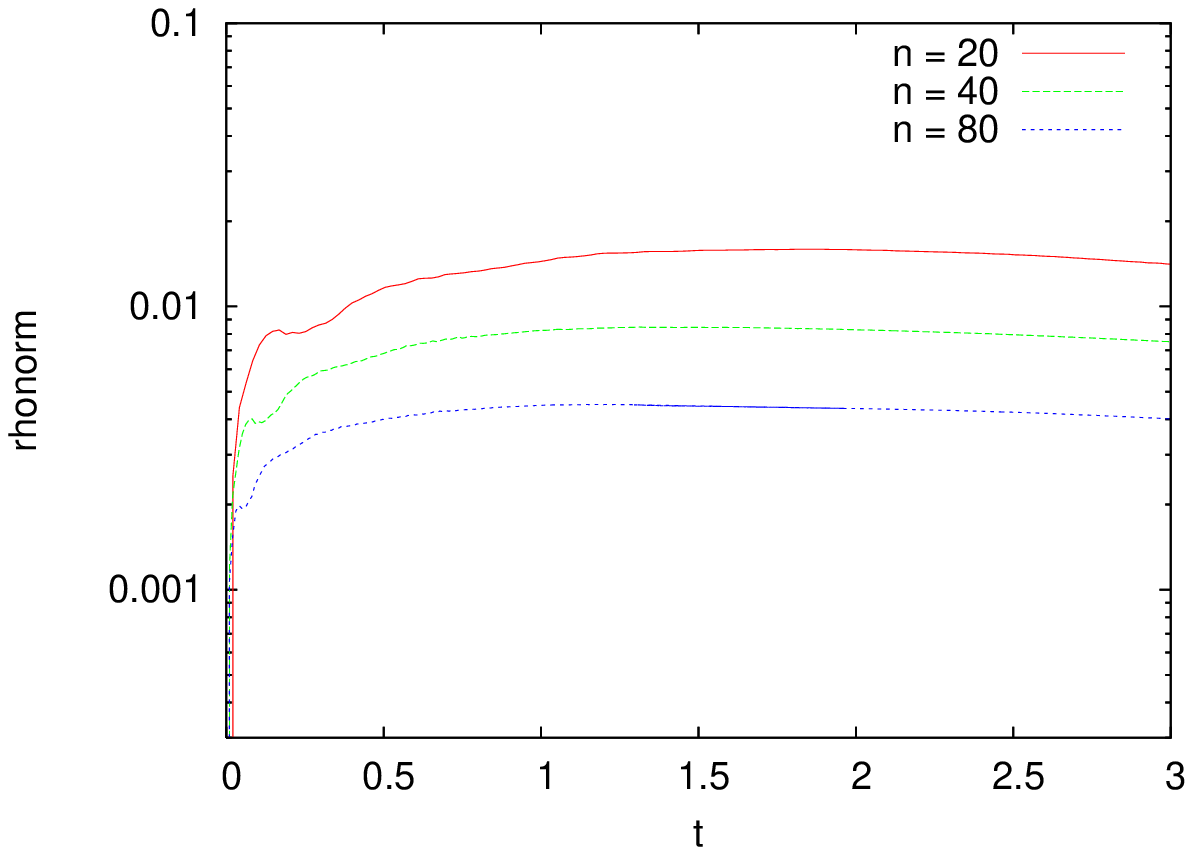} \\
\includegraphics[width=\columnwidth]{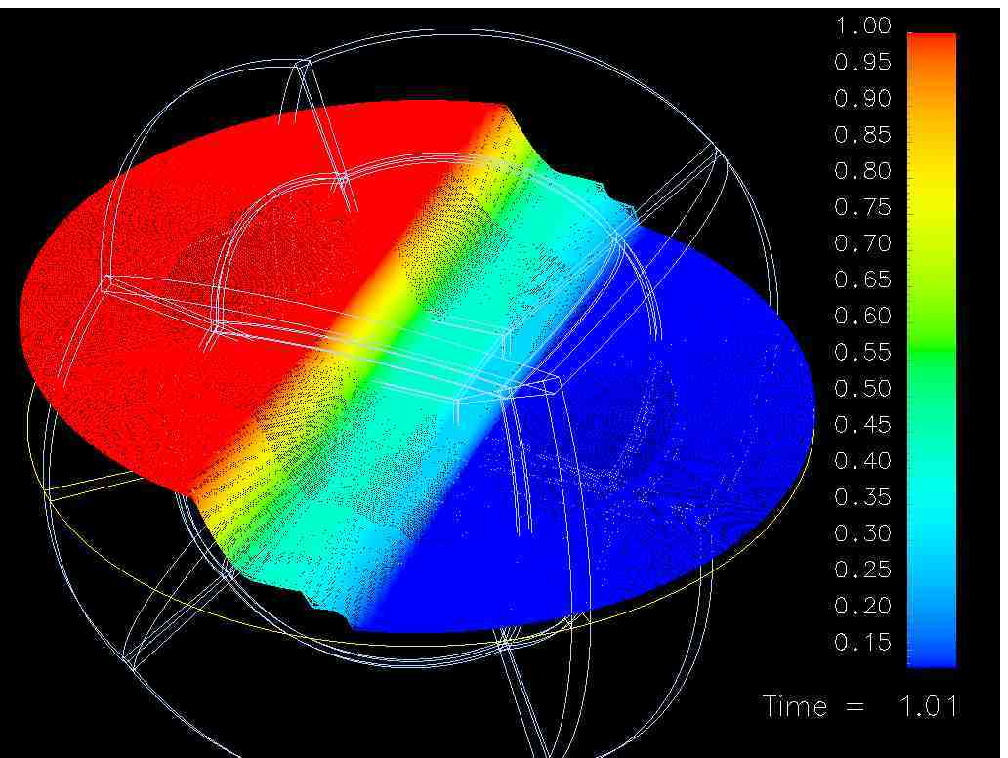} &
\includegraphics[width=\columnwidth]{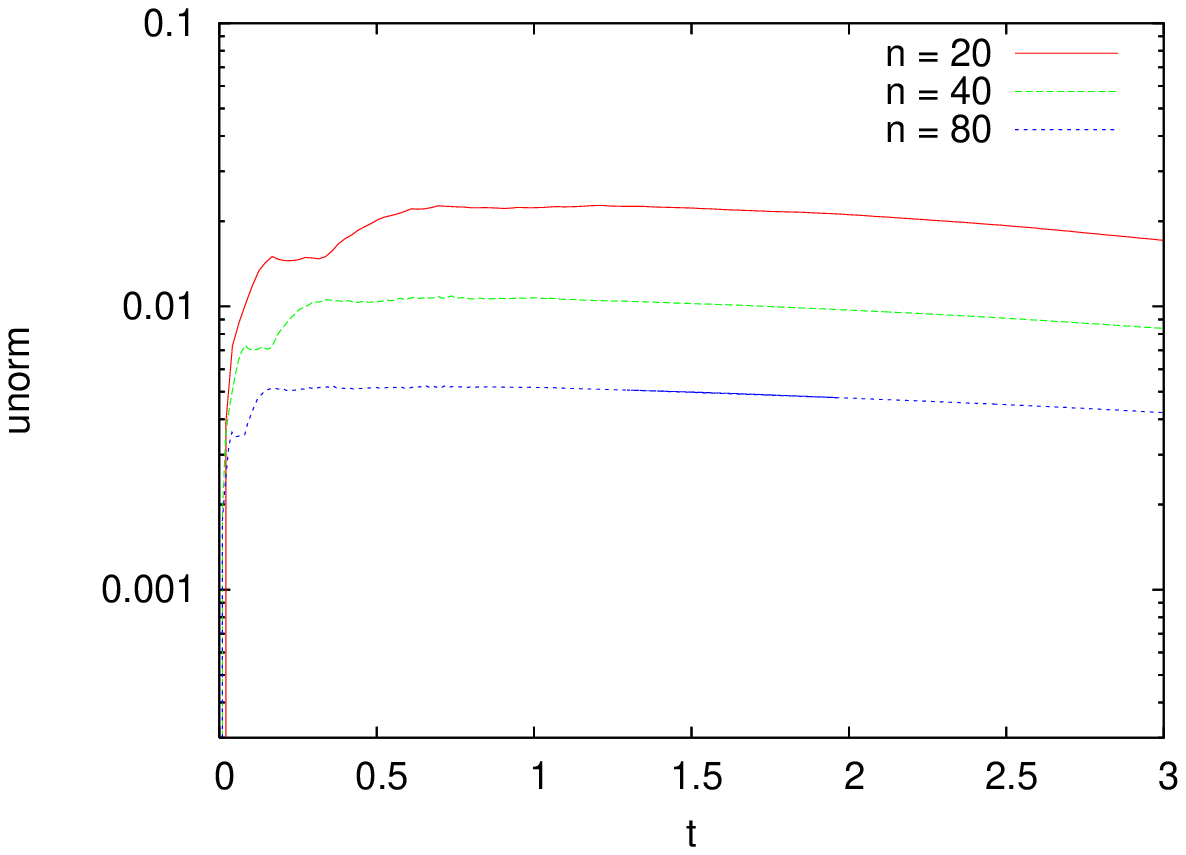} \\
\includegraphics[width=\columnwidth]{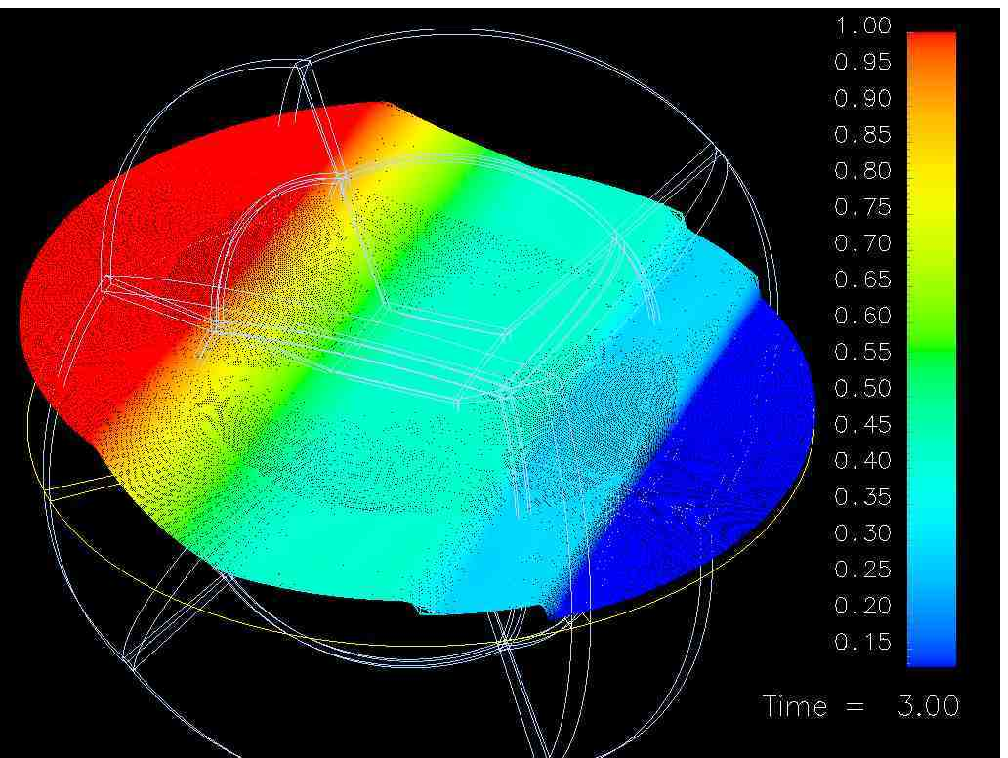} &
\includegraphics[width=\columnwidth]{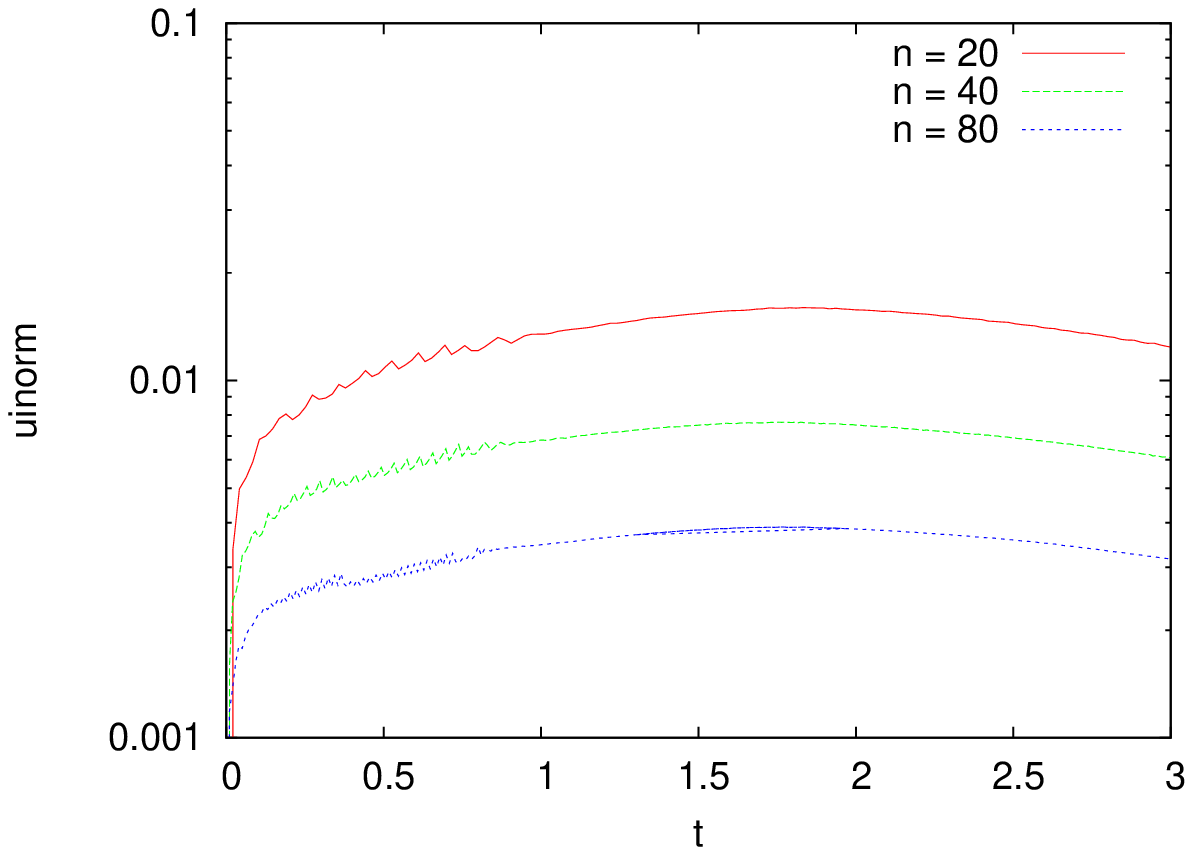} 
\end{tabular}
\caption{Sod test on the thirteen patches system. The left panel shows the
evolution of the density for the case
where each of the patches has resolution $80 \times 80 \times 80$. The
white lines indicate the boundaries of the thirteen patches which are used to
cover the computational domain. The yellow surface is the cut of the 
plane $z = 0$ with the grid boundaries, and the density function is shown
at $z = 0$. The right panel shows convergence in the primitive
variables.}
\label{fig:thirteen_patches_sod}
\end{figure*}

In the same way as in Sections~\ref{sec:sod_six_patches} and 
\ref{sec:sod_seven_patches}, we perform the Sod test on the cubed sphere
thirteen patches system described in Section~\ref{sec:thirteen_patches}.
Since this setup consists of seven inner patches matched to six outer
ones, we have three free parameters available. For purposes of this
particular test, we use $r_0 = 0.5$, $r_1 = 2$ and $r_2 = 3$.
The results are presented in Fig.~\ref{fig:thirteen_patches_sod}, and
show convergence to the exact solution.

\subsection{Uniformly rotating polytrope on the cubed sphere thirteen patches system}
\label{sec:rns}

An important equilibrium solution for general relativistic astrophysics are rotating 
polytropes, since they provide an approximate model for relativistic stars. They
are also particular well-suited to serve as code tests, since they involve a 
non-trivial spacetime geometry even in adapted coordinates \cite{Stergioulas98}, 
and their stability properties are well investigated in the case of stiff 
(polytropic index $\Gamma = 2$) and uniformly rotating solutions.

We constructed a uniformly rotating polytropic solution with the {\tt rns} code 
\cite{Stergioulas95}, assuming a stratification $P = K \rho^\Gamma$ with 
$K = 100$ and $\Gamma = 2$. The central density
is fixed to $\rho_c = 10^{-3}$, and the ratio of polar to equatorial radii
(in terms of the particular gauge choice in {\tt rns} \cite{Stergioulas98}) 
is set to $r_p/r_e = 0.7$, which produces a rapidly rotating model
(see Table~\ref{tab:rns}).
The solution, which is represented on a 
two-dimensional grid in {\tt rns}, is then mapped to every patch, and we apply 
the appropriate tensor coordinate transformations on all
quantities (the four-metric, its derivatives, and the primitive hydrodynamical
variables). 

\begin{table}
\begin{center}
\begin{tabular}{|l|l|l|}
\hline
Polytropic scale & $K$ & $100$ \\ 
Polytropic index & $\Gamma$ & $2$ \\
Central rest-mass density & $\rho_c$ & $10^{-3}$  \\  
Coordinate axis ratio & $r_p/r_e$ & $0.7$ \\ 
ADM mass & $M$ & $1.4906$ \\
Rest mass & $M_0$ & $1.5936$ \\ 
Equatorial proper radius & $R_e$ & $12.322$ \\
Equatorial inverse compactness & $R_e/M$ & $7.7321$ \\  
Angular momentum & $J$  & $1.3192$  \\ 
Normalized angular momentum & $J/M^2$  & $0.5938$  \\  
Kinetic over binding energy & $T/|W|$ &  $7.4792 \cdot 10^{-2}$  \\ 
(see caption) & $\Omega/\Omega_K$ & $0.7536$ \\
\hline
\end{tabular}
\end{center}
\caption{Parameters and integral quantities of the uniformly rotating polytrope
used as a code test for the cubed sphere thirteen patches system. 
The quantities $K$, $\Gamma$, $\rho_c$, and $r_p/r_e$ are parameters.
The quantity $\Omega$ is the angular velocity, while $\Omega_K$ is
the associated Keplerian velocity. Therefore, the mass-shedding sequence is located
at $\Omega_e/\Omega_K = 1$ (cf. also \cite{Stergioulas98}).}
\label{tab:rns}
\end{table}

To specify a multi-patch system, we need to choose resolutions and set
the free parameters $r_0, r_1, r_2$ (see Section~\ref{sec:thirteen_patches})
which correspond to the location of the cube boundary, the spherical boundary
between the seven patches and the six patches system, and the spherical
outer boundary. We use $r_0 = 1$, $r_1 = 4$ and $r_2 = 14$ for these locations
in the test case presented here, but we have also experimented with different
values and obtained very similar results. The number of cells per patch
are set to $n^3$, where $n = 20, 40, 80$. During evolution, we do \emph{not}
enforce the polytropic constraint $P = K \rho^\Gamma$, but rather use
the gamma law $P = (\Gamma - 1) u$. 

\begin{figure}
\includegraphics[width=\columnwidth]{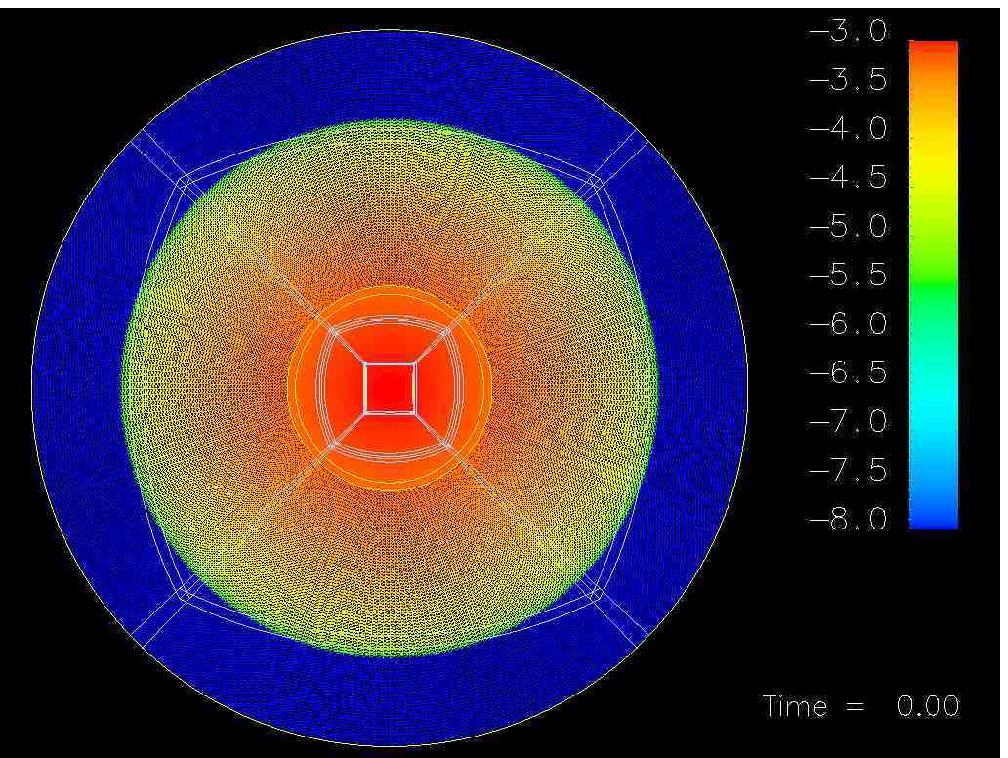}
\includegraphics[width=\columnwidth]{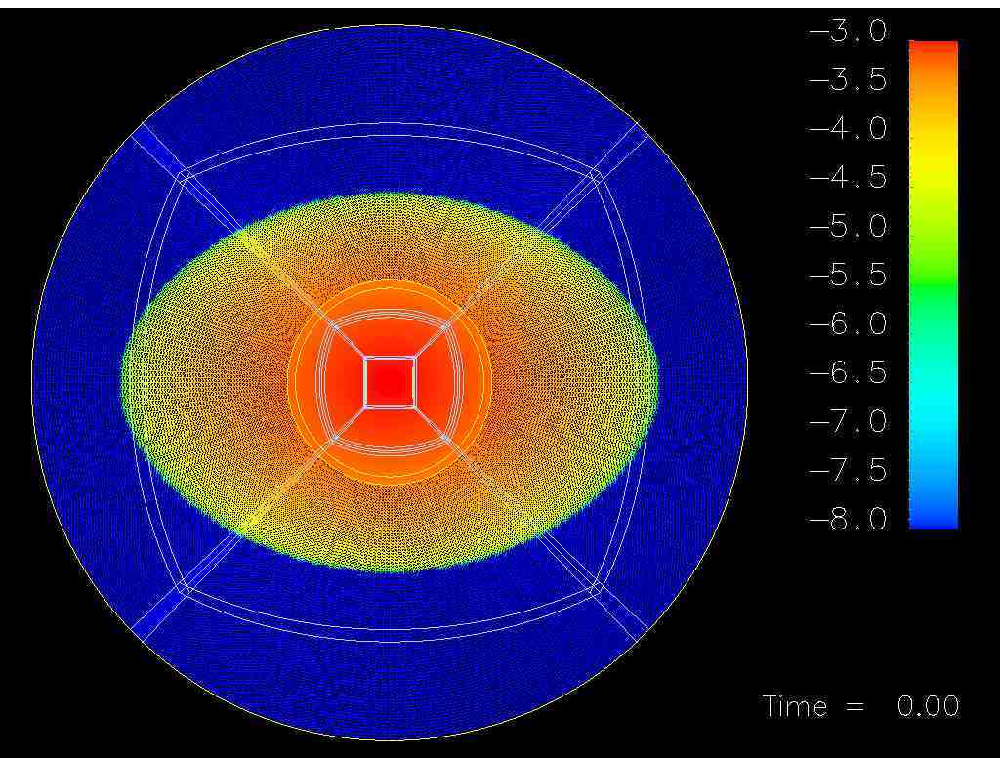}
\caption{Rotating neutron star on the cubed sphere system of thirteen patches.
The initial data is produced with the {\tt rns} code \cite{Stergioulas95}
then mapped to the each patch and transformed to the local coordinate
system. These plots show the logarithm of the density function in the
equatorial plane $z = 0$ (left) and the plane $x = 0$ (right). For visual
clarity, the density is cut below $10^{-8}$; the actual initial density
in the atmosphere is $10^{-10}$.}
\label{fig:rns_rho_t0}
\end{figure}

\begin{figure*}
\psfrag{t}{$t$}
\psfrag{rhoc}{$\rho_c$}
\psfrag{dDerrornorm1}{${||D - D_{exact}||}_1$}
\psfrag{dQterrornorm1}{${||Q_t - (Q_t)_{exact}||}_1$}
\psfrag{dQierrornorm1}{${||Q_i - (Q_i)_{exact}||}_1$}
\psfrag{massintegral}{\hspace{-1.5cm}$(M_0 - (M_0)_{exact})/(M_0)_{exact}$}
\psfrag{angmomintegral}{\hspace{-0.5cm}$(J - J_{exact})/J_{exact}$}
\begin{tabular}{cc}
\includegraphics[width=\columnwidth]{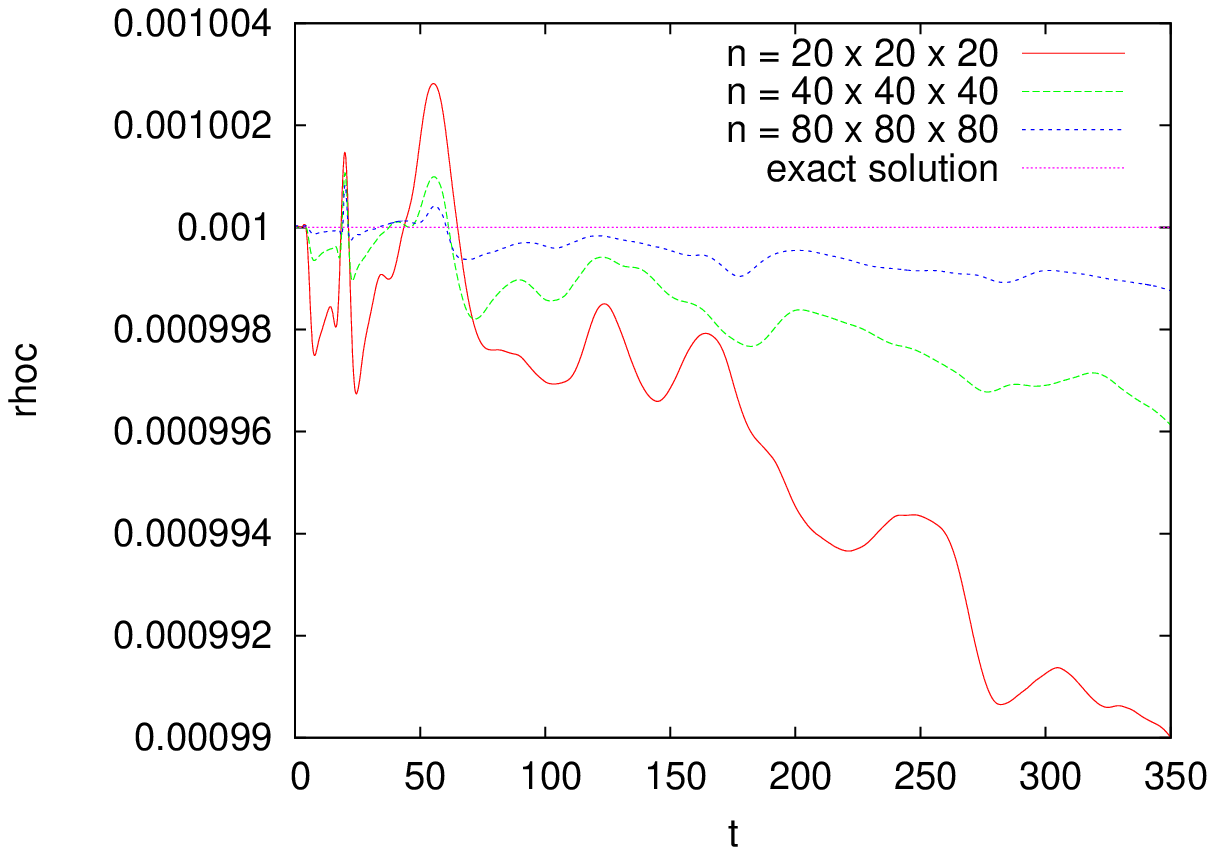} & 
\includegraphics[width=\columnwidth]{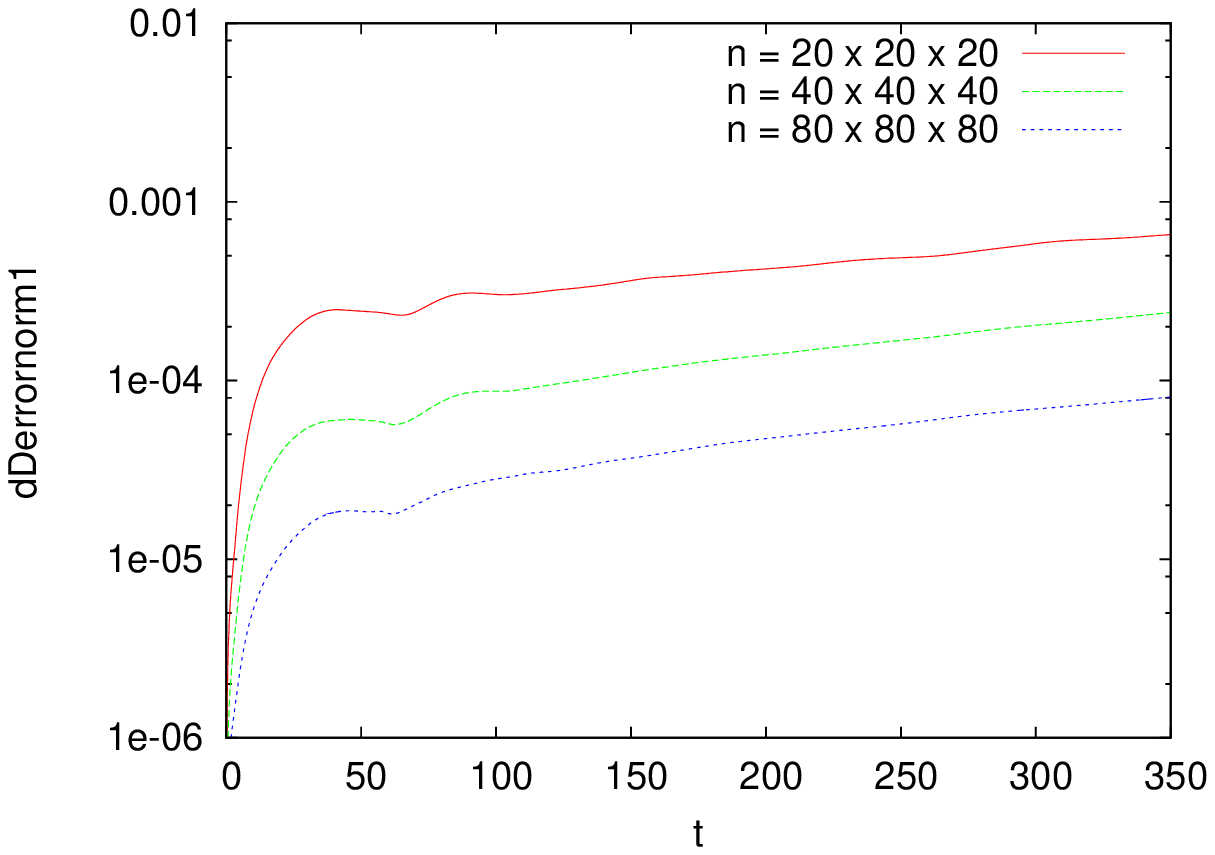} \\
\includegraphics[width=\columnwidth]{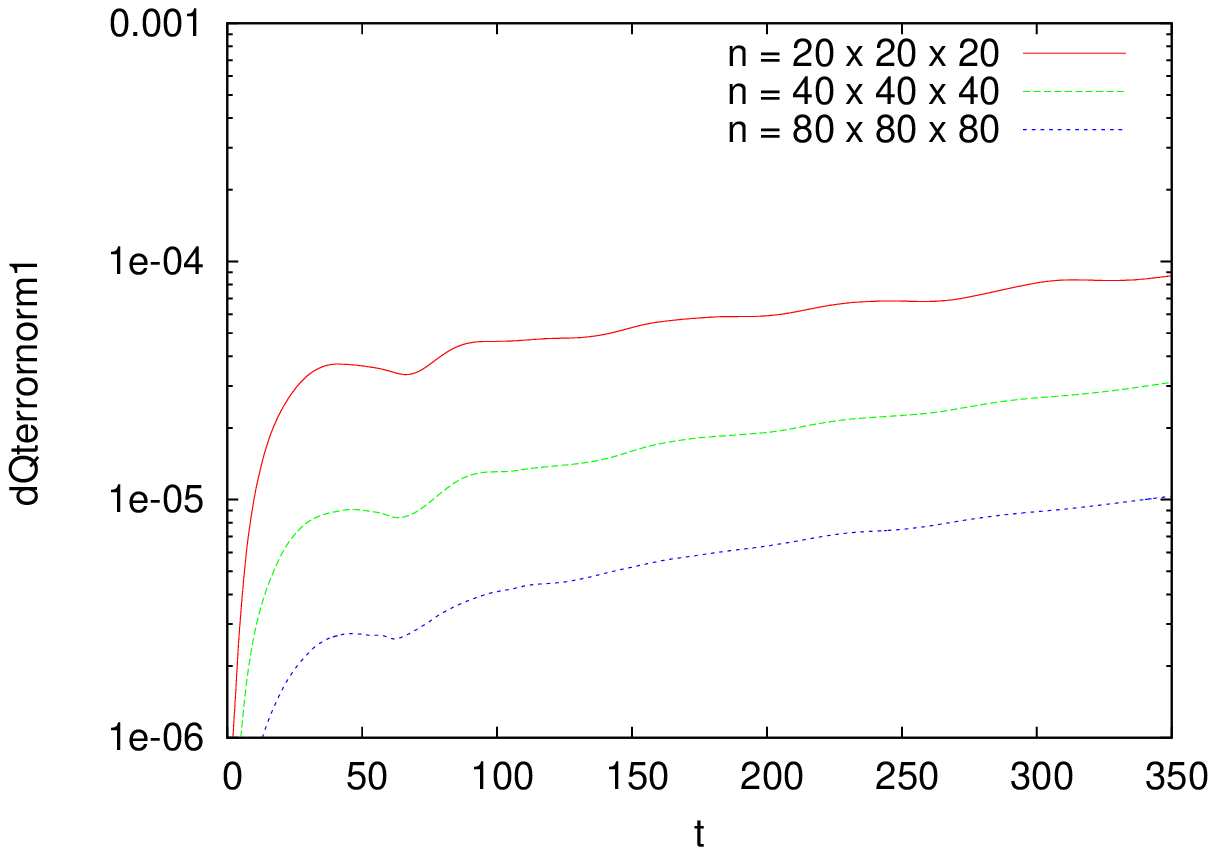} & 
\includegraphics[width=\columnwidth]{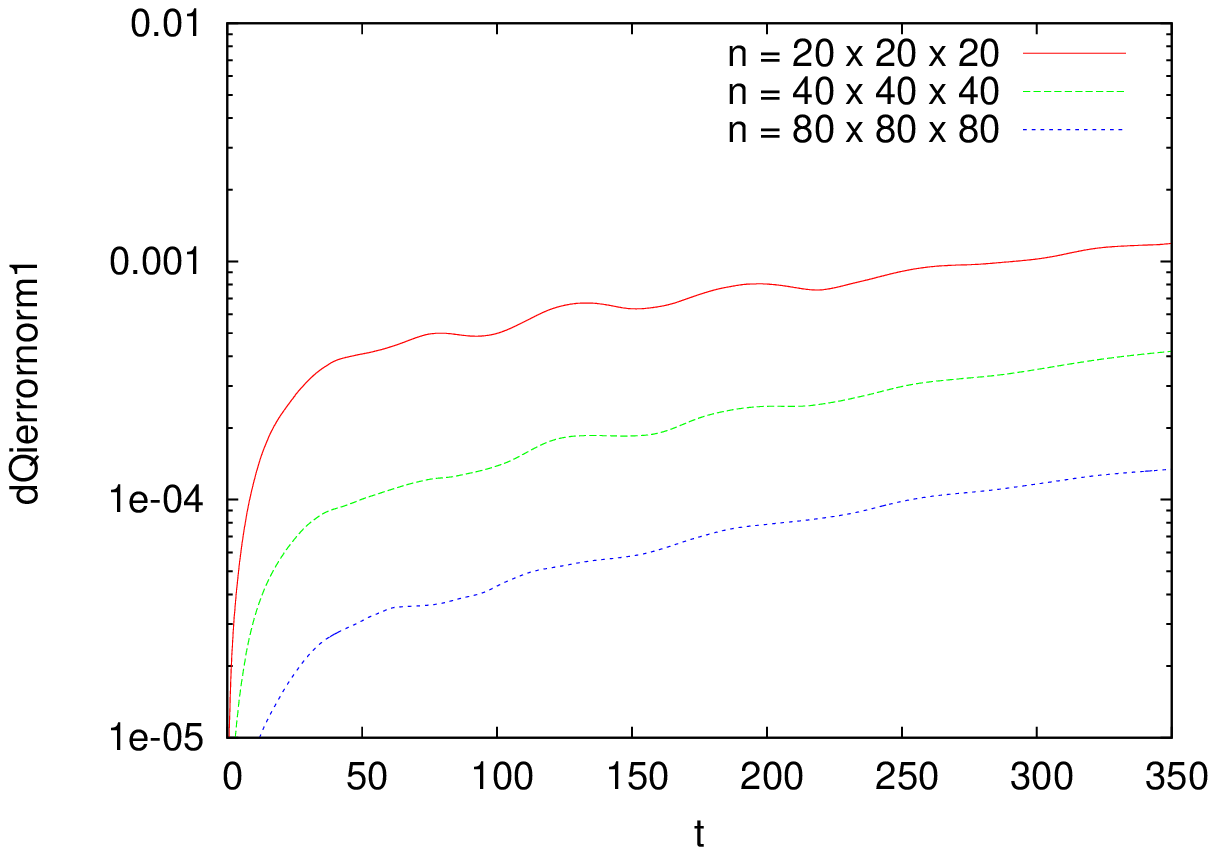} \\
\includegraphics[width=\columnwidth]{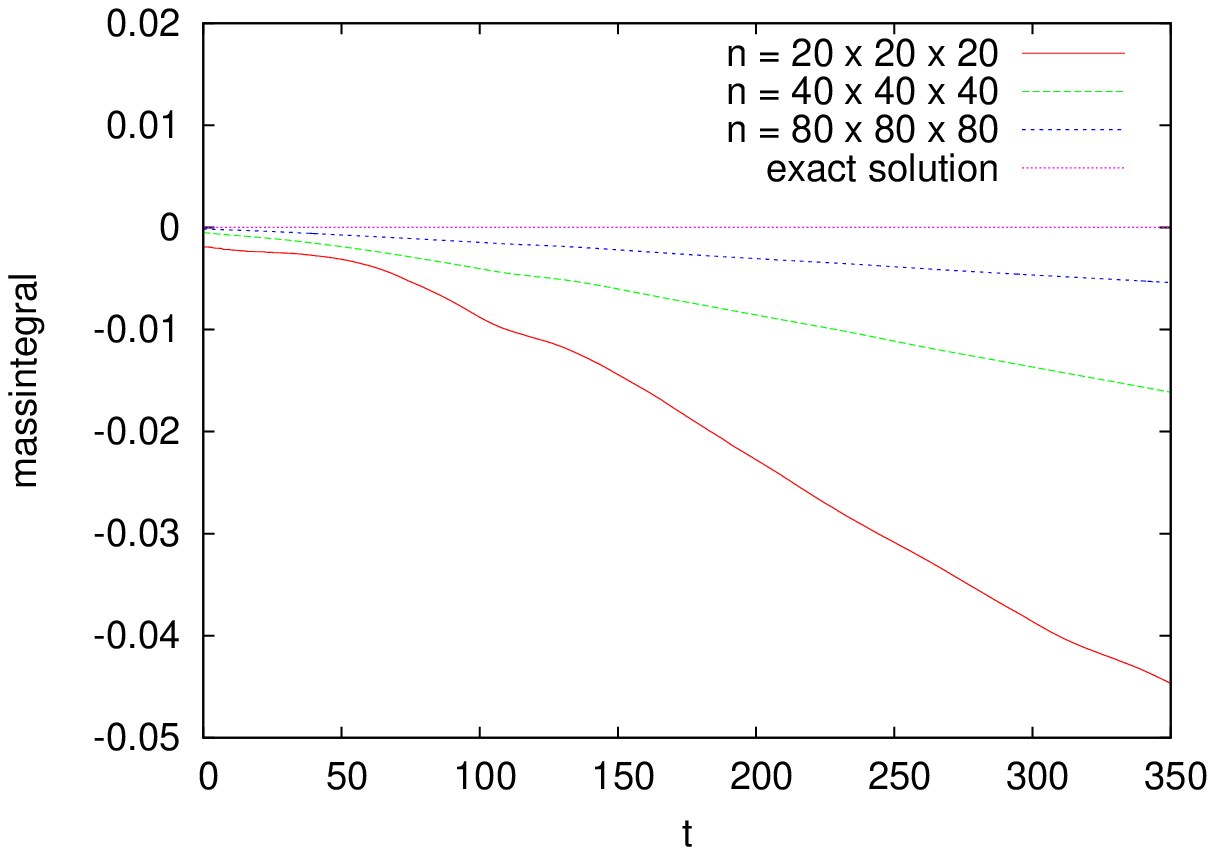} &
\includegraphics[width=\columnwidth]{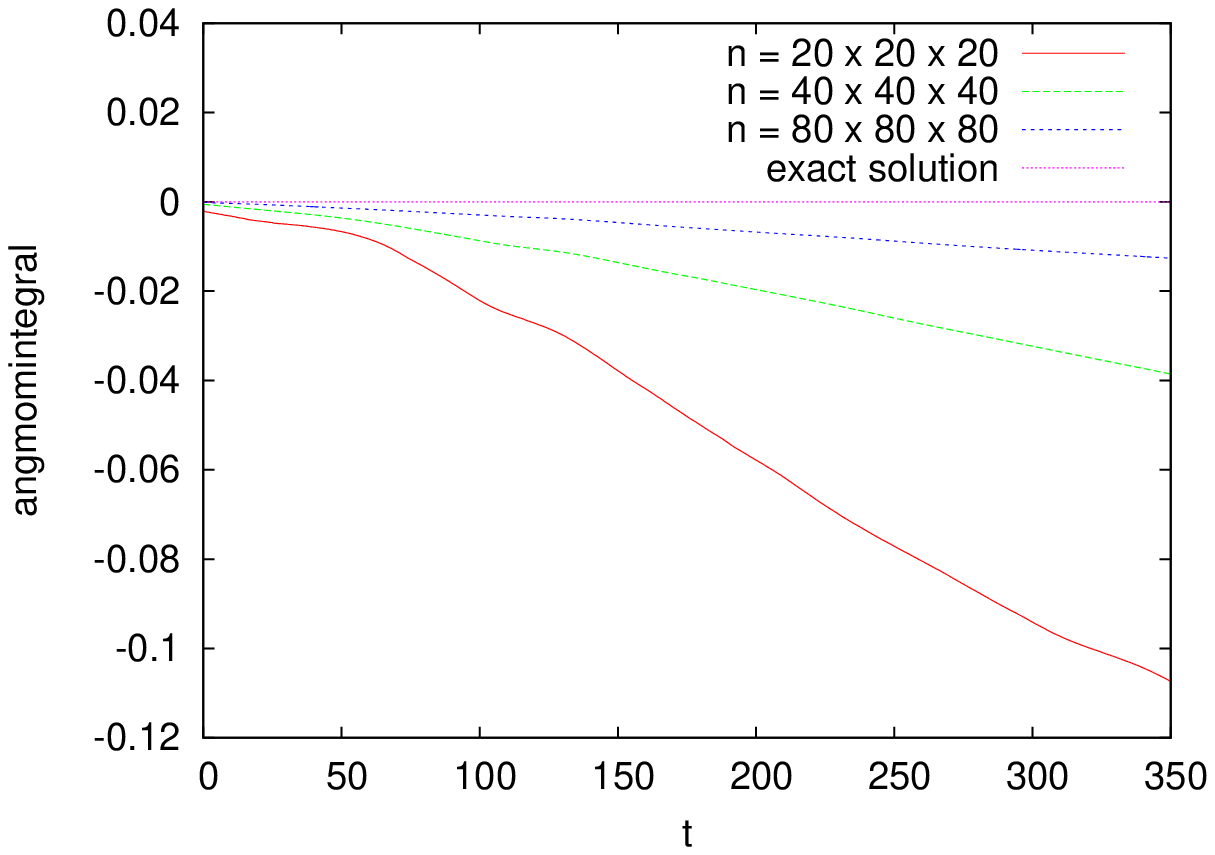}
\end{tabular}
\caption{Evolution of a rotating neutron star on the cubed sphere system 
of thirteen patches with different resolutions. The star is evolved up to
a coordinate time of $350$, which translates into about ten dynamical
times ($t_D = R_e \sqrt{R_e/M} \approx 35.428$). The top left panel shows the
central density, the top right, middle left, and middle right panels show
convergence of the conserved variables, and the last two plots show
the errors in rest mass (bottom left panel) and angular momentum (bottom
right panel). For a definition of these quantities see \cite{Stergioulas98}.}
\label{fig:rns_convergence}
\end{figure*}

Fig.~\ref{fig:rns_rho_t0} shows the initial data mapped to the thirteen patches system.
Note that the density outside the star is not exactly zero, but set to the
atmosphere value (see Section~\ref{sec:discretization}) since our techniques
are unable to handle vacuum-matter interfaces. Fig.~\ref{fig:rns_convergence} shows an evolution
of the polytrope for about 10 dynamical times\footnote{We make use of the common choice
$t_D = R_e \sqrt{R_e/M}$ as a measure of dynamical time, where $R_e$ is the proper equatorial 
circumferential radius and $M$ the ADM mass of the star.}. Note that we use the \emph{conserved}
variables to show convergence: the three-velocity error is dominated by low-density, but fast
material leaving the stellar surface, which does not carry a high amount of momentum, however.
This reflects the approximation we make with using an artificial atmosphere in the first place.
Therefore, convergence tests in the $u^i$ are dominated by noise. 

The evolution of the star
exhibits artifact oscillations and a linear drift in the central density, but converges 
to the stationary exact solution. The average errors acquired by the star per
dynamical time are listed in Table~\ref{tab:rns_errors}.

\begin{table}
\begin{center}
\begin{tabular}{|l|r|r|r|}
\hline
Quantity & $20 \times 20 \times 20$ & $40 \times 40 \times 40$ & 
  $80 \times 80 \times 80$ \\
\hline
Central density & $0.1\%$ & $0.04\%$ & $0.013\%$ \\
Rest mass & $0.45\%$ & $0.16\%$ & $0.054\%$ \\
Angular momentum & $1.07\%$ & $0.38\%$ & $0.13\%$  \\
\hline
\end{tabular}
\end{center}
\caption{Rotating neutron star on the cubed sphere system of thirteen patches. 
This table shows average errors acquired per dynamical time in different 
resolutions.}
\label{tab:rns_errors}
\end{table}

\subsection{Equilibrium accretion torus on a Schwarzschild background}
\label{sec:torus}

This test models a thick accretion disk in the test-fluid limit around a 
black hole. The equilibrium structure of these solutions has been discussed in several
seminal papers by Fishbone and Moncrief \cite{Fishbone76} and Abramowicz, Jaroszy{\'n}ski,
Sikora and Koz{\l}owski \cite{Abramowicz78, Kozlowski78, Jaroszynski80}. For the purposes
of this test, we use a polytropic disk with constant specific angular momentum on a 
Schwarzschild background. 

The details of how to construct these models can be found in the aforementioned publications,
we will only give a quick overview here: Starting with a spacetime with a line
element of the form 

\begin{equation}
ds^2 = g_{tt} dt^2 \, + \, 2 g_{t \phi} dt d\phi \, + \, g_{rr} dr^2
  \, + \, g_{\theta \theta} d\theta^2 \, + \, g_{\phi \phi} d\phi^2, 
\end{equation}
we seek stationary solutions for a rotating fluid given by the energy-momentum tensor 
eqn.~\ref{eqn:fluid_tab}. For the four velocity of the fluid, we assume the form

\begin{equation}
u^\mu = (u^t, 0, 0, u^\phi)^T
\end{equation}
and define the angular velocity $\Omega = u^\phi/u^t$ and specific angular momentum 
$l = -u_\phi/u_t$. From the normalization $u^\mu u_\mu = -1$ we obtain

\begin{equation}
(u_t)^{-2} = g^{tt} - 2 g^{t \phi} l + g^{\phi \phi} l^2.
\end{equation}

From this quantity, and using $l = const$ and the polytropic relation $P = K \rho^\Gamma$, 
we can calculate the specific internal energy 
$\epsilon \equiv u / \rho$ and the density \cite{Kozlowski78}:

\begin{eqnarray}
\epsilon & = & \left( \frac{(u_t)_0}{u_t} - 1 \right) / \Gamma \\
\rho & = & \left( {\frac{\Gamma - 1}{K \epsilon}} \right)^\frac{1}{\Gamma - 1} \nonumber
\end{eqnarray}

Here, $(u_t)_0$ is a free parameter of the solution. The particular parameter values 
we use to construct the torus are $K = 10^{-2}$, $\Gamma = 4/3$, $l = 4.5$, and
$(u_t)_0 = -0.98$. The atmospheric density is set to $10^{-12}$ (compared to a maximal
density of $2 \cdot 10^{-2}$). In addition, we apply a radial coordinate transformation
of the form $r = \exp(\bar{r})$, i.e., a logarithmic grid, to resolve the region close to
the black hole better. The inner and outer boundaries are treated by extrapolation onto
ghost zones as mentioned in Section~\ref{sec:boundaries}. 

We use a cubed sphere six patches system, with $15 \times 15 \times 40$, 
$30 \times 30 \times 80$ and $60 \times 60 \times 160$ cells per patch (remember that 
our convention in eqn.~\ref{eqn:six_patches} implies
that the third local coordinate corresponds to the location of each sphere in the global
coordinate space, whereas the first two coordinates roughly correspond to the angular
directions on the sphere). The free parameters specifying the boundary location are set to
$r_0 = 6$ and $r_1 = 50$.
This places the inner radius outside of the horizon, which is
necessary since there is a coordinate singularity at the horizon in
these coordinates.  This is possible since the background is not
evolved and since there is no matter near the horizon.

\begin{figure*}
\psfrag{t}{$t$}
\psfrag{rhomax}{$\rho_{max}$}
\psfrag{dDerrornorm1}{${||D - D_{exact}||}_1$}
\psfrag{dQterrornorm1}{${||Q_t - (Q_t)_{exact}||}_1$}
\psfrag{dQierrornorm1}{${||Q_i - (Q_i)_{exact}||}_1$}
\psfrag{massintegral}{\hspace{-1.5cm}$(M_0 - (M_0)_{exact})/(M_0)_{exact}$}
\psfrag{angmomintegral}{\hspace{-0.5cm}$(J - J_{exact})/J_{exact}$}
\begin{tabular}{cc}
\includegraphics[width=\columnwidth]{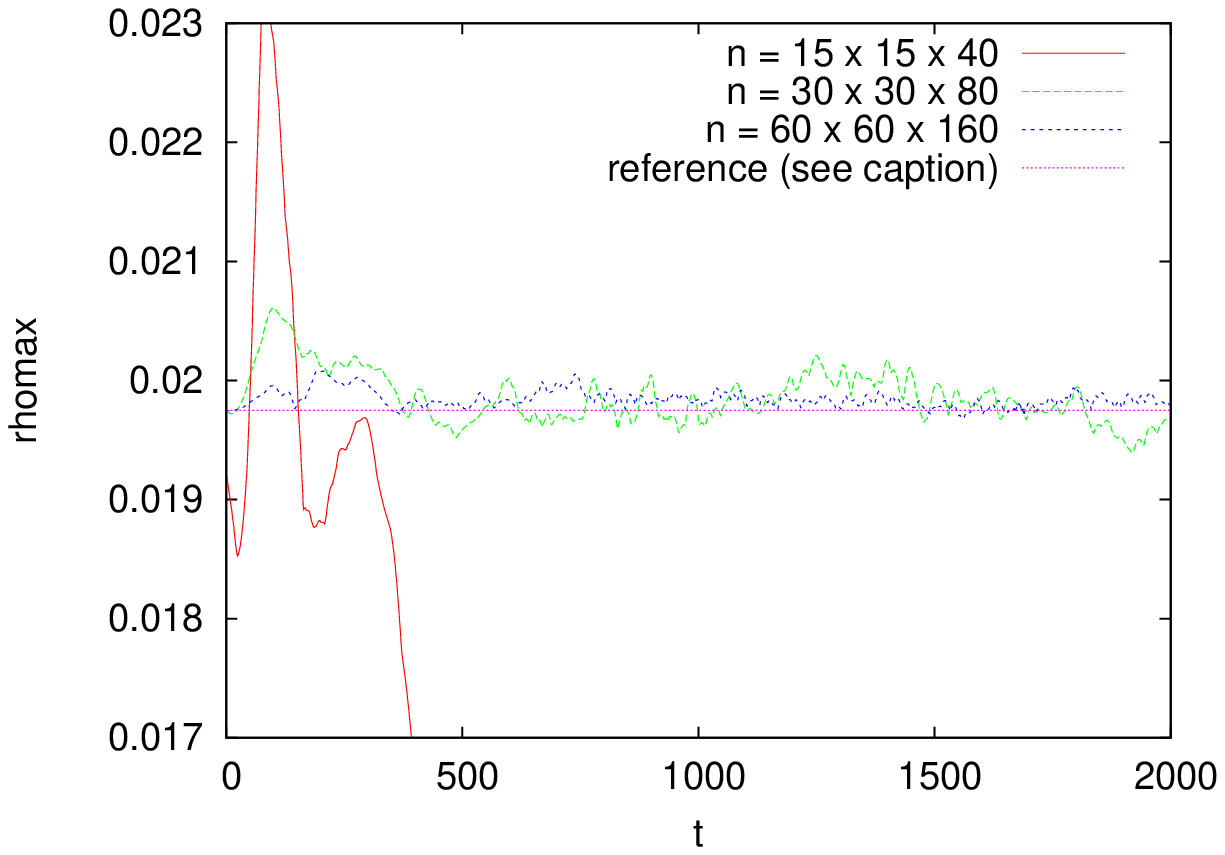} & 
\includegraphics[width=\columnwidth]{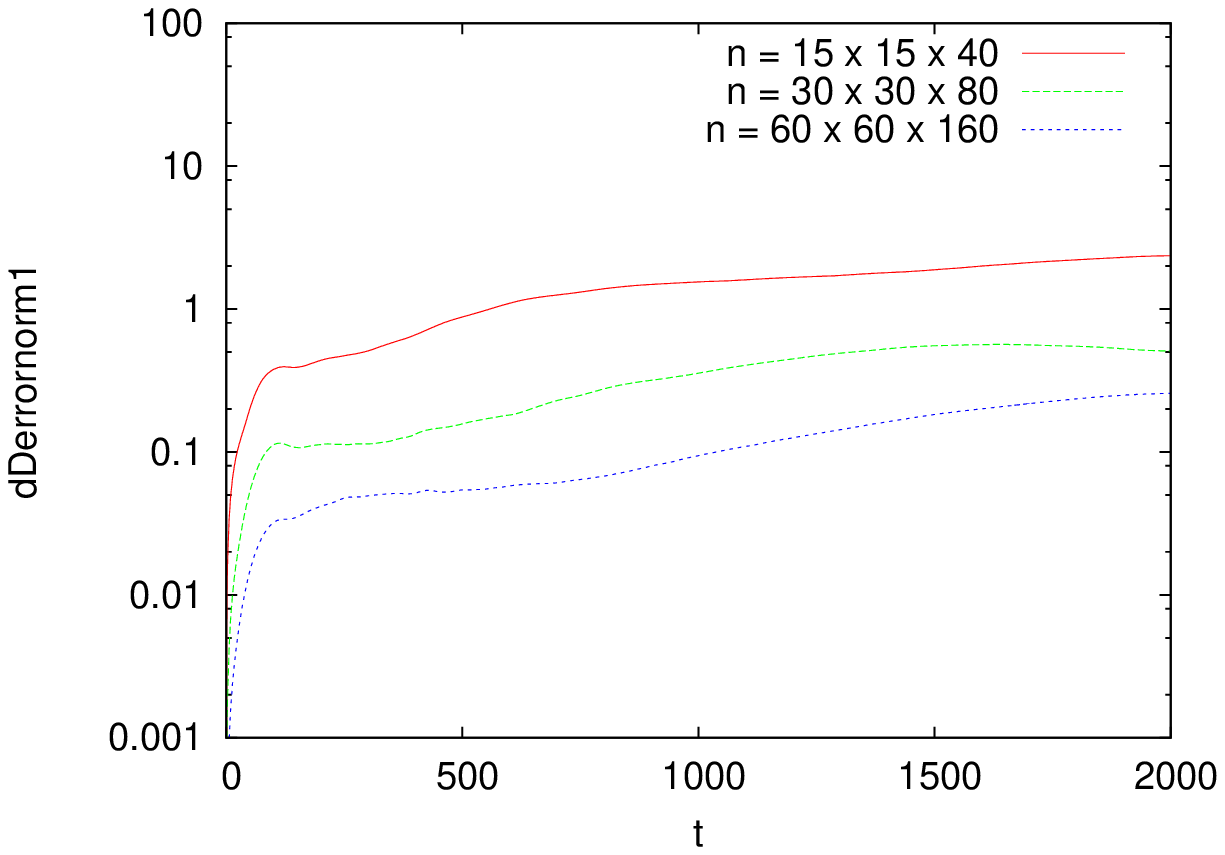} \\
\includegraphics[width=\columnwidth]{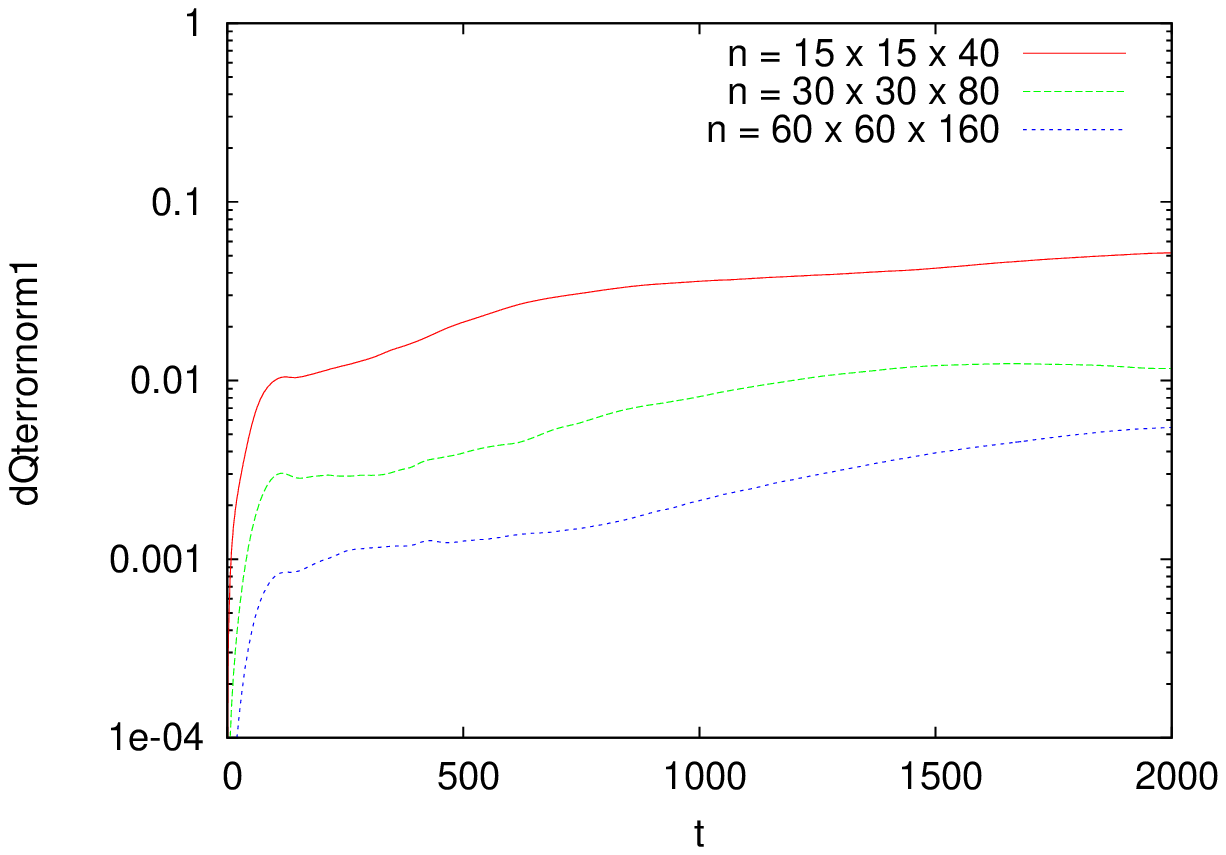} & 
\includegraphics[width=\columnwidth]{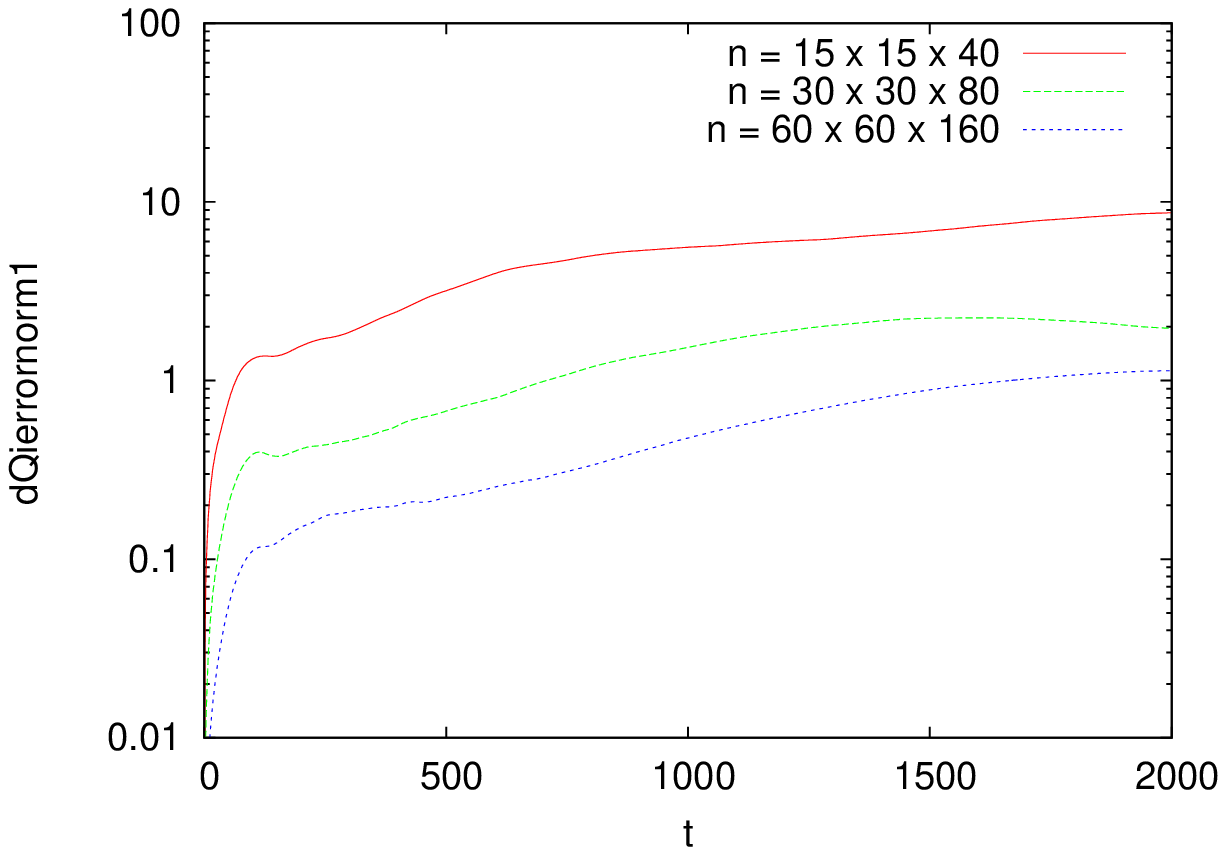} \\
\includegraphics[width=\columnwidth]{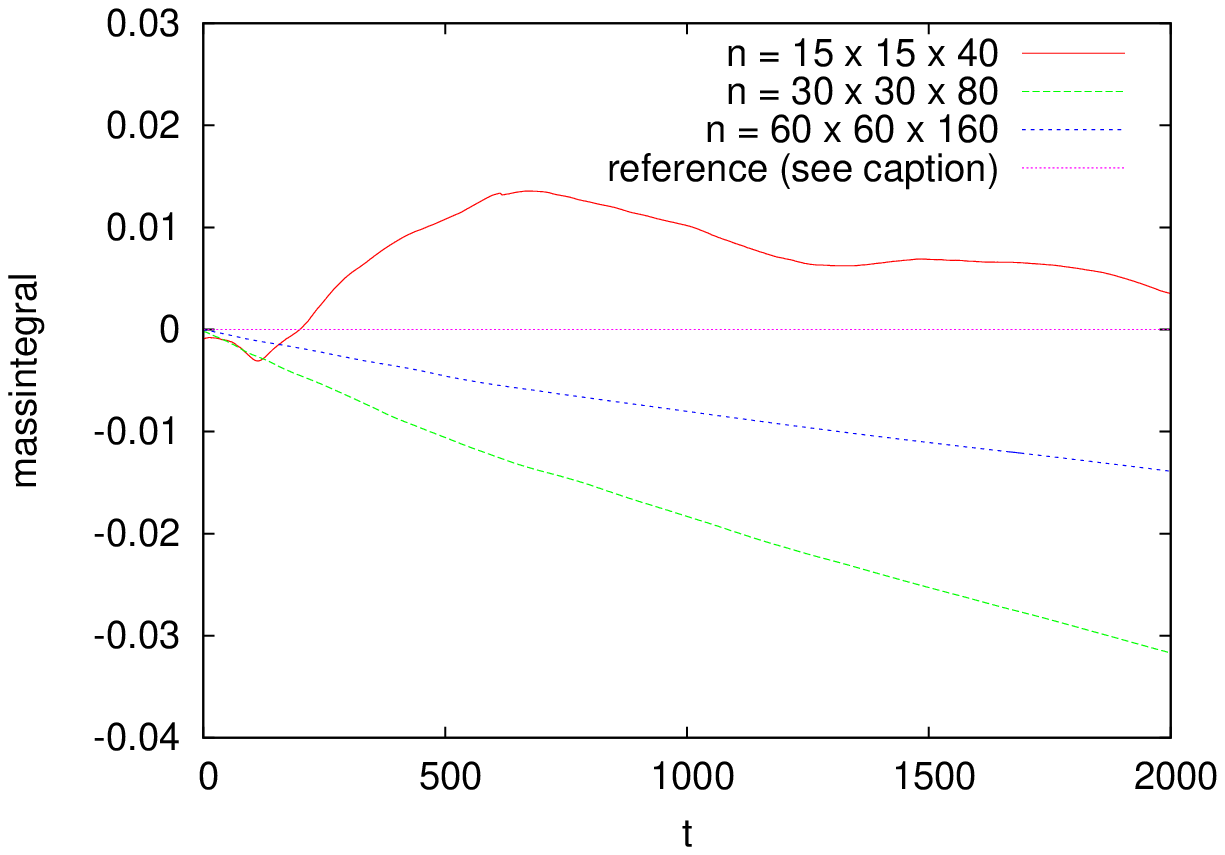} &
\includegraphics[width=\columnwidth]{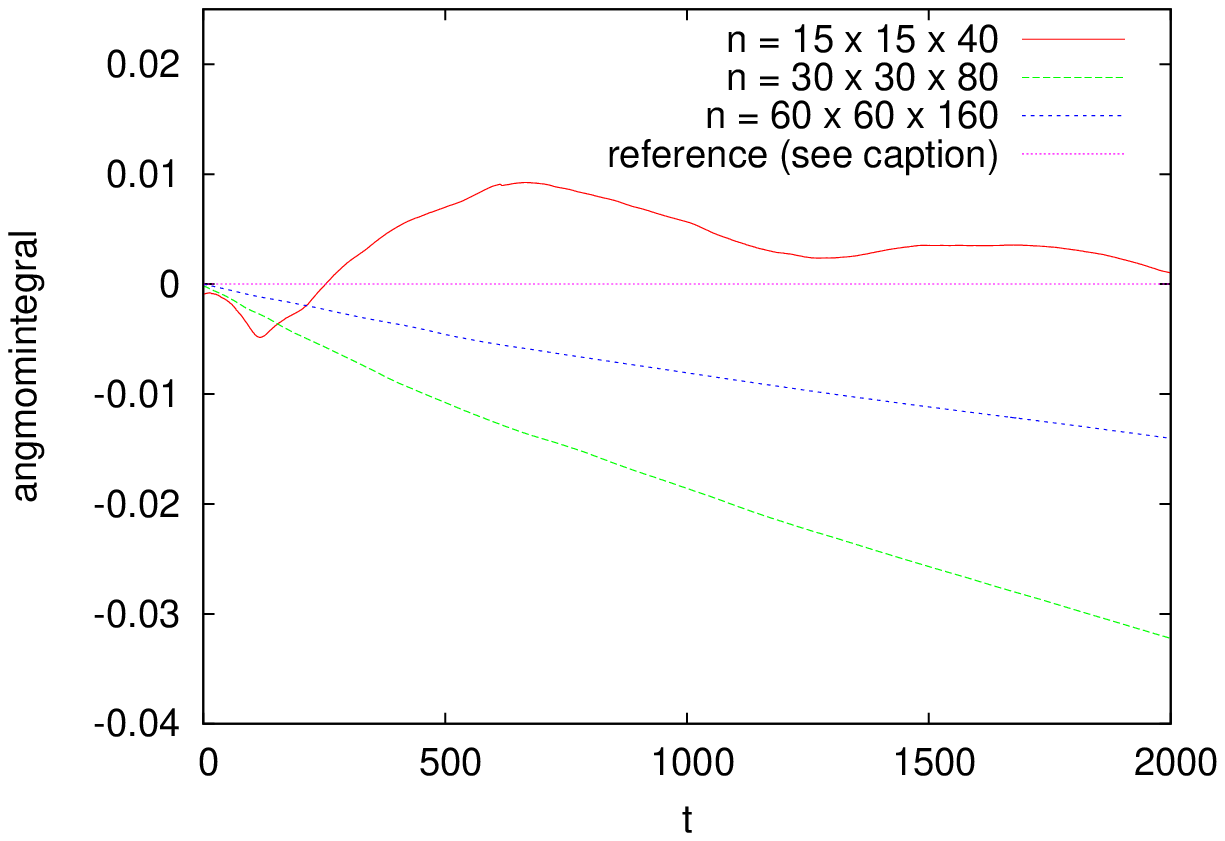}
\end{tabular}
\caption{Equilibrium accretion torus on a Schwarzschild background, using the cubed
sphere six patches system. The disk is evolved up to a coordinate time of $2000$, which translates into about ten 
rotational times at the location of maximal density. The top left panel shows the
maximal density, the top right, middle left, and middle right panels show
convergence of the conserved variables, and the last two plots show
the errors in rest mass (bottom left panel) and angular momentum (bottom
right panel). For a definition of these quantities see \cite{Stergioulas98}. (The ``reference'' solution mentioned in the text is the value
corresponding to the initial data on grid of $120 \times 120 \times 320$ cells
per patch.) Note that
the case $n = 15 \times 15 \times 40$ is clearly under-resolved.}
\label{fig:torus_convergence}
\end{figure*}

The evolution of the torus for about ten rotational times
(in terms of the radius of maximal density) is displayed in Fig.~\ref{fig:torus_convergence}.
The lowest resolution case, $n = 15 \times 15 \times 40$ per patch, is clearly under-resolved.
The error ratio between the higher resolutions is about $0.44$, which is only slightly better 
than first order convergence (Table~\ref{tab:torus_errors} collects the errors for different
resolutions). The numerical techniques we are using are ``at best'' second-order 
accurate, but drop to first order near maxima and at the boundary to the atmosphere. In cases where internal
turbulence is present, as is expected in realistic accretion disks
\cite{Shakura73, Balbus91a}, it will likely dominate the solution errors.

\begin{figure*}
\begin{tabular}{cc}
\includegraphics[width=\columnwidth]{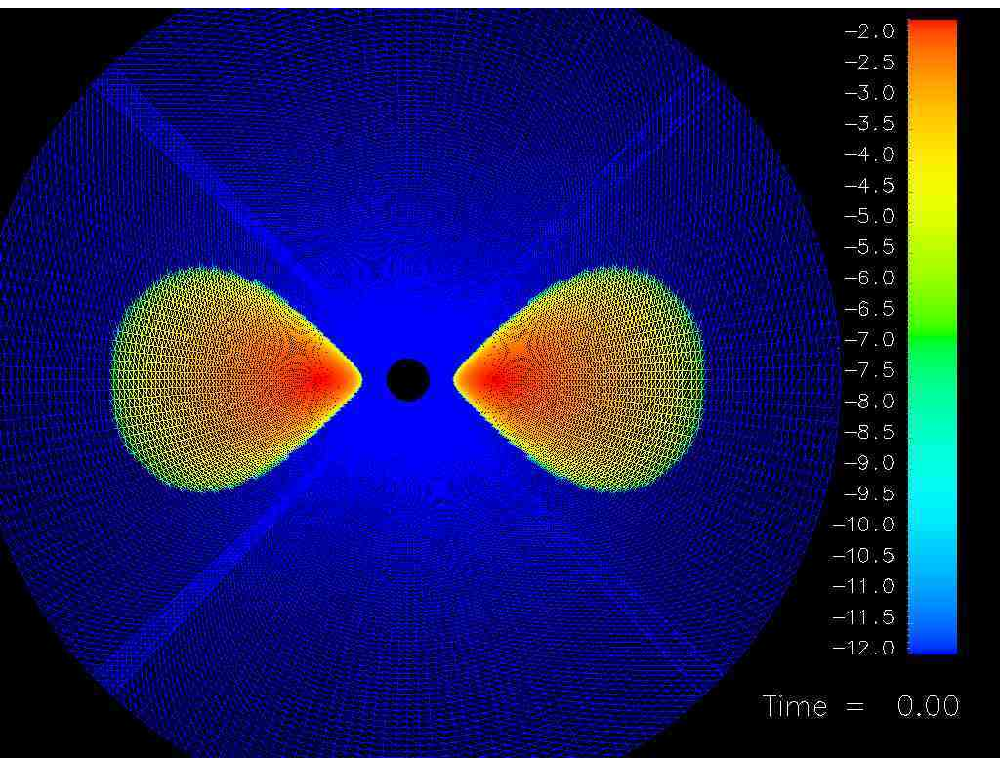} &
\includegraphics[width=\columnwidth]{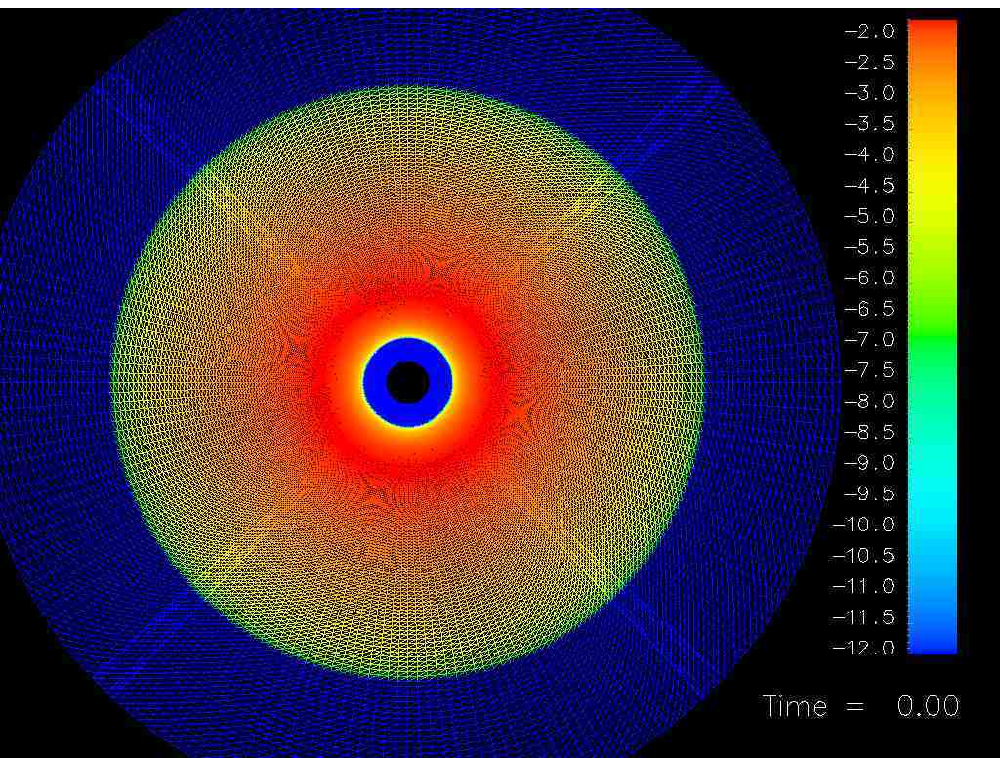} \\
\includegraphics[width=\columnwidth]{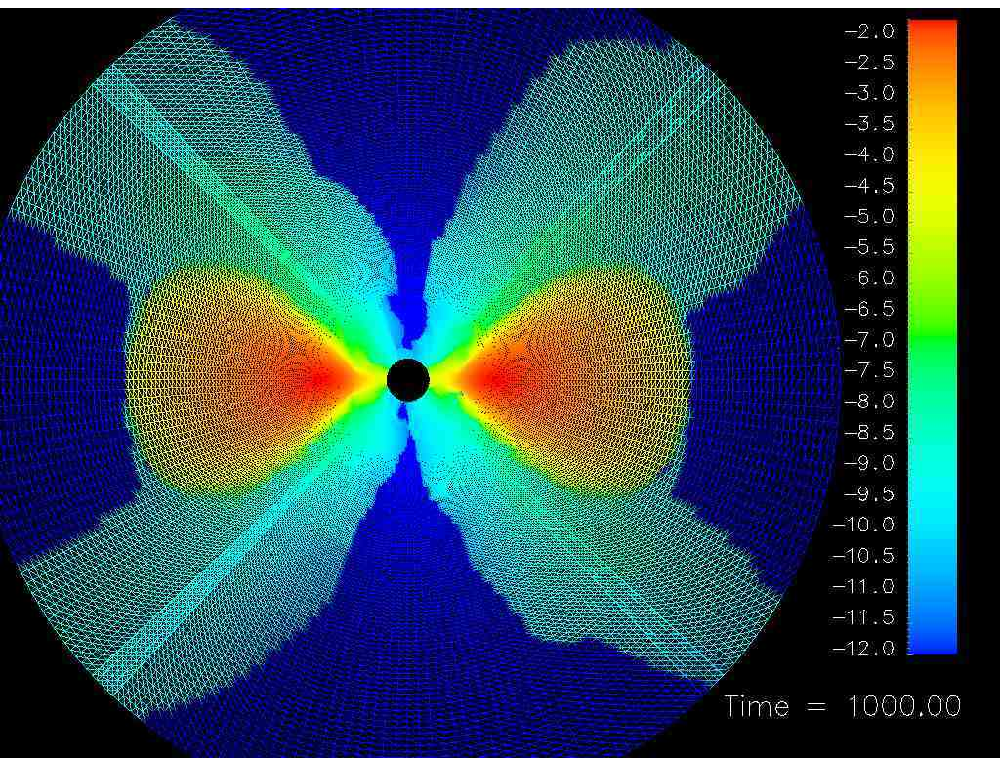} &
\includegraphics[width=\columnwidth]{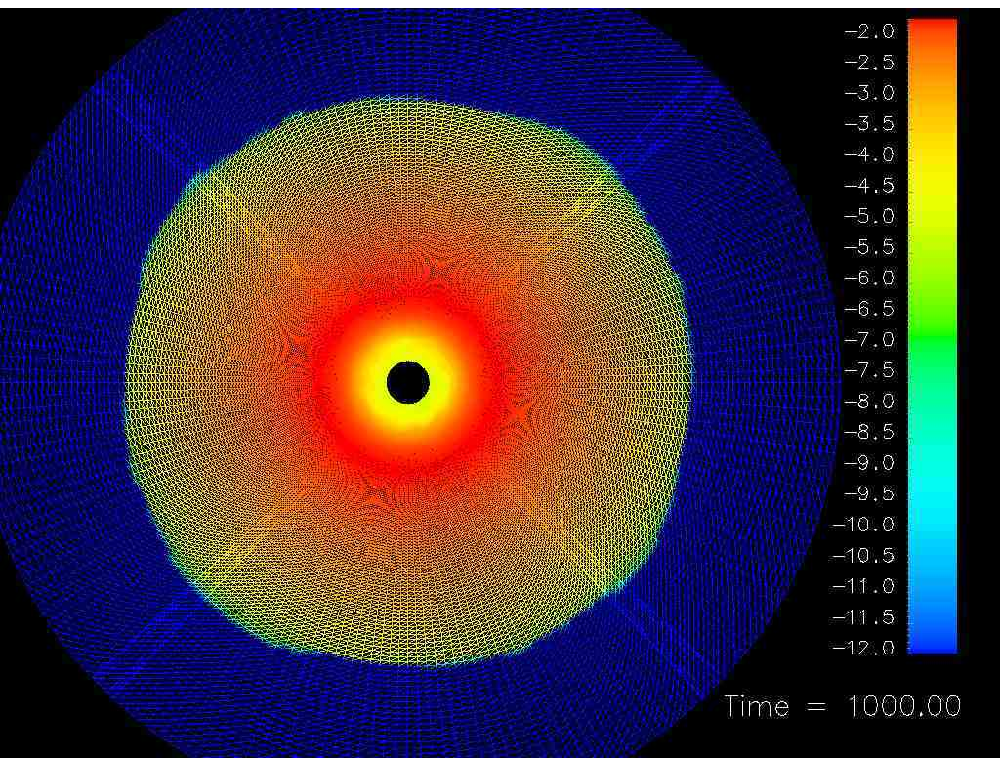} \\
\includegraphics[width=\columnwidth]{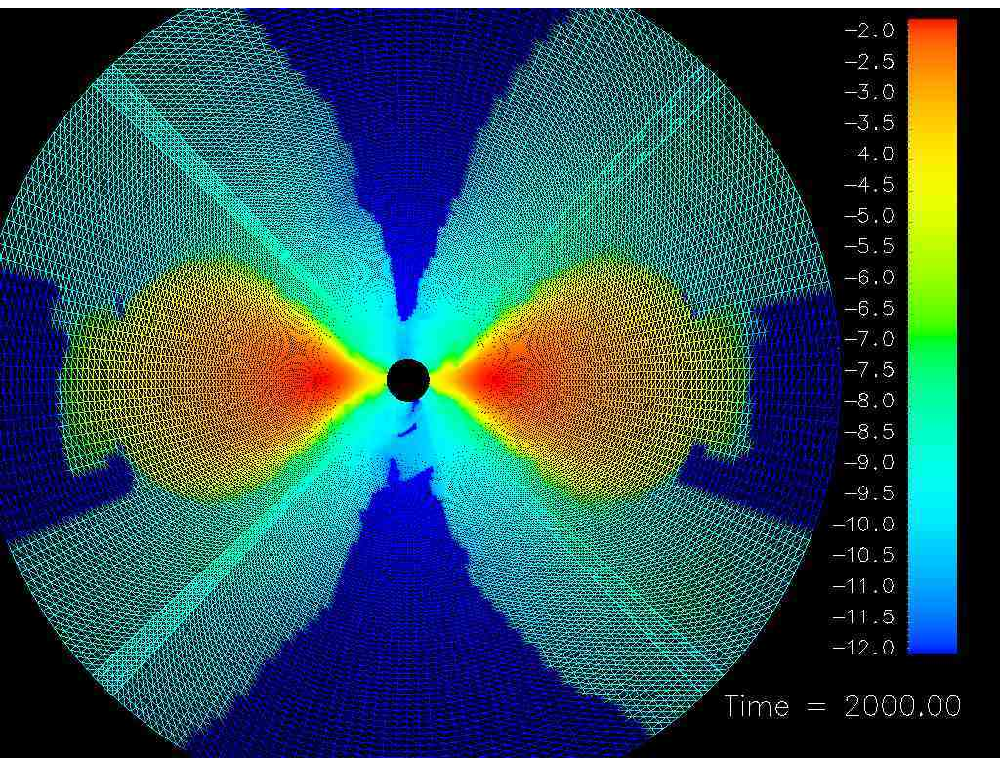} &
\includegraphics[width=\columnwidth]{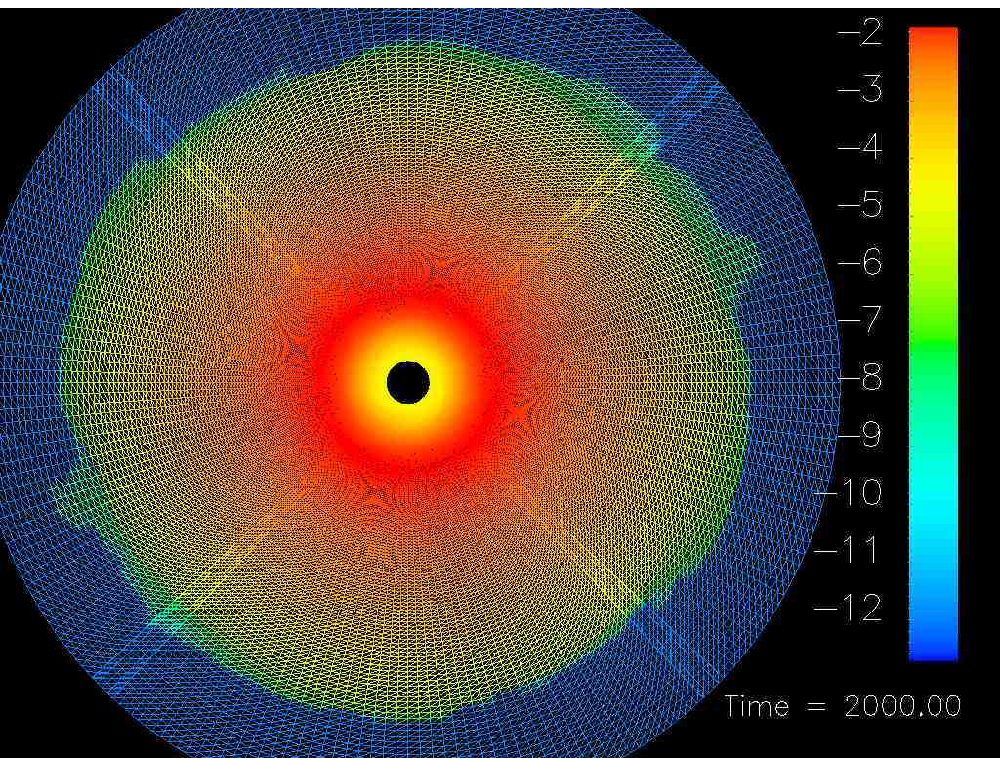} 
\end{tabular}
\caption{Equilibrium accretion torus on a Schwarzschild background, using the cubed
sphere six patches system. The plots show, for the evolution with $60 \times 60 \times 160$
cells per patch, the decadic logarithm of the density for
coordinate times $t = 0$, $1000$ and $2000$ (the rotational time of the torus at
the locus of highest density is about $200$, i.e., we are evolving for about 
$10$ rotational times). The left panel shows the cross-section surface $x = 0$ in terms of the global,
pseudo-Cartesian coordinates, and the right panel provides a view of the equatorial plane $z = 0$.
The system is stationary, so all deviations from the
initial data is an artifact due to the finite resolution, finite precision of
floating point numbers, or the use of an artificial atmosphere (though small numerical errors might
initiate physical instabilities in the disk). A small amount
of material gets unbound and is either accreted into the black hole or forms a
corona. In disks where a real physical effect like turbulence causes redistribution
of angular momentum \cite{DeVilliers03}, similar domains are present, though with much
higher densities (note that the color map covers a range of $10$ orders of magnitude in
density).}
\label{fig:torus_rho}
\end{figure*}

\begin{table}
\begin{center}
\begin{tabular}{|l|r|r|r|}
\hline
Quantity & $15 \times 15 \times 40$ & $30 \times 30 \times 80$ & 
  $60 \times 60 \times 160$ \\
\hline
Central density & $\approx 4.6\%$ & $0.055\%$ & $0.021\%$ \\
Rest mass & $\approx 0.1\%$ & $0.32\%$ & $0.14\%$ \\
Angular momentum & $\approx 0.1\%$ & $0.32\%$ & $0.14\%$  \\
\hline
\end{tabular}
\end{center}
\caption{Equilibrium accretion torus on a Schwarzschild background, using the cubed
sphere six patches system. This table shows average errors acquired per rotational time in different 
resolutions. Note that the lowest resolution case ($n = 15 \times 15 \times 40$ is under-resolved.}
\label{tab:torus_errors}
\end{table}

%% file: conclusions.tex
%%%%%%%%%%%%%%%%%%%%%%%%%%%%%%%%%%%%%%%%%%%%%%%%%%%%%%%%
\section{Summary}
\label{sec:conclusions}
%%%%%%%%%%%%%%%%%%%%%%%%%%%%%%%%%%%%%%%%%%%%%%%%%%%%%%%

The purpose of this paper was to demonstrate that multi-patch techniques can
be employed to use quasi-spherical grids for applications in
general relativistic astrophysics. In addition to a standard set of techniques
for modeling relativistic flows, we use overlapping grid patches with boundary
ghost zones, which are synchronized to data obtained from an interpolation
operation, and a transformation associated with the transition map.

The resulting scheme is able to cleanly transport shock fronts across patch
interfaces and evolve accretion disks with all the advantages of a grid based on
spherical polar coordinates, but without sharing its disadvantages like artificial
outer boundaries due to coordinate singularities, or small time steps imposed by 
the convergence of the latitudinal great circles towards the poles. In addition,
spherical grids admit to choose radial and angular resolutions independently, which
is mirrored in our approach, and makes it possible to use radial grids
with sufficient radial and angular resolution near the black hole horizon.

The main focus of this approach will be general relativistic 
single star and star + accretion disk models, but there is no fundamental reason 
why they could not be employed in binary models as well. In particular, the inspiral
phase could potentially be treated with a reduced solution error compared to Cartesian mesh refinement
setups (see, for example, \cite{Boyle2007}). Another possible application is to model a source with Cartesian mesh refinement
in the domain center, but use a cubed sphere seven or thirteen patches system to provide
outer boundary conditions and propagate gravitational radiation. Clearly, many of these 
scenarios require to solve Einstein's equations coupled to relativistic hydrodynamics
or even magnetohydrodynamics, which will be a topic of this series in the future.